\documentclass[aps,prx,reprint,amsmath,amssymb,longbibliography,floatfix]{revtex4-2}
\DeclareFontShape{OT1}{cmss}{m}{n}{<->cmss10}{}
\DeclareFontShape{OT1}{cmr}{bx}{n}{<->cmbx12}{}
\DeclareFontShape{OML}{cmm}{b}{it}{<->cmmib10}{}
\input{umsb.fd}
\DeclareFontShape{U}{msb}{m}{n}{<->msbm10}{}
\usepackage{graphicx}
\usepackage{bm}
\usepackage{hyperref}
\hypersetup{colorlinks=true,citecolor=blue,linkcolor=blue,urlcolor=blue}
\newtheorem{theorem}{Theorem}
\newtheorem{proposition}{Proposition}

\newtheorem{corollary}{Corollary}
\newenvironment{proof}{\par\noindent\textit{Proof.}\ }{\hfill$\square$\par}
\newcommand{\id}{\operatorname{id}}
\newcommand{\Tr}{\operatorname{Tr}}
\newcommand{\diag}{\operatorname{diag}}
\newcommand{\cD}{\mathcal D}
\newcommand{\cN}{\mathcal N}
\newcommand{\cM}{\mathcal M}
\newcommand{\cH}{\mathcal H}
\newcommand{\U}{\boldsymbol\Upsilon}
\newcommand{\norm}[1]{\left\Vert#1\right\Vert}
\newcommand{\ket}[1]{\lvert#1\rangle}
\newcommand{\bra}[1]{\langle#1\rvert}
\newcommand{\vket}[1]{\lvert#1\rangle\!\rangle}
\newcommand{\vbra}[1]{\langle\!\langle#1\rvert}
\begin{document}
\title{Access Resources and Stability of Error-Correction Capabilities in Continuous Strategic Codes}
\author{He Wang}
\email{dr\_hewang19@163.com}
\begin{abstract}
Continuous quantum error correction depends on which parts of a measurement record remain available to the controller at recovery. We study this temporal access using continuous strategic codes, formulated as open continuous process tensors contracted with causal instruments and conditional recovery. For monitored stabilizer codes, a common-optimal-correction criterion identifies when records can be merged without loss and yields a factorization for separated CSS inference. Terminal decision labels and causal persistent memory carry different costs: five persistent classical states reach the complete-record optimum in a finite-strength counting model, whereas its two-window diffusive counterpart has nine terminal decision labels and requires unbounded memory for exact causal recovery. Uniform lower and constructive upper bounds establish an optimal fidelity deficit of order $m^{-2}$ for $m$ persistent states in the diffusive model. Counting and diffusive stability bounds translate open-process approximation into guarantees for deployed-controller performance and retained access advantages, quantified by explicit quantizers and model-mismatch curves. The same process--instrument framework connects storage with persistent environments and irreversible noise to spacetime interrogation and record-selected paths for a logical $T$ gate. These results specify which record distinctions recovery needs, the memory cost of approximating them, and the logical protection retained under model error.
\end{abstract}
\maketitle

\section{Introduction}
Continuous monitoring and feedback protect an unknown logical state while syndrome information arrives over time. A physical controller must choose which features of that history to retain within its finite memory. We ask which distinctions affect optimal recovery and how their value changes when the noise process is known only approximately. These questions connect controller resources and process-model precision to logical protection.

Continuous syndrome measurement and Hamiltonian feedback provide mechanisms for protecting unknown encoded states~\cite{ahn2002,sarovar2004}. Optimal error tracking, hybrid classical filtering, and reduced filtering models supply realizable controller architectures~\cite{vanhandel2005,mabuchi2009,chase2008}. Practical parity-signal filters compare Bayesian processing, boxcar averaging, and threshold strategies~\cite{mohseninia2020}. Continuous correction also encompasses time-dependent encoded Hamiltonians~\cite{atalaya2021}, non-Markovian decoherence~\cite{oreshkov2007}, and experimental real-time feedback~\cite{livingston2022}. Quantum trajectories describe the associated instruments~\cite{wiseman2014,barchielli2009}, while deterministic feedback equations describe joint system--controller evolution~\cite{rosal2025}. These developments motivate a complementary optimization problem: determine the best correction achievable with a prescribed number of persistent classical states. This cardinality constraint concerns discrete retained labels, rather than the dimension of a continuously valued Bayesian sufficient statistic.

Strategic codes unify spatial encoding, adaptive interrogation, and decoding in a quantum-comb framework~\cite{strategic}. Approximate dynamical codes develop noise-adapted optimization and temporal Petz recovery~\cite{approximate_dynamical}. Continuous process tensors (cPTs), continuous matrix product states (cMPSs), and continuous instruments extend this operational language to interventions of finite duration~\cite{continuum}. Their dilation theory connects abstract continuous processes to finite-environment approximations and supplies a continuum Choi representation~\cite{dilation}. Continuous operations offer a complementary process-level description of non-Markovian monitoring~\cite{costa2025}. In this setting, the complete correction task contracts an open process with a causal instrument and a conditional decoder.

Resource-based perspectives on multi-time processes and temporal access organize restrictions on access and dynamical processing~\cite{berk2021,berk2023,temporalaccess,cprc}. For continuous correction, the quantitative question is how much acquired information must survive until decoding. We formulate this question through syndrome-conditioned optimal correction decisions for monitored stabilizer codes. The resulting compression criterion covers counting (Poisson) monitoring and diffusive (Wiener) monitoring, with a product construction for separated CSS inference. It distinguishes terminal labels from the persistent classical states needed to compute those labels causally. In a finite-strength counting model, increasing persistent memory produces a strict hierarchy. Five persistent states attain the complete-record optimum over arbitrary terminal completely positive trace-preserving (CPTP) recovery maps.

The corresponding diffusive readout reveals a different memory law for the same hidden-error schedule. Nine labels encode every terminal correction decision, yet every fixed finite-state causal controller has a uniformly positive fidelity deficit. Matching lower and constructive upper bounds establish an inverse-square deficit in the persistent-state count, or a memory requirement proportional to the inverse square root of the target accuracy. Explicit Gaussian quantizers connect this distinction to achievable recovery performance. The two modalities thus expose both exact finite saturation and controlled approximation within one continuous strategic-code task.

We also examine the stability of optimized correction under process-model error. Process-tensor characterization and tomography make model-based controller design experimentally accessible~\cite{white2020,white2022}, while tensor representations compress long-range environmental memory~\cite{dowling2024}. The relevant operational distance must respect the allowed processing of the process~\cite{zambon2024}. We establish the required connection between an open cPT coefficient norm and the reference-assisted distance between continuous record instruments. Physical factorial tails and the corresponding Wiener- or Poisson-chaos estimate yield parallel uniform H\"older bounds for fixed-strength diffusive and counting monitoring with fixed response resources, finite memory, and bounded controls. Both optimized entanglement fidelity and half diamond error obey these bounds over the same feasible controller class. The result turns process approximation into a quantitative performance guarantee, complementing channel recovery conditions and dilation continuity~\cite{beny2010,ng2010,stinespring2008}.

The same framework also addresses nonidentity logical operations. Measurement-based holonomic computation generates a logical gate through the measurement path itself~\cite{lanka_holonomic}. Mid-flight steering uses observed syndromes to modify the remaining path~\cite{lanka_steering}, and bosonic-code constructions extend this measurement-geometric approach~\cite{lanka_bosonic}. We incorporate an explicit path controller and restricted terminal recovery into the continuous strategic code. In a measurement-generated logical $T$ task, retained records select subsequent measurement trajectories and improve gate fidelity in both environmental models while the terminal syndrome decoder remains fixed.

The paper follows a single resource question from inference to implementation: which correction decisions must be retained, how much causal memory they require, and how their protection survives model error. Section~II gives the general stabilizer decision criterion and CSS factorization. Section~III resolves exact counting saturation and diffusive approximation in a common hidden-error task. Section~IV turns uniform process stability into deployed-controller and access-advantage guarantees. Sections~\ref{sec:recoveryoptimization} and \ref{sec:numerics} connect these statements to recovery optimization, spacetime interrogation, storage, and a prescribed logical gate. Process--instrument cMPS contractions, temporal Petz recovery, and terminal semidefinite programs supply the common operational boundary~\cite{fletcher2007,approximate_dynamical}.

\section{Continuous strategic codes with accessible records}
\subsection{Open process and recovery task}
Let a finite-dimensional logical system be encoded isometrically into a physical system. The input is independent of the initially prepared hidden environment:
\begin{equation}
V:\cH_L\longrightarrow\cH_S,\qquad
\mathcal E_V(\rho)=V\rho V^\dagger\otimes\sigma_E.
\end{equation}
The environment may retain memory for the full interval $[0,T]$. Quantum combs and process tensors retain the response to interventions within that interval~\cite{combs2009,pollock2018,milz2021}. A cPT is a square-summable family of time-ordered intervention coefficients whose joint temporal smoothings are compatible positive, causal finite combs~\cite{continuum,dilation}. In its open version, the quantum input and output ports are retained, so coefficients are boundary Choi matrices rather than scalar probabilities. A causal continuous instrument is a completely positive, countably additive record-valued intervention whose integrated action is deterministic and whose controls depend only on past records. Its contraction with the open cPT, including the encoding, gives the effective branch channel
\begin{equation}
 \mathcal N_y=(\U^{\partial}\star\mathcal I_y)\circ\operatorname{Ad}_V.
 \label{eq:open-instrument-pairing}
\end{equation}
The star pairs the temporal intervention slots, leaving the quantum boundaries open. We use the cPT's operation-basis Fock coefficients and Wick contraction for the general theory, and cMPS contractions or their equivalent record-conditioned propagation equations for the explicit models. The continuum Choi/cMPO representation is an alternative description~\cite{dilation}, summarized in Appendix~\ref{app:continuum-primer} alongside the cMPS and instrument contractions. For a reference measure $\mu$, write
\begin{equation}
\cN_r(\rho)=\sum_\alpha B_{\alpha,r}\rho B_{\alpha,r}^\dagger,
\quad
\int\sum_\alpha B_{\alpha,r}^\dagger B_{\alpha,r}\,d\mu(r)=I_L.
\label{eq:recordmap}
\end{equation}
The index $\alpha$ includes all inaccessible bath and field outputs. The visible label $r$ may be the full measurement path or a finite persistent state. A continuous strategic code comprises $V$, the causal instrument, the accessible record interface, and measurable conditional CPTP maps $\cD_r$. Its logical channel is
\begin{equation}
\cM=\int\cD_r\circ\cN_r\,d\mu(r).
\end{equation}
We use the unnormalized Choi operator $J(\cN)=(\id\otimes\cN)(\vket{I_L}\vbra{I_L})$, with $\vket{I_L}=\sum_a\ket{a,a}$; numerical branch states are $C_r=J_r/d_L$. Their traces sum to one.

For a common resource class $\mathfrak C_{\mathbf B}$ define
\begin{align}
F^*(\U)&=\sup_{\mathsf C\in\mathfrak C_{\mathbf B}}F_e(\cM_{\U,\mathsf C}),\\
E^*(\U)&=\inf_{\mathsf C\in\mathfrak C_{\mathbf B}}
\tfrac12\norm{\cM_{\U,\mathsf C}-\id_L}_\diamond.
\label{eq:objectives}
\end{align}
Here $F_e(\cM)=\bra\Phi(\id\otimes\cM)(\ket\Phi\bra\Phi)\ket\Phi$, where $\ket\Phi=\vket{I_L}/\sqrt{d_L}$. The two objectives generally require different recovery maps.

\subsection{Target logical tasks and admissible recovery}
\label{sec:target}
The same continuous strategic code can protect a prescribed logical operation rather than the identity. Fix the initial and terminal logical coordinates and a target unitary $U$ before optimizing the strategy. Define $\operatorname{Ad}_U(\rho)=U\rho U^\dagger$ and
\begin{align}
F_U(\mathcal M)&=\bra{\Phi_U}(\id\otimes\mathcal M)(\ket\Phi\bra\Phi)\ket{\Phi_U},\\
\ket{\Phi_U}&=(I\otimes U)\ket\Phi,\\
E_U(\mathcal M)&=\tfrac12\Vert\mathcal M-\operatorname{Ad}_U\Vert_\diamond .
\label{eq:gate-objectives}
\end{align}
The output frame alone does not fix the implementation task: an arbitrary terminal CPTP decoder can append $U$ after a storage recovery. We therefore specify an admissible recovery family $\mathfrak R$ as part of the physical task, independently of the target, and optimize only over causal strategies with $\mathcal D_r\in\mathfrak R$. A fixed syndrome decoder, optionally followed by declared Pauli byproduct removal, is one such family. Its no-error action is the identity in the fixed logical frame. Membership in this family is an additional implementability requirement; the unrestricted Knill--Laflamme criterion below does not supply it.

Exact record-resolved implementation means
\begin{equation}
\mathcal D_y\circ\mathcal N_y=p(y)\operatorname{Ad}_U,
\qquad \mathcal D_y\in\mathfrak R.
\label{eq:gate-branches}
\end{equation}
For an integrated deterministic unitary channel, this branch condition follows from positivity and its rank-one Choi operator. Approximate performance is evaluated on the entire trace-preserving channel by Eq.~\eqref{eq:gate-objectives}, without postselecting favorable records. Storage is the case $U=I_L$. Section~\ref{sec:holonomic} fixes $U$ geometrically from a measurement loop and uses one fixed decoder, so that the terminal recovery cannot supply the gate.

\subsection{Controller memory and process--instrument contraction}
A finite classical controller is represented by diagonal blocks on an auxiliary space $\mathbb C^m$. The open cPT carries the system and inaccessible environment; the continuous instrument carries detector outcomes, controller updates, and conditional controls. Their contraction, rather than either object alone, determines the record-conditioned open evolution. The operation-basis cMPS and dual instrument effect are defined in Appendix~\ref{app:continuum-primer}; their overlap is the process--instrument pairing of Eq.~\eqref{eq:primer-overlap}~\cite{continuum,dilation}. In numerical work the same pairing can be evaluated by propagating Liouville-space blocks.

For a counting instrument, let $M_j^S$ be the monitored system jump with observed outcome $j$, $P_j(c'|c)$ the corresponding stochastic controller update, $q_{c'c}$ an autonomous controller transition rate, and $H_c^S$ a conditional system control. The instrument acts on $\mathcal K_I=\cH_S\otimes\mathbb C^m$ through
\begin{align}
H_I&=\sum_c H_c^S\otimes\ket c\bra c,\\
F^I_{j,c'c}&=\sqrt{P_j(c'|c)}M_j^S\otimes\ket{c'}\bra c,\\
A^I_{c'c}&=\sqrt{q_{c'c}}I_S\otimes\ket{c'}\bra c\quad(c'\ne c).
\end{align}
Collect the monitored jumps and controller transitions into $F_a^I$. This instrument realization has no bath degree of freedom. Define $Q_I=-iH_I-\frac12\sum_a(F_a^I)^\dagger F_a^I$. Its record-averaged Liouville transfer generator is
\begin{align}
\mathbb T_I&=Q_I\otimes I+I\otimes\overline{Q_I}
+\sum_aF_a^I\otimes\overline{F_a^I}.
\label{eq:transfer}
\end{align}
Here $Q_I$ acts on state amplitudes, whereas $\mathbb T_I$ acts on vectorized density operators in the convention of Appendix~\ref{app:continuum-primer}. Equation~\eqref{eq:transfer} sums detector outcomes while preserving controller blocks; individual record branches give the corresponding selective instrument.

The process cMPS instead acts on the SE Liouville auxiliary space of dimension $(d_Sd_E)^2$, with open quantum boundaries. Its drift represents $\mathcal L_{SE}$ and its insertion matrices represent $\mathcal E_\nu^S\otimes\id_E$. Hamiltonian dynamics gives $Q_p=-iH_{SE}\otimes I+iI\otimes H_{SE}^*$. The process and dual instrument effect pair through the overlap transfer in Eq.~\eqref{eq:primer-overlap}. A regular realization evaluates this pairing by propagating their generators on the shared system with controller blocks. The bath is traced at the terminal boundary, leaving the accessible label open to form the recovery Choi branches $C_r$.

Explicitly, the retained controller blocks $\rho_c$ obey
\begin{equation}
\begin{split}
 \dot\rho_c={}&\mathcal L_{SE}(\rho_c)-i[H_c^S,\rho_c]
 -\tfrac12\sum_j\{(M_j^S)^\dagger M_j^S,\rho_c\}\\
 &+\sum_{j,b}P_j(c|b)M_j^S\rho_b(M_j^S)^\dagger\\
 &+\sum_{b\ne c}(q_{cb}\rho_b-q_{bc}\rho_c),
\end{split}
\label{eq:contracted-controller-blocks}
\end{equation}
where system operators are lifted by $I_E$ and a logical reference is a spectator. The environment enters through $\mathcal L_{SE}$, while all other terms specify the instrument and controller. This propagation evaluates the cPT--instrument contraction; Appendix~\ref{app:continuum-primer} connects the operation-basis coefficients to detector-field records~\cite{continuum,dilation}. Section~\ref{sec:stability} treats the corresponding diffusive and counting contractions.

Time-dependent measurement paths fit the same open contraction. Let $P(t)=W(t)VV^\dagger W(t)^\dagger$ and monitor $L_{a,c}(t)=\sqrt{\kappa_a}g_{a,c}(t)$, where $g_{a,c}(t)$ are prescribed rotated stabilizers selected by the persistent state $c$. The instrument effects and Liouville transfer generator are then time dependent and time ordered; the underlying environment cPT is specified independently. The frame $W(t)$ specifies the measured observables; it is not a unitary drive applied to the state. A deterministic loop requires no outcome memory. A path controller instead updates $c$ from accessible detector symbols and selects the next segment. Its state count, response clock, path amplitudes, and terminal decoder remain explicit strategy resources.

\subsection{Record-conditioned correction criterion}
\begin{proposition}[Accessible-record criterion]
Consider a standard Borel record space with a measurable, integrable finite-dimensional Choi density. Arbitrary measurable conditional CPTP recovery achieves exact correction for all logical inputs if and only if, almost everywhere,
\begin{equation}
B_{\alpha',r}^\dagger B_{\alpha,r}
=\lambda_{\alpha'\alpha}(r)I_L.
\label{eq:KL}
\end{equation}
\end{proposition}
This is the continuous record version of the strategic-code criterion~\cite{strategic} and the Knill--Laflamme condition~\cite{knill1997}. For necessity, set $R_r=J(\cD_r\circ\cN_r)\succeq0$ and $Q_\perp=I-\vket{I_L}\vbra{I_L}/d_L$. Exact recovery gives
\begin{equation}
 \int\Tr(Q_\perp R_r)\,d\mu(r)=0.
\end{equation}
The integrand is measurable and nonnegative, so it vanishes almost everywhere. Positivity then gives $R_r=p(r)\vket{I_L}\vbra{I_L}$ with measurable $p(r)\geq0$. Thus $\cD_r\circ\cN_r=p(r)\id_L$ on one common full-measure set. On each positive-weight branch, trace preservation of $\cD_r$ makes $\cN_r/p(r)$ a channel. The usual Knill--Laflamme necessity applies to that channel. If $p(r)=0$, then $\Tr J(\cN_r)=\Tr R_r=0$. Positivity of $J(\cN_r)$ implies that the branch map vanishes. For sufficiency, write $A_r=\cN_r(I_L)$ and let $P_r$ be its support. A measurable recovery is
\begin{align}
\cD_r(X)={}&\sum_\alpha B_{\alpha,r}^\dagger A_r^{-1/2}XA_r^{-1/2}B_{\alpha,r}\nonumber\\
&+\Tr[(I-P_r)X]\tau_L.
\label{eq:petz}
\end{align}
The inverse is the support inverse and $\tau_L$ is fixed. Under Eq.~\eqref{eq:KL}, diagonalizing $\lambda(r)$ shows that Eq.~\eqref{eq:petz} recovers $p(r)\rho$, where $p(r)=\sum_\alpha\lambda_{\alpha\alpha}(r)$. The positive matrix square root is continuous. Reshaping its columns into Kraus operators therefore gives a measurable Kraus field without choosing eigenvectors. The support projection and support inverse are Borel matrix functions, so the completed recovery is measurable. This construction connects the criterion to transpose and temporal Petz recovery~\cite{barnum2002,approximate_dynamical}.

If only a compressed label is accessible, all histories with that label belong to the same CP map. Their Kraus operators must satisfy the cross terms in Eq.~\eqref{eq:KL}. Keeping a record of either $I$ or $Z$ permits conditional reversal; discarding that record produces dephasing. Access is therefore part of the correction problem itself.

\subsection{Recovery decisions and causal memory}
\label{sec:general-memory}
Three memory resources enter continuous recovery. A terminal label is a
measurable function of the complete record; it specifies which decoder is
applied after monitoring. A persistent classical state must instead be updated as observations arrive. For window records $Y_1,\ldots,Y_J$, its updates
are $C_j=f_j(C_{j-1},Y_j)$, with at most $m$ persistent states at every boundary.
Detector registers are cleared before decoding, and the persistent state is
additional to the declared detector interface. Write $m_{\rm label}^*$ for
the smallest number of terminal labels attaining $F_{\rm path}$, and
$F_m^{\rm causal,*}$ for the supremum over the specified causal controllers.
Define
\begin{align}
 m_{\rm causal}^*&=\inf\{m:F_m^{\rm causal,*}=F_{\rm path}\},\nonumber\\
 m_\varepsilon^*&=\inf\{m:F_{\rm path}-F_m^{\rm causal,*}\leq\varepsilon\}.
 \label{eq:three-memory-resources}
\end{align}
The infimum of an empty set is infinity. These definitions distinguish
the information needed at decoding from the states needed to compute it
causally, and from the memory needed at a specified recovery accuracy.

For a general $[[n,k,d]]$ stabilizer code, this distinction has an explicit
decision-theoretic form. Assume classical stochastic Pauli error histories
and fixed quantum nondemolition (QND) stabilizer monitoring. Specifically,
on each monitoring segment, if $E$ is the accumulated Pauli error and
$M_y$ a monitoring Kraus operator, we require
\begin{equation}
 M_yEV=c_y(E)EV.
 \label{eq:qnd-stabilizer-assumption}
\end{equation}
Thus monitoring acts as a scalar on each transported code space.
The Pauli law may have temporal correlations. The terminal physical
output retains its orthogonal syndrome sectors, and arbitrary conditional
CPTP recovery is allowed. No intermediate quantum control changes this
Pauli-history model. Let $w_{s,\ell}(y)$ be the joint density of record $y$,
terminal syndrome $s\in\{0,1\}^{n-k}$, and logical Pauli class
$\ell\in\mathcal P_k$, where $|\mathcal P_k|=4^k$ and
$\int\sum_{s,\ell}w_{s,\ell}(y)\,d\mu(y)=1$.

\begin{theorem}[Stabilizer record compression]
\label{thm:general-pauli-memory}
Under these assumptions, the optimal complete-record entanglement fidelity
and that for a finite measurable partition $\mathcal A$ are
\begin{align}
 F_{\rm path}&=\int\sum_s\max_\ell w_{s,\ell}(y)\,d\mu(y),\nonumber\\
 F_{\mathcal A}&=\sum_{A\in\mathcal A}\sum_s
       \max_\ell\int_A w_{s,\ell}(y)\,d\mu(y).
 \label{eq:general-pauli-score}
\end{align}
Pauli corrections attain both optima over arbitrary CPTP decoders. A cell
$A$ is lossless precisely when, for each syndrome $s$, there exists one
logical class $\ell_{A,s}$ maximizing $w_{s,\ell}(y)$ for almost every
$y\in A$. With unique maximizers, $m_{\rm label}^*$ is the number of
syndrome-conditioned decision profiles realized on sets of positive
probability. The optimal half diamond errors equal $1-F_{\rm path}$ and
$1-F_{\mathcal A}$, respectively.
\end{theorem}
The maximization is performed separately in each terminal syndrome sector.
Thus the physical output supplies side information that need not be stored
in the controller. The theorem specializes the common-decoder criterion of
Theorem~\ref{thm:accessbridge}; that result treats general open processes
and supplies the access-gap stability estimate. Appendix~\ref{app:general-memory}
proves the Pauli reduction, handles tied decisions, and characterizes causal
realizability for finite observation alphabets. The latter requires
agreement under all future continuations, which can demand more states
than the final decision profiles alone.

\begin{corollary}[Separated CSS inference]
\label{cor:css-memory}
For a CSS code, suppose the $X$- and $Z$-error histories are independent,
their observations are separated, and the joint weights factor as
\begin{equation}
 w_{s_X,s_Z,\ell_X,\ell_Z}(y_X,y_Z)
 =w^X_{s_X,\ell_X}(y_X)w^Z_{s_Z,\ell_Z}(y_Z).
 \label{eq:css-posterior-factorization}
\end{equation}
Then the full-record fidelity factors, $F_{\rm path}=F_XF_Z$.
If the component decisions are unique and their record support is a product,
the exact terminal-label count is
\begin{equation}
 m_{\rm label,CSS}^*=m_{\rm label,X}^*m_{\rm label,Z}^*.
 \label{eq:css-label-product}
\end{equation}
If the component exact causal optima have attaining finite-state
controllers, their product is an attaining controller with at most
$m_{\rm causal,X}^*m_{\rm causal,Z}^*$ states. More generally, product
controllers with component deficits $\varepsilon_X,\varepsilon_Z$ achieve
a total deficit at most $\varepsilon_X+\varepsilon_Z$.
\end{corollary}
Equation~\eqref{eq:css-posterior-factorization} is the physical independence
condition: it is satisfied by separated independent bit- and phase-error
monitoring, while a general correlated $Y$-error law requires the joint
decision problem. These results apply to both counting and diffusive
readout through their respective weights $w$.

\section{Strict recovery advantages from finite records and memory}
\label{sec:capacity}
\subsection{A finite-strength continuous model}
Consider the three-qubit repetition code $V\ket0=\ket{000}$, $V\ket1=\ket{111}$ and stabilizers $S_1=Z_1Z_2$, $S_2=Z_2Z_3$. Three independent hidden flips, each of probability $0<p<1/2$, act respectively on qubits 1, 2, and 3 before, between, and after two monitoring windows. Window $j$ monitors $S_j$ for time $\tau>0$ through
\begin{align}
L_{j,\pm}&=\sqrt\kappa P_{j,\pm},\quad
P_{j,\pm}=\tfrac12(I\pm S_j),\\
K&=\kappa\tau,\qquad z=e^{-K}.
\end{align}
Within a window no other system dynamics acts. Finite-duration conditional rotations against three retained bath bits realize the hidden flips.

Contracting the counting-field cMPS over no event or at least one event of either sign gives
\begin{equation}
\begin{gathered}
M_{j,0}=\sqrt z I,\quad
M_{j,\pm}=\sqrt{1-z}P_{j,\pm},\\
\sum_rM_{j,r}^\dagger M_{j,r}=I.
\end{gathered}
\label{eq:window}
\end{equation}
For these definite-syndrome histories, event times and multiplicities convey no further information given the three-valued event. Retaining both window events is consequently sufficient for the full continuous record. The open Choi branches are
\begin{align}
C_{r_1r_2}&=\tfrac12\sum_{\bm e}w_{\bm e}
\vket{A_{\bm e,r_1r_2}}\vbra{A_{\bm e,r_1r_2}},\\
A_{\bm e,r_1r_2}&=X_3^{e_3}M_{2,r_2}X_2^{e_2}M_{1,r_1}X_1^{e_1}V,\\
w_{\bm e}&=p^{|\bm e|}(1-p)^{3-|\bm e|}.
\label{eq:capacitychoi}
\end{align}

\begin{theorem}[Strict record advantage]
Set $q_0=3p^2-2p^3$. With the same encoding, hidden flips, monitoring strength, and complete terminal physical output, let $E_j^{\mathrm{acc},*}$ denote the optimal half diamond error at access level $j$: no accessible record, one window event, or both window events for $j=0,1,2$, respectively. Then
\begin{equation}
 E_j^{\mathrm{acc},*}=z^j q_0,\qquad j=0,1,2.
\label{eq:recordhierarchy}
\end{equation}
Writing $F_j^{\mathrm{acc},*}=1-E_j^{\mathrm{acc},*}$, both improvements are strict for $p\in(0,1/2)$ and $0<K<\infty$.
\label{thm:records}
\end{theorem}
\begin{proof}
The terminal syndrome is $s=(e_1\oplus e_2,e_2\oplus e_3)$. Conditional on the accessible record $r$ and syndrome $s$, the logical Choi block is a mixture of two orthogonal Bell states, with weights $W_{r,s,0}$ and $W_{r,s,1}$. For any logical CPTP recovery $\mathcal R$ with Kraus operators $K_\beta$, set $f_P=F_e(\mathcal R\circ\operatorname{Ad}_P)$ for $P\in\{I,X,Y,Z\}$. The qubit Pauli trace identity gives
\begin{equation}
\begin{gathered}
f_P=\frac14\sum_\beta|\Tr(K_\beta P)|^2,\\
\sum_{P\in\{I,X,Y,Z\}}|\Tr(K_\beta P)|^2
=2\Tr(K_\beta^\dagger K_\beta).
\end{gathered}
\end{equation}
so trace preservation, $\sum_\beta K_\beta^\dagger K_\beta=I$, implies $\sum_P f_P=1$; also $f_P\geq0$. Thus $W_{r,s,0}f_I+W_{r,s,1}f_X\leq\max_\ell W_{r,s,\ell}$, attained by the corresponding Pauli correction. The syndrome blocks are orthogonal, so their optima add. Hence
\begin{equation}
F^*=\sum_{r,s}\max_\ell W_{r,s,\ell},\qquad
E^*=\sum_{r,s}\min_\ell W_{r,s,\ell}.
\label{eq:belloptimum}
\end{equation}
The resulting logical channel is a flip with probability $E^*$. A maximally entangled input gives its matching diamond-distance lower bound, and the convex unitary decomposition gives the upper bound. Any recovery has half diamond error at least its entanglement infidelity. Thus Eq.~\eqref{eq:belloptimum} is globally optimal for both criteria. A sign in either window and the terminal syndrome determine all three hidden bits. Only the absence of a sign retains the original ambiguity, giving Eq.~\eqref{eq:recordhierarchy}.
\end{proof}

\subsection{An exact persistent-memory hierarchy}
\begin{table}[!htbp]
\caption{Exact optima at $p=0.1$ and $K=\log5$. The persistent-state count is additional to the common detector event latch, which is cleared before terminal decoding.}
\label{tab:capacity}
\begin{ruledtabular}
\begin{tabular}{ccc}
Persistent states & Optimal $F_e$ & Optimal half diamond error\\
1 & 0.97200 & 0.02800\\
2 & 0.98720 & 0.01280\\
3 & 0.99600 & 0.00400\\
4 & 0.99744 & 0.00256\\
5 or more & 0.99888 & 0.00112
\end{tabular}
\end{ruledtabular}
\end{table}
Fix a common three-state detector event latch. It delivers and clears one window event at each of two prescribed response boundaries. The varied resource is the number $m$ of persistent classical states across the boundaries:
\begin{equation}
c_1=f_1(r_1),\qquad c_2=f_2(c_1,r_2),\qquad |\mathcal C|\leq m.
\label{eq:updates}
\end{equation}
The terminal decoder sees $c_2$ and the complete three-qubit output, after the detector latch has been cleared. The encoding, two monitoring windows, response boundaries, and detector latch are fixed; the optimized resources are the causal updates and conditional terminal CPTP recovery. No online drive acts in this model. A common finite update delay can be inserted during an idle interval without changing the statistics.

\begin{theorem}[Optimal finite-memory recovery]
Let $F_m^{\mathrm{mem},*}$ and $E_m^{\mathrm{mem},*}$ denote the optimal entanglement fidelity and half diamond error with at most $m$ persistent states. For every $0<p<1/2$ and $0<K<\infty$, optimization over causal rules in Eq.~\eqref{eq:updates} and arbitrary terminal CPTP recovery gives
\begin{equation}
\begin{gathered}
 F_1^{\mathrm{mem},*}<F_2^{\mathrm{mem},*}<F_3^{\mathrm{mem},*}\\
 <F_4^{\mathrm{mem},*}<F_5^{\mathrm{mem},*}=F_{\rm path},
 \\
 F_{\rm path}=1-z^2(3p^2-2p^3).
\end{gathered}
 \label{eq:memory-general-hierarchy}
\end{equation}
Five is the smallest persistent-state count attaining the complete-record optimum. For every memory size $E_m^{\mathrm{mem},*}=1-F_m^{\mathrm{mem},*}$, and all $m\geq5$ attain the same optimum. At $p=1/10$ and $K=\log5$, the exact values are given in Table~\ref{tab:capacity}.
\label{thm:memory}
\end{theorem}
The strict hierarchy follows from incompatible recovery decisions: the five records $00,0+,0-,+0,-0$ require five distinct, uniquely optimal correction patterns over the terminal syndromes. Any partition into fewer than five labels merges conflicting decisions; splitting a conflicting class strictly increases its score. With at least three states the first event can be stored faithfully, so the second update realizes any partition of the nine event pairs. A two-state causal construction gives the first strict gain. Appendix~\ref{app:capacity} proves these statements throughout the stated parameter range and gives positive lower bounds on the adjacent gains.

Appendix~\ref{app:capacity} gives the attaining causal maps, exact integer weights, and exhaustive optimization underlying Table~\ref{tab:capacity}. Linearity in each update kernel makes deterministic causal rules sufficient.

\begin{figure*}[t]
\includegraphics[width=\textwidth]{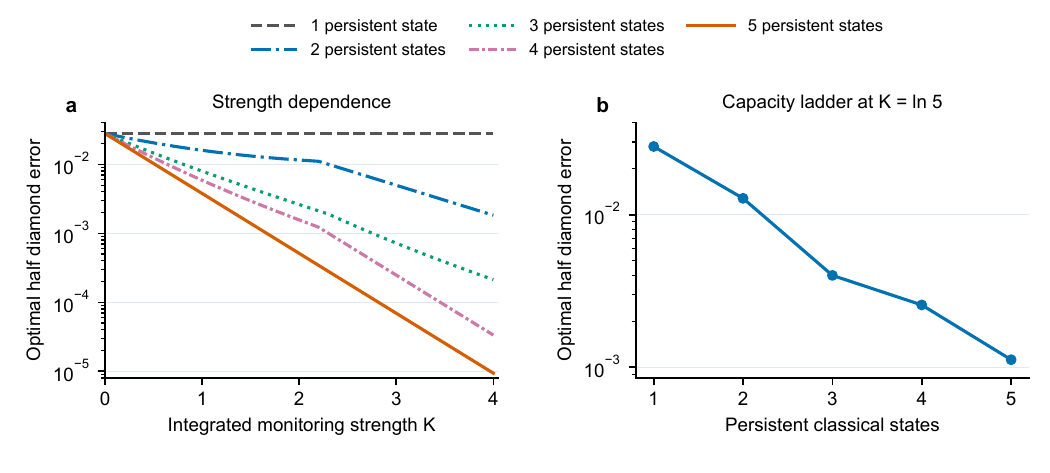}
\caption{Recovery as a function of temporal access at fixed repetition encoding and hidden-flip probability $p=0.1$. The monitoring windows, detector latch, and response boundaries are common to all protocols. Curves compare retained window events with optimized persistent-memory budgets; the five-state value attains the complete-record benchmark.}
\label{fig:capacity}
\end{figure*}

Figure~\ref{fig:capacity} evaluates this hierarchy across finite monitoring strengths. At $K=\log5$, the optimal three-state recovery has fidelity 0.996, exceeding 0.9944 from retaining a single window event. Resource monotonicity follows from simulating a coarse record with a fine record at the same budget~\cite{temporalaccess}; the correction-pattern argument establishes strictness and the exact five-state threshold for this fixed resource interface.

\subsection{Diffusive monitoring and causal memory requirements}
\label{sec:diffusive-memory}
Keep the repetition encoding, three hidden flips, two window locations,
and complete terminal physical output of Sec.~\ref{sec:capacity}. Replace
the counting instrument in window $j$ by efficient diffusive monitoring
with $L_j=\sqrt{\kappa_j}S_j$, duration $\tau_j$, and
$K_j=\kappa_j\tau_j>0$. The syndrome is constant inside each window.
Its integrated current, normalized to unit noise variance, is therefore
a sufficient window statistic:
\begin{equation}
\begin{gathered}
 Z_1|\bm e\sim N((-1)^{e_1}\mu_1,1),\\
 Z_2|\bm e\sim N((-1)^{e_2}\mu_2,1),\qquad
 \mu_j=2\sqrt{K_j}.
\end{gathered}
 \label{eq:diffusive-two-window}
\end{equation}
The two noises are independent conditional on the hidden flips.
This specifies the readout interface: a window register delivers its real
value at its response boundary; only the declared persistent state
persists between boundaries and reaches the terminal decoder. Window
registers are cleared after each update. Measurable processing of the
current window value is allowed; online quantum storage and drive are
absent. The terminal syndrome is extracted from the physical output at
recovery. Counting and diffusive monitoring thus share the hidden-error task, while
each monitoring instrument has its explicitly specified response interface.

Write $q=1-p$, $h_p=\log(q/p)$, $x=2\mu_1Z_1$, and $y=2\mu_2Z_2$.
For each syndrome choose the representative with $e_2=0$ and compare it
with its complement. Define the posterior log odds with the representative
probability in the numerator and its complement probability in the denominator:
\begin{equation}
 \begin{array}{ll}
 L_{00}=3h_p+x+y,&L_{01}=h_p+x+y,\\
 L_{10}=h_p-x+y,&L_{11}=-h_p-x+y.
 \end{array}
 \label{eq:nine-decision-odds}
\end{equation}
\begin{theorem}[Diffusive causal memory]
\label{thm:diffusive-memory}
For $0<p<1/2$ and $0<K_1,K_2<\infty$ in
Eq.~\eqref{eq:diffusive-two-window}, exactly nine terminal labels attain
the complete-record optimum. Its fidelity is
\begin{align}
 F_{\rm path}^{\rm diff}
 &=S(q^3,p^3,M)+3S(pq^2,p^2q,M),\nonumber\\
 M&=\sqrt{\mu_1^2+\mu_2^2},\nonumber\\
 S(a,b,M)&=a\Phi\!\left(M+\frac{\log(a/b)}{2M}\right)\nonumber\\
 &\quad+b\Phi\!\left(M-\frac{\log(a/b)}{2M}\right).
 \label{eq:diffusive-full-score}
\end{align}
There are constants $c,C>0$, depending only on $p,K_1,K_2$, such that
\begin{align}
 F_{\rm path}^{\rm diff}-F_m^{\rm causal,*}\geq\frac{c}{m^2}
 &\quad(m\geq1),\nonumber\\
 F_{\rm path}^{\rm diff}-F_m^{\rm causal,*}\leq\frac{C}{m^2}
 &\quad(m\geq9).
 \label{eq:diffusive-memory-bounds}
\end{align}
Consequently $m_{\rm causal}^*=\infty$ and
$m_\varepsilon^*=\Theta(\varepsilon^{-1/2})$ as
$\varepsilon\downarrow0$, at fixed nondegenerate model parameters.
The lower bound is uniform over measurable randomized finite-state
controllers and arbitrary terminal CPTP recovery. Pauli decoding attains
the corresponding fidelities with half diamond error $1-F$.
\end{theorem}
The two parallel thresholds in $x+y$ and the two in $y-x$ give nine
positive-probability decision regions. Causal computation requires a finer
first-window representation: every finite compression leaves uncertainty
about where the second-window observation will cross a decision boundary.
Appendix~\ref{app:diffusive-memory} turns this uncertainty into a uniform
one-dimensional quantization lower bound and supplies the matching
likelihood-quantization construction. For an explicit controller, divide
$u=(1+e^{-2\mu_1Z_1})^{-1}$ into $B$ equal bins and retain the bin index.
At the second boundary, the bin likelihoods and $Z_2$ determine a
syndrome-conditioned decision profile. The four binary thresholds are
ordered in two pairs, so this update needs at most nine final states.
The scheme uses at most $\max\{B,9\}$ persistent states. Its achieved
fidelity follows from Gaussian interval probabilities, with the second
readout integrated analytically; it is an attainable value within the
controller class.

\begin{figure*}[t]
\includegraphics[width=\textwidth]{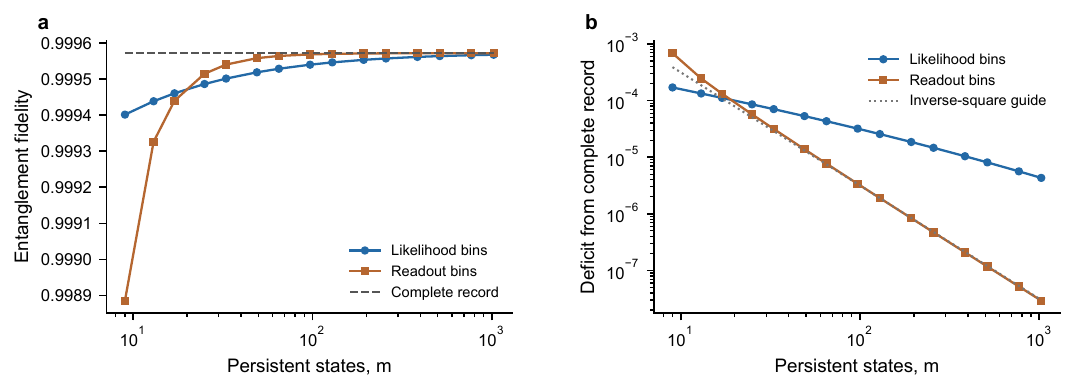}
\caption{Causal memory cost of diffusive recovery at $p=0.1$ and
$K_1=K_2=1$. (a) Achievable entanglement fidelities for uniform likelihood
bins and uniform first-readout bins, compared with the complete-record
value. (b) Their deficits from that value on logarithmic axes. Each scheme
is evaluated at 15 first-stage bin counts; the persistent-state budget is
$m=\max\{B,9\}$. Curves report achievable scores for the specified
quantizers; the inverse-square guide indicates the optimal exponent in
Theorem~\ref{thm:diffusive-memory}. The first-readout grid has infinite tail cells and uniformly
spaced internal edges on $[-10,10]$.}
\label{fig:diffusive-memory}
\end{figure*}

At $p=0.1$ and $K_1=K_2=1$, the full-record fidelity is
$0.9995724010$. Figure~\ref{fig:diffusive-memory} compares likelihood bins
with equal-width first-readout bins, using 15 memory budgets. The plotted
deficits are measured from the same analytic full-record benchmark.
Nine likelihood bins achieve $0.9994018490$; 129 readout bins achieve
$0.9995705053$, a deficit of $1.90\times10^{-6}$.
Likelihood bins give the higher fidelity at the smallest plotted budgets.
The readout-bin scheme overtakes them by $m=25$ and approaches the
inverse-square guide at large budgets. For example, increasing the
readout-bin budget from 513 to 1025 reduces the deficit by a factor
of 4.01. These curves show how quantizer geometry sets the practical
memory cost, while Eq.~\eqref{eq:diffusive-memory-bounds} fixes the
optimal asymptotic exponent.
The averaged dephasing generators have coefficients $\kappa/2$ for the
counting instrument and $\kappa$ for the diffusive instrument at equal
$\kappa$, as detailed in Appendix~\ref{app:stability}.

\section{Uniform stability under open-process approximation}
\label{sec:stability}
\subsection{The boundary needed for recovery}
The process dilation theorem establishes density of physical realizations in a closed process Fock topology~\cite{dilation}. Fix a unit-norm Pauli left--right operation basis on a qubit register of dimension $d=2^n$, with $q=d^4-1$ nonidentity labels, as in Appendix~\ref{app:stability}. Let $J^{(n)}_{\bm\nu}(\bm t)$ be the open input-output Choi coefficient for ordered insertions. Closing the input with a fixed state $\rho_*$ and the output with the trace gives scalar coefficients and their Fock norm:
\begin{equation}
\begin{split}
 u^{(n)}_{\bm\nu}(\bm t)
 &=\Tr\!\left[(\rho_*^{\mathsf T}\otimes I_o)
 J^{(n)}_{\bm\nu}(\bm t)\right],\\
 \Vert\U^{\rm cl}\Vert_{\Gamma_{\rm cl}}^2
 &=\sum_{n\geq0}\sum_{\bm\nu\in[q]^n}
 \int_{\Delta_n}|u^{(n)}_{\bm\nu}(\bm t)|^2\,d\bm t.
\end{split}
\label{eq:closednorm}
\end{equation}
Here $\Delta_n$ is the ordered time simplex and $u^{(0)}=1$ for a normalized process. This is the scalar closed-process topology in the stated operation basis.

Unknown-state recovery retains both quantum boundaries. To see the distinction, take an unencoded qubit, a bath bit initially in $I_B/2$, and
\begin{equation}
 H_\delta(t)=\frac{\pi}{2\delta}|1\rangle\langle1|_B\otimes X,
 \qquad T-\delta<t<T,
\end{equation}
with zero Hamiltonian before that interval. Every realization has a finite bath and a bounded Hamiltonian, with pulse peak proportional to $\delta^{-1}$. Use the same fixed $\rho_*$ and terminal trace for this process and the identity process. When every insertion precedes $T-\delta$, the final trace erases the pulse. Each scalar coefficient has modulus at most one, so only time tuples intersecting the final interval contribute to the difference. Their simplex volume gives
\begin{equation}
\begin{split}
\Vert\U_\delta^{\rm cl}-\U_{\id}^{\rm cl}\Vert_{\Gamma_{\rm cl}}^2
&\leq4\left(e^{qT}-e^{q(T-\delta)}\right)\\
&\leq4q\delta e^{qT}\longrightarrow0.
\end{split}
\end{equation}
The open terminal channel remains $(\rho+X\rho X)/2$, whose optimal entanglement fidelity is $1/2$ and optimal half diamond error is $1/2$. The corresponding identity-process values are 1 and 0. This fixes an elementary access class and shows why a closed norm does not by itself control its recovery task.

Retaining the boundary Choi coefficients in the same operation basis gives the open weighted norm
\begin{equation}
\norm{\U}_{\Gamma_s^\partial}^2
=\sum_{n\geq0}s^n\sum_{\bm\nu\in[q]^n}
\int_{\Delta_n}\norm{J^{(n)}_{\bm\nu}(\bm t)}_2^2\,d\bm t.
\label{eq:opennorm}
\end{equation}
Here $\norm{\cdot}_2$ is the Hilbert--Schmidt norm. The zero sector is the open no-intervention channel Choi operator, which retains the pulse's action in the example above. Open physicality means a compatible family of causal, normalized finite combs under temporal averaging, including consistency under identity insertion; positivity of unrelated margins is insufficient. No environment dimension is fixed in this definition.

\subsection{Diffusive and counting process--instrument contractions}
Take real diffusive channels with bounded, traceless monitoring operators. For each deterministic control program, the instrument is a Wick contraction with drift coefficients $a_0,a_\nu$ and noise coefficients $B_{\nu\alpha}$. The bounds are
\begin{equation}
\operatorname{Re}a_0\leq a_0^*,\quad
\sum_{\nu\ne0}|a_\nu|^2\leq a^2,\quad
\norm{B_\perp}_{\rm op}\leq b,\quad B_{0\alpha}=0.
\label{eq:coeffbounds}
\end{equation}
Appendix~\ref{app:stability} specifies the operation basis, the dual coefficients, and the resulting normalization factors; for the traceless monitoring operators used here, the resulting dual coefficient satisfies $B_{0\alpha}=0$.

For diffusive monitoring, let $\Lambda_{\rm W}$ be the nonidentity labels supporting its drift and noise coefficients. Set $q_{\rm act}^{(\rm W)}=|\Lambda_{\rm W}|$. For counting monitoring, let $\Lambda_{\rm P}$ contain the supported nonidentity labels. Include label $0$ whenever some $B_{0\alpha}\ne0$, and set $q_{\rm act}^{(\rm P)}=|\Lambda_{\rm P}|$. A specified instrument library can use fewer labels than the full operation basis.

The identification of this formal contraction with a legal instrument is established constructively. On a time cell of length $h$, set $A=-iH-\tfrac12\sum_\alpha L_\alpha^\dagger L_\alpha$ and take independent Gaussian increments of variance $h$. The local Kraus density
\begin{equation}
K_h=(I+hA+\textstyle\sum_\alpha L_\alpha\Delta W_\alpha)
(I+h^2A^\dagger A)^{-1/2}
\label{eq:weakK}
\end{equation}
satisfies $\mathbb E[K_h^\dagger K_h]=I$ exactly. Thus its contraction with each compatible finite comb is CP and normalized. Expanding in Gaussian chaos, separating same-cell collisions, and using the physical factorial tail gives convergence to the continuous Wick contraction in integrated Choi trace norm. The positive cone and normalization pass to this limit. The limiting contraction therefore defines the open continuous instrument used in the recovery task. Appendix~\ref{app:stability} gives the construction and bounds.

For a counting instrument, choose reference intensities with $0<\lambda_-\leq\lambda_\alpha(t)\leq\lambda_+<\infty$. These are reference-measure parameters, independent of the physical click rates. Define $\mathcal C_\alpha=\operatorname{Ad}_{J_\alpha}/\lambda_\alpha$, where $\operatorname{Ad}_{J}(X)=JXJ^\dagger$. Write $\mathcal G_0(X)=-i[H,X]-\frac12\sum_\alpha\{J_\alpha^\dagger J_\alpha,X\}$ and $\mathcal G=\mathcal G_0+\sum_\alpha\operatorname{Ad}_{J_\alpha}$. Relative to independent reference Poisson processes, the record density has no-click drift $\mathcal G_0+\sum_\alpha\lambda_\alpha\operatorname{id}$ and click map $\mathcal C_\alpha$. Its compensated form has drift $\mathcal G$ and noise maps $\mathcal B_\alpha=\mathcal C_\alpha-\operatorname{id}$ against $d\widetilde N_\alpha$. Expand both maps in the operation basis of Appendix~\ref{app:stability}, including the identity coefficient. Let $\Lambda$ be the set of nonidentity labels supporting any $a_\nu$ or $B_{\nu\alpha}$ in the allowed instrument library, and define
\begin{equation}
b_{\rm P}^2=\operatorname*{ess\,sup}_t
\left\Vert\bigl(\sqrt{\lambda_\alpha(t)}B_{\nu\alpha}(t)\bigr)_{\nu\in\{0\}\cup\Lambda,\alpha}\right\Vert_{\rm op}^2,
\label{eq:poisson-coefficients}
\end{equation}
with $r_{\rm P}=1+b_{\rm P}^2$.
The Poisson multiple-integral isometry gives the same weighted contraction structure, with $r_{\rm P}$ replacing $1+b^2$. Physical identification uses one continuous Poisson reference space for every mesh. In each cell, the no-click event and the first-click mark select a normalized CP cell instrument. Higher clicks remain unused reference coordinates. Cellwise Poisson projection and a global squared-moment estimate prove convergence of its Choi density in $L^2$. Positivity and normalization pass to this common-path limit. Appendix~\ref{app:stability} gives the construction and its finite-memory feedback extension.

\subsection{Uniformity over finite-state feedback}
Fix response times $0=t_0<\cdots<t_K=T$. On interval $j$ the control is the deterministic program $H_{j,c_{j-1}}(t)$ chosen by the left-end persistent state. At its end a fixed finite detector alphabet supplies $z_j$, and $c_j=f_j(c_{j-1},z_j)$. Fix at most $m$ classical states, a uniform control amplitude, the monitoring strength, and the detector interface. Encodings and conditional terminal CPTP recovery may vary within the common class. The response clock specifies the hardware timing model.

For either modality $\chi\in\{\mathrm W,\mathrm P\}$, let $\mu_\chi$ denote the common Wiener or Poisson reference measure and use the completely bounded operational distance
\begin{equation}
\begin{split}
d_{\rm rec}^{(\chi)}(\cN,\widetilde\cN)=\tfrac12\sup_{\rho_{SR}}\int
\bigl\Vert[(\cN_y-\widetilde\cN_y)\otimes\id_R]\\
\hspace{6em}(\rho_{SR})\bigr\Vert_1d\mu_\chi(y).
\end{split}
\end{equation}
The full path supplies a common mathematical representation in either case. Only the declared finite detector events reach the controller. Coarse graining and recovery are CPTP postprocessing of this output.

We state the theorem for qubit-register boundary spaces, with $d_i=2^{n_i}$ and $d_o=2^{n_o}$, using the Pauli-string operation basis defined in Appendix~\ref{app:stability}.

\begin{theorem}[Uniform process stability]
Let $\U$ and $\widetilde\U$ be compatible open physical cPTs. They share
boundary spaces, an operation basis, and the fixed resource class. Set
$D=d_i d_o$ and $C_\partial=d_i$; the class has at most
$N_{\mathbf B}\leq m^K$ distinct frozen programs.

The modality-dependent weights are
\begin{equation*}
\begin{aligned}
r_{\rm W}&=1+b^2, & r_{\rm P}&=1+b_{\rm P}^2,\\
C_{\chi,0}&=\exp(a_0^*T+a^2T/2).
\end{aligned}
\end{equation*}
Here $\chi\in\{{\rm W},{\rm P}\}$ indexes the diffusive and counting interfaces. Define
\begin{align}
\sup_{\mathsf C\in\mathfrak C_{\mathbf B}}
d_{\rm rec}^{(\chi)}(\cN_{\U,\mathsf C},\cN_{\widetilde\U,\mathsf C})
&\leq A_{\mathbf B}^{(\chi)}\norm{\U-\widetilde\U}_{\Gamma_1^\partial}^{1/2},\\
A_{\mathbf B}^{(\chi)}&=\sqrt{DN_{\mathbf B}C_\partial/2}\,C_{\chi,0}
e^{r_\chi^2q_{\rm act}^{(\chi)}T/4}.
\label{eq:uniformbound}
\end{align}
For the common feasible class in Eq.~\eqref{eq:objectives}, the change in either optimized objective is bounded by the corresponding right-hand side. The proof is common after replacing Wiener chaos by Poisson chaos and the reference record measure accordingly.
\label{thm:stability}
\end{theorem}
The access-gap theorem, its target-aware extension, and the nesting monotonicity of memory and record interfaces are modality-independent. They use only convexity and positive homogeneity of the recovery support function, CPTP contractivity, and the nesting of the feasible causal strategy classes. Petz recovery and the terminal SDP likewise act on the resulting branch Choi matrices and therefore apply without change to diffusive and counting records. In a gate task, a retained click can trigger a later path switch, while a retained diffusive symbol can select the corresponding continuation; both are instances of the same history-to-path control mechanism.
The proof combines physicality with a weighted instrument estimate. Unit-norm operation insertions into a finite comb have completely bounded trace norm at most one, giving
\begin{equation}
\norm{\U_{\Lambda_\chi}^{(n)}}_2^2\leq C_\partial^2(q_{\rm act}^{(\chi)}T)^n/n!.
\end{equation}
For $\Delta=\U-\widetilde\U$, Cauchy--Schwarz across sectors yields
\begin{equation}
\norm{\Delta_{\Lambda_\chi}}_{\Gamma_{r_\chi}^\partial}
\leq\sqrt{2C_\partial}e^{r_\chi^2q_{\rm act}^{(\chi)}T/4}
\norm\Delta_{\Gamma_1^\partial}^{1/2}.
\end{equation}
For modality $\chi$, the corresponding Wiener or Poisson contraction satisfies
\begin{equation}
\norm{\mathsf T_\chi\Delta}_{L^2(\mu_\chi;\mathrm{HS})}
\leq C_{\chi,0}\norm{\Delta_{\Lambda_\chi}}_{\Gamma_{r_\chi}^\partial}.
\end{equation}
The Wiener case is the Gaussian estimate and the counting case is its Poisson-chaos counterpart; both are proved in Appendix~\ref{app:stability}. A feedback protocol decomposes into frozen programs on disjoint events of the common reference path. Cauchy--Schwarz over those events gives the factor $\sqrt{N_{\mathbf B}}$. The Choi trace-norm bound then gives Eq.~\eqref{eq:uniformbound}, including reference-assisted inputs. Pairing the same strategies before optimization proves stability of both optimal values.

The estimate is uniform in hidden environment dimension and in every admissible controller in the fixed resource class. Open-process convergence thus controls the best achievable recovery performance throughout that class. An operational error target $\varepsilon$ is certified whenever $\delta<(\varepsilon/A_{\mathbf B}^{(\chi)})^2$. A strict access gap $G$ survives whenever $\delta<(G/(2A_{\mathbf B}^{(\chi)}))^2$. This H\"older bound provides a uniform worst-case guarantee across the entire controller class. The optimized cutoff below uses the full factorial tail, while a realization-level Lipschitz bound can give tighter estimates for specified physical models. The response clock and detector interface remain fixed within each modality.

For a specified error $\delta=\Vert\Delta\Vert_{\Gamma_1^\partial}$, retaining the physical factorial tail explicitly gives the optimized cutoff modulus
\begin{equation}
 \omega(\delta)=\inf_{n\geq0}\left[
 r_\chi^n\delta^2+4d_i^2\sum_{j>n}\frac{(r_\chi q_{\rm act}^{(\chi)}T)^j}{j!}
 \right]^{1/2}.
 \label{eq:cutoff-modulus}
\end{equation}
The same operational and optimized-value bounds hold with
$\frac12\sqrt{DN_{\mathbf B}}C_{\chi,0}\omega_\chi(\delta)$.
This modulus optimizes the separation between low sectors controlled by $\delta$ and high sectors controlled by physicality. Both coefficient-based estimates use the modality-specific active labels $\Lambda_\chi$ and the number of distinct frozen programs in the declared instrument library. The response count $K$ specifies controller decisions; time-integration grid refinement leaves it fixed. Optimizing the Wick splitting also gives a direct coefficient-norm Lipschitz bound when $b^2<1$ in the stated normalization. For one qubit with $L=\sqrt{0.1}\,Z$, $T=1$, and zero drive, that bound is $0.000911$ at $\delta=10^{-3}$.

When two models share a bounded system--environment realization, a complementary Lipschitz estimate uses their initial bath-state trace distance and integrated joint-generator diamond distance directly [Eq.~\eqref{eq:dilation-lipschitz}]. It applies uniformly to the same causal instruments and terminal interfaces, with the bath and measurement record retained during the comparison. Thus coefficient errors certify open-process approximations through Theorem~\ref{thm:stability} and Eq.~\eqref{eq:cutoff-modulus}, while realization errors certify specified physical model perturbations through Eq.~\eqref{eq:dilation-lipschitz}. As a direct access-gap test, the bath-prior example in Appendix~\ref{app:accessbridge} starts with gap $\Phi(1)-1/2=0.34134$; changing the bath prior by $0.1$ gives the exact gap $0.24134$. The realization-level estimate guarantees a remaining gap of at least $0.34134-2(0.1)=0.14134$, while the open-cPT coefficient error and its operational bound are evaluated for the same process pair. Appendix~\ref{app:stability} proves these estimates.

The implementation-level Lipschitz estimate and the fixed-process
discretization limit are not specific to Gaussian records. The Lipschitz
argument uses only a common system--environment dilation, a hybrid
generator telescope, and CPTP contractivity, so it covers both Wiener and
Poisson instruments. The discretization proof uses the corresponding local
weak instrument (Gaussian increment or no-click/click update) and the same
global contraction estimate; its local error constants are modality
dependent, while the optimized-value conclusion is the same.

\subsection{Access gaps within one continuous process}
The counting hierarchy and the diffusive stability theorem meet at the level of an open cPT instrument. Fix an encoding and a terminal quantum interface, and contract an open physical cPT with a continuous monitoring instrument. Let $C_y=J(\mathcal N_y)/d_L$ be its normalized-input, unnormalized-record Choi density under the common reference measure $\mu$. For a terminal CPTP recovery $\mathcal D$, write $\ell_{\mathcal D}(C_y)=F_e(\mathcal D\circ\mathcal N_y)$ and $h(C)=\max_{\mathcal D}\ell_{\mathcal D}(C)$. The support function $h$ is convex and positively homogeneous. A finite record interface partitions the full path into fixed Borel events $E_z$, with $C_z=\int_{E_z}C_y\,d\mu(y)$.

\begin{theorem}[Access gap and stability]
\label{thm:accessbridge}
For an open physical cPT with an integrable record Choi density,
\begin{equation}
F_{\rm path}=\int h(C_y)\,d\mu(y)\geq
F_{\rm label}=\sum_z h(C_z).
\label{eq:accessjensen}
\end{equation}
Equality holds precisely when every positive-probability label class admits one terminal CPTP decoder that maximizes $\ell_{\mathcal D}(C_y)$ for almost every path in that class. More generally, for nested finite causal record interfaces with fine labels $z$ and coarse labels $u=g(z)$, the gap is exactly
\begin{align}
F_{\rm fine}-F_{\rm coarse}
&=\sum_u\sum_{z:g(z)=u}h(C_z)\nonumber\\
&\quad-\sum_u h\!\left(\sum_{z:g(z)=u}C_z\right)\geq0.
\label{eq:finite-access-gap}
\end{align}
It is strictly positive exactly when at least one positive-probability coarse class has no decoder optimal for all its fine branches. Now take any two nested, fixed-resource access classes $\mathfrak C_0\subseteq\mathfrak C_1$ covered by Theorem~\ref{thm:stability}, including a common terminal interface. Write $G_F(\U)=F_{\mathfrak C_1}^*(\U)-F_{\mathfrak C_0}^*(\U)$ and $G_E(\U)=E_{\mathfrak C_0}^*(\U)-E_{\mathfrak C_1}^*(\U)$, where the subscript denotes the feasible strategy class. If $\delta=\Vert\U-\widetilde\U\Vert_{\Gamma_1^\partial}$, then both gaps are nonnegative and
\begin{equation}
|G_F(\U)-G_F(\widetilde\U)|,\quad
|G_E(\U)-G_E(\widetilde\U)|
\leq 2A_{\mathbf B}\sqrt\delta .
\label{eq:access-gap-stability}
\end{equation}
Here $A_{\mathbf B}$ is the common bound for the larger class. Thus every strict access advantage remains strict throughout the open-cPT neighborhood $\delta<(G/(2A_{\mathbf B}))^2$, where $G$ is its gap at the reference process.
\end{theorem}
The first statement characterizes strictness for arbitrary continuous records and finite coarse graining, without specifying a noise model. Appendix~\ref{app:accessbridge} gives the measurable-selection argument for the full-path optimum and the almost-everywhere equality criterion. The second statement applies to optimized encoders, controllers, and decoders in any nested fixed-resource classes satisfying the stability theorem; its proof and a diffusive witness are given in the same appendix.

For the witness, use the same three-qubit repetition encoding as in Sec.~\ref{sec:capacity}. Let $D$ denote its data register and $X_L=X_1X_2X_3$ its logical flip. A sensor qubit $A$ and a retained bath bit $B$ evolve under a bounded controlled $X_LX_A$ Hamiltonian for a finite preparation interval. A bath bit of value one flips both the logical state and the sensor; the bath starts in the equal classical mixture. During the following interval of length $t_m>0$, monitor the sensor continuously through the traceless operator $L=\sqrt\kappa Z_A$, with $\kappa>0$. The fixed terminal interface discards $A$ and $B$, leaving the physical data register and the declared classical record for recovery. For the integrated Wiener record $y$, the open-cPT contraction onto the physical data output is
\begin{align}
\mathcal N_y(\rho)&=\tfrac12g_+(y)V\rho V^\dagger\nonumber\\
&\quad+\tfrac12g_-(y)X_LV\rho V^\dagger X_L,\\
g_\pm(y)&=e^{\pm2\sqrt\kappa y-2\kappa t_m}.
\label{eq:diffusive-witness-instrument}
\end{align}
Hold the encoding, preparation, monitoring, and terminal physical interface fixed, with zero online drive, and optimize the terminal decoder. With the record discarded, optimal entanglement fidelity is $1/2$. A single retained bit, the sign of $y$, attains the full-path optimum $\Phi(2\sqrt{\kappa t_m})>1/2$, where $\Phi$ is the standard normal distribution function. The corresponding optimal half diamond errors are $1/2$ and $\Phi(-2\sqrt{\kappa t_m})$. Both strict gaps equal $\Phi(2\sqrt{\kappa t_m})-1/2$, and Eq.~\eqref{eq:access-gap-stability} preserves them under sufficiently small compatible open-cPT perturbations within these fixed access classes. This is a continuous diffusive instrument and a finite-memory controller on the same cPT.

Under the reference Wiener measure, $\mathcal N_y$ is an unnormalized branch map; normalization gives the state conditioned on that record. Appendix~\ref{app:continuum-primer} relates this representation to the conditional stochastic master equation. In the adaptive gate example, a retained counting event selects a later measurement path.

\begin{figure*}[t]
\includegraphics[width=\textwidth]{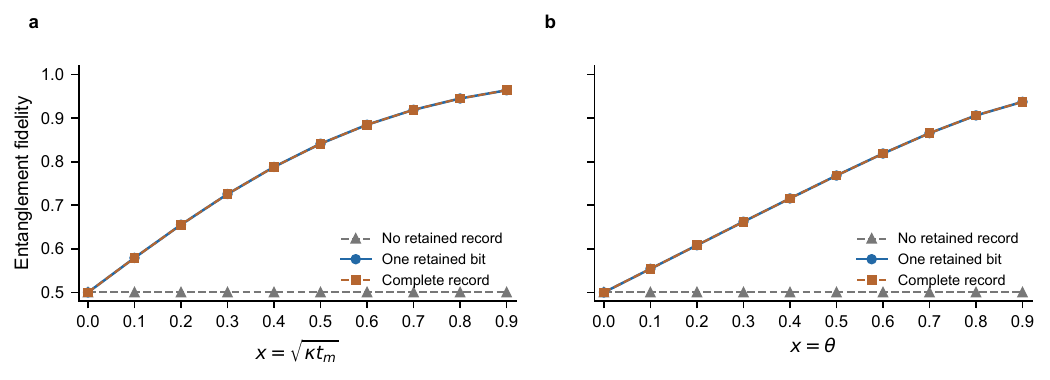}
\caption{Parallel access witnesses for diffusive and counting instruments.
Each panel uses ten matched operating points and the same three interfaces:
discarded record, one retained bit, and complete record.  The horizontal
coordinate is the dimensionless monitoring signal ($\sqrt{\kappa t_m}$ for
the diffusive panel and the counting separation parameter $\theta$ for the
counting panel).  The strict one-bit gain is a property of the record
interface; in the binary likelihood-ratio witness the one-bit and
complete-record curves coincide.}
\label{fig:stability-modalities}
\end{figure*}

The counting counterpart keeps the same encoding, equal sector priors,
and terminal interface. Replace the diffusive monitoring operator by
$J_{\rm cnt}=\sqrt{\gamma_+}|0\rangle\langle0|_A
+\sqrt{\gamma_-}|1\rangle\langle1|_A$, with
$\gamma_\pm=\bar\gamma(1\pm\theta)$ and $0\leq\theta<1$.
Conditional on the sensor sector, the total count $N$ is Poisson with mean
$\gamma_\pm t_m$ and is sufficient for the complete counting record.
Writing $P_\pm(n)=e^{-\gamma_\pm t_m}(\gamma_\pm t_m)^n/n!$, the
likelihood-ratio bit and its fidelity are
\begin{align*}
 b(N)&=\mathbf1\!\left\{
 N\log\frac{1+\theta}{1-\theta}-2\bar\gamma\theta t_m\geq0\right\},\\
 F_{\rm bit}=F_{\rm path}
 &=\frac12\sum_{n=0}^{\infty}\max\{P_+(n),P_-(n)\}.
\end{align*}
The discarded-record value remains $1/2$. Figure~\ref{fig:stability-modalities}(b)
uses $t_m=1$ and $\bar\gamma=2$ in the chosen time units, with
$\theta=0,0.1,\ldots,0.9$. Every nonzero signal gives a strict one-bit gain
in both modalities. The shared memory and terminal interface identify this
gain as a consequence of temporal access.

\subsection{Target-aware access and stability}
\begin{corollary}[A fixed logical task]
\label{cor:target}
Fix $U$ and a nonempty compact family $\mathfrak R$ of CPTP decoders, common to all records and independent of the process approximation. In Theorem~\ref{thm:accessbridge}, replace $h$ by
\begin{equation}
 h_{U,\mathfrak R}(C)=\max_{\mathcal D\in\mathfrak R}
 \Tr\!\left[\ket{\Phi_U}\bra{\Phi_U}(\id\otimes\mathcal D)(C)\right].
\end{equation}
The access inequality and its common-optimal-decoder equality criterion remain valid. Over any common strategy family covered by Theorem~\ref{thm:stability}, the optimized $F_U$ and $E_U$ change by at most $A_{\mathbf B}\sqrt\delta$, and their nested-class gaps change by at most $2A_{\mathbf B}\sqrt\delta$.
\end{corollary}
The family may be nonconvex, including a fixed decoder or a finite set of permitted Pauli corrections. Linearity, compactness, and CPTP contractivity establish the result (Appendix~\ref{app:targetholonomy}). Storage and gate tasks therefore share access and stability structure with their respective physical recovery constraints. Adaptive steering changes subsequent measurements, so its monotonicity uses nested causal strategy families encompassing those different instruments.

Deterministic time-dependent monitoring is already allowed by the coefficient estimates in Appendix~\ref{app:stability}. The same frozen-program proof also covers controller-selected measurement segments if their operators are bounded and traceless, all programs obey common bounds in Eq.~\eqref{eq:coeffbounds}, and they use the same number of real quadratures and fixed causal readout functions on a common reference Wiener space. Each frozen controller history fixes both $H(t)$ and $L_a(t)$; the number of histories and hence the factor $m^K$ are unchanged.

\subsection{Certified performance under model mismatch}
\label{sec:certified-deployment}
The stability estimates translate a controller designed on a nominal
process into a guarantee on the actual process. Let $\U_0$ be the nominal
model, $\U_1$ the true model, and fix the same instrument and recovery
resources in both. Suppose a uniform modulus $\eta$ satisfies
\begin{equation}
 |F(\U_1,\mathsf C)-F(\U_0,\mathsf C)|\leq\eta
 \quad\hbox{for every feasible strategy }\mathsf C.
 \label{eq:deployment-uniform-modulus}
\end{equation}
Theorem~\ref{thm:stability}, the cutoff modulus, or the common-realization
estimate in Eq.~\eqref{eq:dilation-lipschitz} supplies such an $\eta$.
Their operational-distance bounds also control the strategywise half
diamond error by the same modulus.

\begin{corollary}[Deployment guarantees]
\label{cor:deployment}
A fixed nominal controller $\mathsf C_0$, including its terminal decoder,
has actual fidelity at least $F(\U_0,\mathsf C_0)-\eta$.
If it is $\varepsilon_{\rm opt}$-optimal in its nominal resource class,
then
\begin{equation}
 F^*(\U_1)-F(\U_1,\mathsf C_0)
 \leq 2\eta+\varepsilon_{\rm opt}.
 \label{eq:deployment-regret}
\end{equation}
For a smaller access class $\mathfrak C_0\subseteq\mathfrak C_1$
and a fixed controller $\mathsf C\in\mathfrak C_1$,
\begin{equation}
 F(\U_1,\mathsf C)-F^*_{\mathfrak C_0}(\U_1)
 \geq F(\U_0,\mathsf C)-F^*_{\mathfrak C_0}(\U_0)-2\eta.
 \label{eq:deployed-access-advantage}
\end{equation}
The analogous cost inequalities hold for half diamond error.
\end{corollary}
The first statement is Eq.~\eqref{eq:deployment-uniform-modulus} for one
strategy. Taking its supremum gives the optimized-value bound; applying
these two bounds proves Eqs.~\eqref{eq:deployment-regret} and
\eqref{eq:deployed-access-advantage}. The latter certifies a specified
feasible controller directly from its achieved nominal fidelity.

For the two-window models, a concrete model error changes the hidden-flip
probability from the design value $p$ to $p'$, keeping the monitoring
instrument and controller fixed. The entire process is a mixture of the
same eight hidden histories, so a modulus uniform in all strategies is
\begin{equation}
\begin{split}
 \eta(p,p')=\frac12\sum_{\bm e\in\{0,1\}^3}
 \Bigl|&p^{|\bm e|}(1-p)^{3-|\bm e|}\\
       &-(p')^{|\bm e|}(1-p')^{3-|\bm e|}\Bigr|.
\end{split}
 \label{eq:hidden-prior-tv}
\end{equation}
It is the trace-distance error of a common classical bath preparation,
and therefore a direct instance of Eq.~\eqref{eq:dilation-lipschitz}.
It holds for both counting and diffusive response interfaces.

At $p=0.1$ and $p'=0.101$, $\eta=0.002427301$. The fixed counting
five-state controller at $K=\log5$ has actual fidelity $0.9988583041$;
its certified lower bound is $0.9964526990$. Its advantage over optimal
discarded-record recovery is $0.0274007021$, certified to be at least
$0.0220253980$. The fixed 129-state diffusive readout-bin controller at
$K_1=K_2=1$ has actual fidelity $0.9995641149$, certified to be at least
$0.9971432043$. Its actual and certified access advantages are
$0.0281065129$ and $0.0227159033$, respectively.

\begin{figure*}[t]
\includegraphics[width=\textwidth]{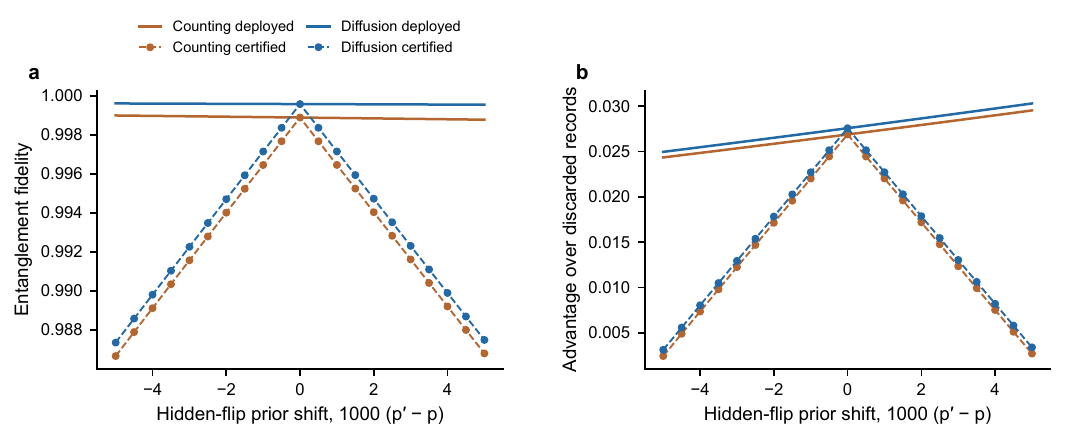}
\caption{Performance retained under hidden-error model mismatch.
Controllers are designed at $p=0.1$ and deployed unchanged at 21 values
$p'\in[0.095,0.105]$. Counting monitoring uses the exact five-state controller at
$K=\log5$; diffusive monitoring uses 129 uniform readout bins at $K_1=K_2=1$.
(a) Actual deployed fidelities and the lower certificates $F_{\rm nominal}-\eta$.
(b) Actual advantages over optimal discarded-record recovery and the
certificates $G_{\rm nominal}-2\eta$. Solid curves are actual values;
dashed curves with markers are certified lower bounds. The total-variation
modulus in Eq.~\eqref{eq:hidden-prior-tv} applies uniformly to both
monitoring interfaces.}
\label{fig:deployment}
\end{figure*}

Figure~\ref{fig:deployment} evaluates 21 true-model priors with the
nominal updates and decoders held fixed. Each curve uses the instrument
strength and controller resource stated above.
Appendix~\ref{app:diffusive-memory} gives the Gaussian deployment score.
The certificates concern performance and access advantages; an integer
exact-saturation threshold is a separate structural property.

\section{Recovery and constrained strategy optimization}
\label{sec:recoveryoptimization}
\subsection{Approximate recovery of open cPT branches}
Contracting the cPT with an encoding and a causal instrument leaves a quantum channel with an accessible classical output. Approximate dynamical QEC supplies two complementary tools at this boundary: temporal Petz recovery and alternating optimization of encoders, checks, and decoders~\cite{approximate_dynamical}. Both act on the finite-dimensional record-conditioned Choi blocks produced by this contraction.

Equation~\eqref{eq:petz}, with support completion, supplies a CPTP candidate on every branch, attaining exact recovery under Eq.~\eqref{eq:KL}. Its computed overlap with $C_r$ gives a constructive baseline and an initialization for decoder optimization. Hardware-restricted tasks additionally impose membership in the declared recovery family.

For fixed $C_r=J(\mathcal N_r)/d_L$, unrestricted terminal recovery is a convex subproblem. With reference--physical indices $(a,i)$ and logical-output--physical-input decoder Choi ordering,
\begin{align}
[W_r]_{(a,i),(b,j)}&=\overline{[C_r]_{(a,i),(b,j)}}/d_L,\\
\max_{D_r\succeq0,\,\Tr_LD_r=I_B}\Tr(W_rD_r)
&=\min_{I_L\otimes Y_r\succeq W_r}\Tr Y_r.
\label{eq:SDP}
\end{align}
This Choi ordering gives the conjugate $\overline C_r=C_r^{\mathsf T}$ in the objective. The channel-adapted decoder program~\cite{fletcher2007} applies separately to the effective branches of a strategic code~\cite{strategic,approximate_dynamical}. A target unitary changes the linear objective while preserving the CPTP constraints. The optimum bounds every restricted decoder at the same encoding, instrument, and access interface.

Unrestricted CPTP recovery is closed under logical-unitary composition. Substitution of $\mathcal D'_r=\operatorname{Ad}_{U^\dagger}\circ\mathcal D_r$ gives $F^*_{U,\mathrm{CPTP}}=F^*_{I,\mathrm{CPTP}}$ and $E^*_{U,\mathrm{CPTP}}=E^*_{I,\mathrm{CPTP}}$ at a fixed pre-recovery instrument. Thus the unrestricted comparison measures recoverable information, while the admissible family $\mathfrak R$ defines physical gate implementation.

\subsection{Finite-memory and hardware-constrained design}
For a fixed discretized process, convex Choi positivity and causal normalization constraints give semidefinite subproblems for alternating channel and instrument blocks~\cite{approximate_dynamical}. Exact block updates produce a nondecreasing fidelity sequence; joint optimization generally remains nonconvex under finite cardinality, isometric encoding, or parameterized-path constraints. The open cPT contraction supplies each feasible controller's branches for comparison between realizable recovery and the unrestricted terminal SDP.

Theorem~\ref{thm:stability} controls optimized task values over a common feasible class. Stability of a selected optimizer additionally requires regularity, such as the nondegeneracy hypotheses used in approximate dynamical QEC~\cite{approximate_dynamical}.

Table~\ref{tab:petzcase} compares three recoveries for the same independent-$X$ storage branches at $\kappa T=8$. The support-completed transpose map gives a constructive Petz baseline. A fixed syndrome instrument with controller-conditioned Pauli-frame selection supplies a restricted recovery, and the complete-output SDP bounds their common interface.

With zero drive, independent $X$ noise and syndrome monitoring preserve a direct sum of two-Bell-state logical mixtures. Equation~\eqref{eq:belloptimum} then makes the syndrome-conditioned Pauli frame an unrestricted optimum, explaining the equality in the monitoring row. The feedback drive creates coherences between syndrome sectors, allowing the complete-output SDP to use additional quantum information.

\begin{table}[!htbp]
\caption{Numerical approximate-recovery case at $\kappa T=8$. The encoding, controller, noise rates, and terminal interface are fixed across columns. Pauli frame denotes the fixed syndrome instrument with controller-conditioned Pauli-frame selection; SDP is unrestricted recovery.}
\label{tab:petzcase}
\begin{ruledtabular}
\begin{tabular}{lccc}
Protocol & Petz & Pauli frame & SDP\\
\hline
monitoring & 0.96362939 & 0.98033980 & 0.98033980\\
with feedback & 0.97706456 & 0.98758096 & 0.98768745
\end{tabular}
\end{ruledtabular}
\end{table}

\section{Spacetime interrogation and continuous recovery}
\label{sec:numerics}
\subsection{A common code, controller, and recovery class}
We use the repetition code above for a persistent four-qubit interacting bath and for independent irreversible $X$ flips. The controller monitors the three nontrivial syndrome projectors with $J_s=\sqrt\kappa P_s$. Its thirteen states comprise one idle state and four stages per syndrome: two waiting and two driven stages. A click in the idle state selects the syndrome cycle; clicks during a busy cycle do not restart it. The response model and drive amplitude remain fixed throughout the strength scan; Appendix~\ref{app:numerics} gives their parameters.

We compare four protocols: terminal recovery without monitoring; monitoring with the final label discarded; monitoring with the label retained and zero drive; and monitoring with drive and label retained. The two no-drive monitored protocols share the same dynamics. Each protocol independently optimizes a terminal CPTP decoder on the complete three-qubit output and its permitted label.

The cPT--instrument contraction propagates the logical reference with the system, bath, and controller. Tracing the bath at the final boundary produces the branches $C_c$, to which Eq.~\eqref{eq:SDP} applies. Appendix~\ref{app:limits} establishes the supporting approximation results, and Appendix~\ref{app:numerics} specifies the numerical methods.

\subsection{Persistent bath and irreversible flips}
The four-qubit bath begins in $\ket{0000}$ and evolves under
\begin{align}
H_4={}&\tfrac12\sum_{j=1}^4\Omega_j Z_{E_j}
+J_E\sum_{j=1}^3X_{E_j}X_{E_{j+1}}\nonumber\\
&+g\sum_{i=1}^3X_iX_{E_i}+g_{\rm sh}\sum_{i=1}^3X_iX_{E_4}
\end{align}
The parameters are
\begin{gather}
\bm\Omega=(0.8,1.0,1.2,1.4),\\
(g,g_{\rm sh},J_E)=(0.35,0.12,0.4).
\end{gather}
The Markovian benchmark has $L_i=\sqrt{0.1}X_i$. With no feedback, each flip history remains in a definite syndrome sector. For its reference-system state $\rho_{\bm e}$,
\begin{equation}
\mathcal D[J_s](\rho_{\bm e})=0.
\label{eq:nobackaction}
\end{equation}
The ensemble state is a convex mixture of such histories, so nonselective monitoring leaves it unchanged. Any improvement from monitoring in this model must use the retained record or feedback. In the interacting bath, monitoring can also change coherent propagation between syndrome sectors.

\begin{figure*}[t]
\includegraphics[width=\textwidth]{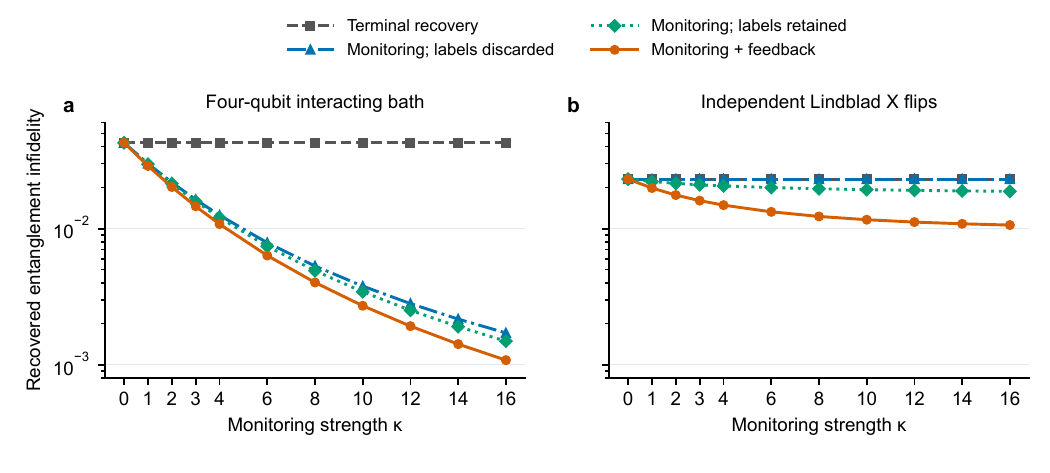}
\caption{Recovered entanglement infidelity for the same three-qubit code under (a) a persistent four-qubit bath and (b) independent Lindblad $X$ flips. The eleven matched monitoring strengths share a fixed persistent-state budget, drive amplitude, and response times. The four protocols distinguish nonselective backaction, accessible memory, and drive, with separately optimized terminal recovery. In (b), terminal-only and discarded-label curves coincide by Eq.~\eqref{eq:nobackaction}.}
\label{fig:benchmark}
\end{figure*}

Figure~\ref{fig:benchmark} identifies two distinct sources of protection. In the interacting bath, monitoring improves recovery even when its label is discarded, consistent with suppression of coherent error propagation. In the independent-flip model, the discarded-label and terminal-only curves coincide throughout the scan. All monitoring gains there come from retained records and feedback. The bath gains therefore combine modified physical evolution with actionable information, whereas the independent-flip gains isolate the value of temporal access. Table~\ref{tab:benchmark} gives the shared reference point $\kappa=8$.

\begin{table}[!htbp]
\caption{Optimal terminal entanglement fidelities at $\kappa=8$. The same physical-output recovery class is used for all protocols in each noise model.}
\label{tab:benchmark}
\begin{ruledtabular}
\begin{tabular}{lcc}
Protocol & Four-qubit bath & Independent $X$\\
Terminal recovery & 0.95719345 & 0.97684516\\
Monitoring, discard label & 0.99468023 & 0.97684516\\
Monitoring, retain label & 0.99508582 & 0.98033980\\
Monitoring and feedback & 0.99597156 & 0.98768745
\end{tabular}
\end{ruledtabular}
\end{table}
Relative to the retained-label, no-drive protocol, feedback reduces entanglement infidelity by $18.02\%$ for the bath and $37.37\%$ for independent flips. At $\kappa=16$ the corresponding fidelities reach 0.99891762 and 0.98935985. The scan shows positive drive gains at every nonzero tested strength under the fixed response model.

Recovery restricted by a terminal syndrome measurement also preserves the positive feedback gain at $\kappa=8$. Complete-output recovery additionally retains inter-syndrome coherence that such a measurement could erase.

\subsection{A two-round spacetime interrogation protocol}
\label{sec:spacetime-numerics}
The finite-strength model of Sec.~\ref{sec:capacity} connects continuous monitoring to spacetime interrogation in strategic codes~\cite{strategic}. Successive windows measure $Z_1Z_2$ and $Z_2Z_3$, retaining their events for terminal recovery. This common repetition encoding supports a different schedule and error placement from the autonomous feedback benchmarks.

Exact contraction of the three-event window instruments gives the optimal fidelities
\begin{equation}
 \begin{gathered}
 F_0^{\mathrm{acc},*}=1-q_0,\qquad F_1^{\mathrm{acc},*}=1-e^{-K}q_0,\\
 F_2^{\mathrm{acc},*}=1-e^{-2K}q_0,
 \end{gathered}
 \label{eq:spacetime-validation}
\end{equation}
where $q_0=3p^2-2p^3$ and the access levels are those of Theorem~\ref{thm:records}. Keeping both windows doubles the exponential decay rate of the residual error relative to one window. Table~\ref{tab:spacetime} gives the $p=0.1$ slice.

\begin{table}[!htbp]
\caption{Two-round spacetime interrogation at hidden-flip probability $p=0.1$. Entries are the analytic optimal entanglement fidelities with complete physical-output recovery. The final row uses the exact projective instrument.}
\label{tab:spacetime}
\begin{ruledtabular}
\begin{tabular}{cccc}
$K$ & Discarded & First window & Both windows\\
0 & 0.97200000 & 0.97200000 & 0.97200000\\
1 & 0.97200000 & 0.98969938 & 0.99621061\\
2 & 0.97200000 & 0.99621061 & 0.99948716\\
4 & 0.97200000 & 0.99948716 & 0.99999061\\
8 & 0.97200000 & 0.99999061 & 0.999999997\\
$\infty$ & 0.97200000 & 1 & 1
\end{tabular}
\end{ruledtabular}
\end{table}

\begin{figure*}[t]
\includegraphics[width=\textwidth]{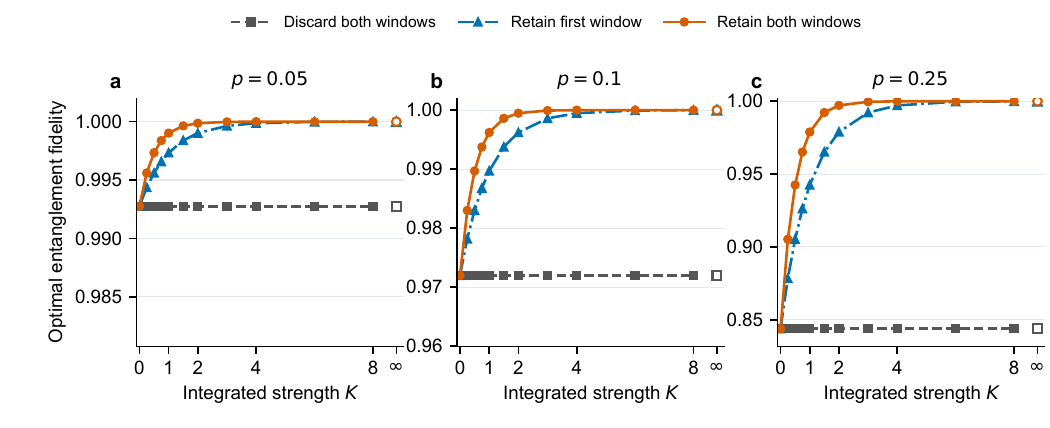}
\caption{Two-round spacetime interrogation for three hidden-flip probabilities. Each panel compares discarded records, the first-window record, and both-window records over measurement strength. Lines show the analytic optima; filled markers denote finite-strength values and open markers the exact projective instrument. Panels use separate vertical scales.}
\label{fig:spacetime}
\end{figure*}

The records distinguish histories with the same terminal syndrome: $\bm e=(1,1,0)$ and $(0,0,1)$, for example, have opposite signs in the first check. In the projective limit, the no-click operator vanishes and the instrument becomes the discrete interrogation protocol with exact recovery. Figure~\ref{fig:spacetime} displays this approach for three hidden-flip probabilities, separating temporal access from noise strength.

\section{A measurement-path gate as a continuous strategic code}
\label{sec:holonomic}
Measurement-induced holonomies provide a concrete nonidentity task for the same open cPT contraction. Rotating the measured stabilizers transports a degenerate code space by the quantum Zeno effect~\cite{lanka_holonomic}. Record-dependent steering can change the remaining path after a detected excursion~\cite{lanka_steering}. Bosonic-code realizations show that the same measurement-geometric principle extends beyond finite-dimensional repetition codes~\cite{lanka_bosonic}. Here the holonomic construction is an application of the framework: the open process supplies the physical dynamics, the time-dependent instrument specifies the path, and the accessible record and decoder define the logical task.

Let $P_0=VV^\dagger$, $P(t)=W(t)P_0W(t)^\dagger$, $W(0)=I$, and $P(T)=P_0$. In the ideal Zeno limit, the gate in fixed endpoint coordinates is
\begin{equation}
\begin{split}
U_{\rm geo}={}&V^\dagger W(T)V\,
\mathcal P\exp\!\left[-\int_0^T A(t)\,dt\right],\\
A(t)={}&V^\dagger W(t)^\dagger\dot W(t)V .
\end{split}
\label{eq:holonomy}
\end{equation}
The endpoint factor distinguishes nontrivial holonomy from a trivial integrated connection in nonperiodic frames. Only the measured operators are varied physically; $W(t)$ is a frame parameterization.

For a minimal example, take $V\ket0=\ket{000}$, $V\ket1=\ket{111}$ and adapt the repetition-code path of Ref.~\cite{lanka_holonomic}:
\begin{align}
H&=Z_1Z_2Z_3,\quad R=X_3,\quad \phi(t)=2\pi t/T,\\
W_\theta(\phi)&=e^{i\theta\phi H/(2\pi)}e^{i\phi R},\\
g_1(\phi)&=Z_1Z_2,\quad
 g_2(\phi)=W_\theta(\phi)Z_2Z_3W_\theta(\phi)^\dagger .
\label{eq:minimal-path}
\end{align}
Continuous monitoring with $L_a=\sqrt\kappa g_a$ and no environmental noise or applied Hamiltonian gives, after discarding the measurement records,
\begin{equation}
\dot\rho=\kappa\sum_{a=1}^2[g_a(t)\rho g_a(t)-\rho].
\label{eq:holonomic-lindblad}
\end{equation}
Equation~\eqref{eq:holonomy} gives $U_{\rm geo}=e^{i\theta Z_L}$; choosing $\theta=-\pi/8$ yields the logical $T_L=\diag(1,e^{i\pi/4})$ up to global phase. Every strength and path uses the same standard repetition-code decoder, defined in Eq.~\eqref{eq:fixed-gate-decoder}.

The open reference-assisted channel calculation gives $F_{T_L}=0.933454$, $0.981409$, and $0.995217$ at $\kappa T=256$, $1024$, and $4096$, respectively, without environmental noise. Frozen measurements and zero monitoring then leave the logical identity, whose target fidelity is $\cos^2(\pi/8)=0.853553$. Weak rotating monitoring can perform below that control: finite-strength backaction, not just the endpoint geometry, determines the implemented channel.

\begin{figure*}[t]
\includegraphics[width=\textwidth]{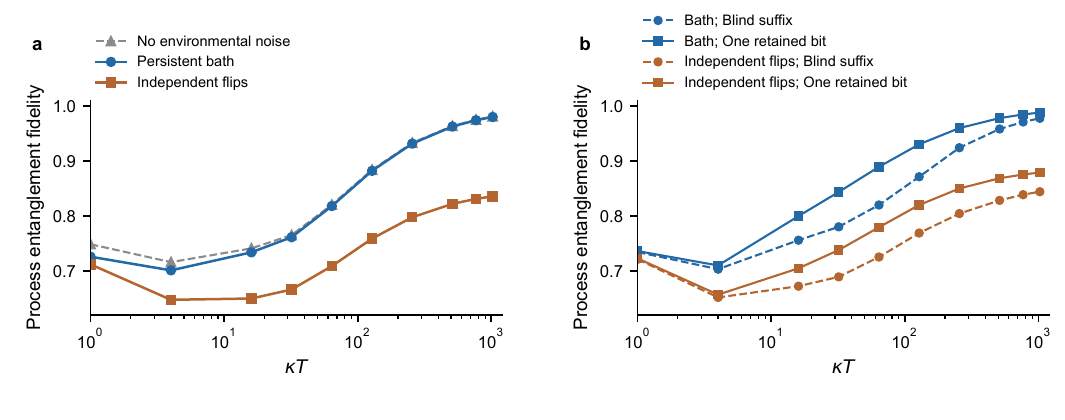}
\caption{A logical $T$ task with fixed terminal syndrome correction. (a) The fixed-speed loop under no environmental noise, a persistent four-qubit bath, or independent flips. The logarithmic axis omits the zero-strength baseline. (b) A common pause/readout window and suffix library compare one-bit path control with the best record-blind continuation. Markers denote target entanglement fidelities; lines connect computed points or paired policies. The two panels use different time schedules.}
\label{fig:holonomic}
\end{figure*}

\paragraph{The same gate in two environments.}
We next add either the persistent four-qubit bath or the independent $X$ flips of Sec.~\ref{sec:numerics}, leaving the loop and terminal decoder unchanged. The measurement path generates the gate while the specified noise dynamics remains active; in the bath case this includes the joint system--environment Hamiltonian. At $\kappa T=256$, the gate fidelities are $0.932111$ and $0.797997$ for the bath and independent flips, respectively; at $1024$ they are $0.980579$ and $0.836102$ [Fig.~\ref{fig:holonomic}(a)]. These channel-level results distinguish coherent environmental evolution from irreversible flips under the same measurement path.

\paragraph{A record-controlled continuation.}
To use the path as a strategic-code resource, we insert a common detector window from $7T/16$ to $9T/16$ halfway around the loop. During this window the measured operators are held fixed. We read the first signed click of parity detector 2, including a no-click outcome, using
\begin{equation}
J_{2,s}=\sqrt{2\kappa}\,P_{2,s},\qquad
P_{2,s}=(I+s g_2)/2,\quad s=\pm1.
\label{eq:gate-click}
\end{equation}
Since $\sum_s\mathcal D[J_{2,s}]=\kappa\mathcal D[g_2]$, the readout preserves the averaged monitor. At the response boundary, a causal controller compresses this three-symbol record into at most $m$ persistent states and selects a continuation from a common nine-element library. These paths vary the measured operators according to the timing and amplitudes in Appendix~\ref{app:noisyholonomy}.

Let $f(y,b)$ be the target Choi overlap of record branch $y$ followed by suffix $b$. For this single-response architecture, the optimum over the declared finite library with at most $m$ persistent states is
\begin{equation}
F_m^{\mathrm{gate},*}=\max_{\substack{S\subseteq\mathcal B\\|S|\le m}}
       \sum_y\max_{b\in S} f(y,b).
\label{eq:gate-policy-optimum}
\end{equation}
Thus $F_1^{\mathrm{gate},*}$ optimizes the record-blind suffix. One retained bit selects different continuations after positive and negative clicks, increasing fidelity in both environments throughout Fig.~\ref{fig:holonomic}(b). At $\kappa T=256$, the gains over the best blind suffix are $0.0358$ for the bath and $0.0457$ for independent flips. Table~\ref{tab:holonomic-memory} gives the benchmarks at $64$ and $256$. A single retained bit captures the path-selection gain at the reported precision. The budget counts persistent states after the common acquisition boundary.

\begin{table}[t]
\caption{Record-driven path control with the same counting detector and fixed terminal decoder. $F_1^{\mathrm{gate},*}$ is the best record-blind suffix in the declared library; $F_2^{\mathrm{gate},*}$ permits one retained bit after the common readout boundary. These are finite-library policy optima.}
\label{tab:holonomic-memory}
\begin{ruledtabular}
\begin{tabular}{lrrrr}
Environment & $\kappa T$ & $F_1^{\mathrm{gate},*}$ & $F_2^{\mathrm{gate},*}$ & Gain\\
Independent flips & 64 & 0.725565 & 0.779704 & 0.054140\\
Independent flips & 256 & 0.804649 & 0.850349 & 0.045700\\
Persistent bath & 64 & 0.820252 & 0.889746 & 0.069494\\
Persistent bath & 256 & 0.924542 & 0.960321 & 0.035779\\
\end{tabular}
\end{ruledtabular}
\end{table}

The retained path register makes Eq.~\eqref{eq:gate-policy-optimum} a comparison of nested causal instruments on the same open cPT. A positive gain without environmental noise identifies finite-strength measurement excursions as a source of actionable path information. Its persistence in both noisy models establishes controller-record memory as a gate resource. Appendix~\ref{app:noisyholonomy} gives the paths and record-conditioned channels; Appendix~\ref{app:gate-numerics} specifies their propagation.

\section{Discussion}
Temporal access acquires a coding meaning through the correction decisions that retained records make possible. The stabilizer theorem identifies these decisions across code families and monitoring modalities. Its causal continuation criterion determines which distinctions must survive before the final observation. The five-state counting threshold and the nine-label diffusive example make this distinction operational. In the latter, a finite terminal decision set accompanies an infinite exact causal-memory requirement, with an inverse-square law governing finite-memory approximation. These results complement temporal-access and multi-time resource theories~\cite{temporalaccess,berk2021,berk2023} with exact task values, attaining controllers, and approximation costs.

For controller design, the question is which record distinctions deserve persistent states. Continuous filters and their reduced models address how syndrome information is processed~\cite{vanhandel2005,mabuchi2009,chase2008,sarovar2004,mohseninia2020,atalaya2021}. The hierarchy identifies the best recovery at a prescribed persistent-state count. Detector benchmarks~\cite{denhez2012} and parity readout from two-body meter couplings~\cite{halaski2026} provide concrete interfaces for such comparisons. The storage examples separate measurement backaction, retained information, and online drive within those interfaces. Joint dynamical control and coding~\cite{ddft} offers a complementary route to protection.

Open-process stability turns this memory allocation into a guarantee under model approximation. The general H\"older bound applies uniformly to both diffusive and counting interfaces, while the factorial-tail modulus gives a process-specific scale. A common physical realization supplies a complementary Lipschitz estimate in bath preparation and generator errors. Their deployment corollary guarantees the fidelity of a fixed designed controller, bounds its loss relative to true-model optimization, and certifies its remaining advantage over a smaller access class. The explicit prior-perturbation curves show how the counting five-state controller and a finite-state diffusive quantizer retain protection under the same type of model error.

The open cPT description carries code-family recovery decisions into models with persistent environmental correlations, where the same process--instrument boundary controls their stability. For separated CSS inference, independent bit- and phase-error processing supplies an explicit product controller and an additive approximation budget.

Channel recovery programs and approximate correction conditions act on the effective branches after cPT contraction~\cite{fletcher2007,beny2010,ng2010}. Approximate dynamical codes additionally optimize intermediate checks~\cite{approximate_dynamical}. Here these methods complement a comparison over whole access classes: a realizable decoder supplies achievable performance, and unrestricted recovery bounds the information available at that interface.

For logical gates, retained records can change the subsequent instrument as well as the terminal recovery. The holonomic example makes this distinction physical: a fixed syndrome decoder accompanies record-selected geometric continuations, whose gain persists without environmental noise. The same target-aware cPT construction accommodates the ancilla-augmented correctable paths of Ref.~\cite{lanka_steering}. Continuous strategic codes thereby connect temporal information storage to the design of the measurement path itself.

\section{Conclusion and outlook}
We have quantified temporal access in continuous strategic codes through the correction decisions that records support, the causal memory needed to retain them, and the protection preserved under process-model error. The stabilizer compression theorem characterizes lossless records by a common optimal correction in each terminal syndrome sector. Separated CSS inference gives a product structure. The two-window task exhibits two exact resource regimes: counting reaches its complete-record optimum with five persistent states, while diffusive monitoring has nine terminal decision labels and an optimal causal approximation deficit of order $m^{-2}$.

Open cPTs and their cMPS instrument contractions connect these resource statements to physical continuous monitoring and correlated environments. Parallel counting and diffusive H\"older bounds and computable error moduli control both optimized correction objectives under fixed-resource feedback. The deployment guarantees convert those moduli into retained fidelity, bounded design suboptimality, and certified access advantages. The fixed-target extension incorporates prescribed logical operations, with record-selected measurement paths improving a logical $T$ task at a common fixed decoder.

Memory should retain distinctions that change future optimal recovery, while the desired protection sets the required process-model precision. Larger stabilizer families and longer observation sequences offer settings for code-specific causal state minimization and likelihood quantization. Joint design of memory allocation and measurement paths, detector inefficiency, and correlated-noise characterization connect this principle to implementation; multiqubit gates and fault-tolerant circuits extend it to continuous encoded computation.

\appendix
\section{Proof and certificates for the finite-memory theorem}
\label{app:capacity}

\subsection{Strict hierarchy throughout the parameter range}
Let $\mathcal R=\{0,+,-\}^2$ and use the probability weights $W_{r,s,\ell}$ of Eq.~\eqref{eq:belloptimum}. For a record class $A$ define
\begin{equation}
 v_W(A)=\sum_s\max_\ell\sum_{r\in A}W_{r,s,\ell},\qquad
 d_{r,s}=W_{r,s,0}-W_{r,s,1}.
\end{equation}
Every partition $\mathcal P$ has score $\sum_{A\in\mathcal P}v_W(A)$. A class retains the full-record score precisely when, at each syndrome, its nonzero $d_{r,s}$ all have the same sign. Indeed, its loss is
\begin{equation}
 \frac12\sum_s\left(\sum_{r\in A}|d_{r,s}|-
 \left|\sum_{r\in A}d_{r,s}\right|\right).
 \label{eq:class-conflict-loss}
\end{equation}
For $0<p<1/2$ and $0<z<1$, the following five records have nonzero decision differences at all four syndromes. Their maximizing logical classes are
\begin{equation}
\begin{array}{c|cccc}
 r & 00&01&10&11\\\hline
 00&0&0&0&0\\
 0+&0&0&0&1\\
 0-&1&1&1&0\\
 +0&0&0&1&0\\
 -0&1&1&0&1
\end{array}
 \label{eq:five-decision-patterns}
\end{equation}
To obtain these entries, the record $00$ leaves the prior comparison of each error string with its complement. The other four records reveal respectively $e_2=0,e_2=1,e_1=0,e_1=1$; the terminal syndrome then determines the remaining bits. Any two rows differ in at least one syndrome. Consequently, a partition into at most four classes contains a conflicting pair and has positive loss in Eq.~\eqref{eq:class-conflict-loss}. Five classes suffice: the map displayed below groups $++$ with $00$, $-+$ with $0+$, and $+-,--$ with $0-$. At every supported syndrome these additional records have the same maximizing logical class as their group. This proves that exactly five labels retain the full-record optimum.

Strict improvement at each intermediate size follows by splitting a conflicting class. If its differences at syndrome $s$ have both signs, put
\begin{equation}
 A_+=\{r\in A:d_{r,s}>0\},\qquad A_-=A\setminus A_+.
\end{equation}
Replacing $A$ by $A_+,A_-$ increases its syndrome-$s$ score by
\begin{equation}
 \min\!\left\{\sum_{r\in A_+}d_{r,s},
 -\sum_{r\in A_-}d_{r,s}\right\}>0,
 \label{eq:strict-splitting-gain}
\end{equation}
and cannot lower any other syndrome score. Let $\widehat F_m^{\mathrm{mem}}$ denote the optimum over all partitions into at most $m$ labels. Finiteness of the record alphabet gives a maximizing partition, so this splitting proves $\widehat F_{m+1}^{\mathrm{mem}}>\widehat F_m^{\mathrm{mem}}$ for $m<5$. For $m\geq3$, a causal controller stores the first ternary event and realizes any such final partition; hence $F_m^{\mathrm{mem},*}=\widehat F_m^{\mathrm{mem}}$. Also $F_2^{\mathrm{mem},*}\leq\widehat F_2^{\mathrm{mem}}<\widehat F_3^{\mathrm{mem}}=F_3^{\mathrm{mem},*}$. Finally a two-state controller retains whether $r_2=-$. Using the standard majority correction otherwise, and its record-conditioned optimum on that negative branch, yields
\begin{equation}
 F_2^{\mathrm{mem},*}-F_1^{\mathrm{mem},*}\geq(1-z)p^2(2-p)>0.
 \label{eq:first-memory-margin}
\end{equation}
This construction is causal because it ignores $r_1$.

The same argument gives explicit margins for the remaining steps. The smallest absolute decision difference among the five rows of Eq.~\eqref{eq:five-decision-patterns} is
\begin{equation}
 \lambda(p,z)=\min\{z^2p(1-p)(1-2p),\ z(1-z)p^3\}>0.
 \label{eq:memory-margin}
\end{equation}
A class containing two of those rows has a split gain at least $\lambda$. Thus
\begin{equation}
 F_{m+1}^{\mathrm{mem},*}-F_m^{\mathrm{mem},*}\geq\lambda(p,z),\qquad m=2,3,4.
 \label{eq:adjacent-memory-margins}
\end{equation}
Conditional Pauli corrections attain every partition score, and the Bell bound applies to every CPTP decoder. The resulting logical flip has half diamond error $1-F_m^{\mathrm{mem},*}$, proving the hierarchy for both objectives. On every compact subset of the stated parameter domain these margins have a positive minimum.

The proof also covers unequal positive finite window strengths $K_1,K_2$, with $z_j=e^{-K_j}$. Their positive weights preserve all five decision patterns and the same attaining map. The complete-record fidelity becomes $1-z_1z_2(3p^2-2p^3)$, the first-step lower bound becomes $(1-z_2)p^2(2-p)$, and the other adjacent gains are bounded below by
\begin{equation}
\begin{split}
 \lambda(p,z_1,z_2)=\min\{&z_1z_2p(1-p)(1-2p),\\
 &z_1(1-z_2)p^3,\ (1-z_1)z_2p^3\}>0.
\end{split}
\end{equation}

\subsection{Reduction to integer weights}
For a hidden error string $\bm e=(e_1,e_2,e_3)$ let
\begin{equation}
s(\bm e)=(e_1\oplus e_2,e_2\oplus e_3),\qquad
\ell(\bm e)=\mathbf 1_{|\bm e|\geq2}.
\end{equation}
At $p=1/10$ the error weights have denominator $1000$. At $K=\log5$, each window has probability $1/5$ for no click and total probability $4/5$ for a signed click event after summing click times and multiplicity. Conditional on a hidden history, only the sign agreeing with its stabilizer has nonzero probability. The joint integer weight for the nine event pairs, four syndromes, and two logical classes is therefore
\begin{equation}
n_{r,s,\ell}=25000\,W_{r,s,\ell}\in\mathbb Z.
\end{equation}
The complete integer array is given in Table~\ref{tab:integerweights};
its entries sum to 25000. Thus the optimization below is an exact integer
addition-and-comparison calculation.
\begin{table*}[t]
\caption{Complete joint integer weights at the theorem point. Columns
are ordered by terminal syndrome $s$ and logical class $\ell$; the
probabilities are the listed entries divided by 25000.}
\label{tab:integerweights}
\begin{ruledtabular}
\begin{tabular}{crrrrrrrr}
Event pair & $(00,0)$ & $(00,1)$ & $(01,0)$ & $(01,1)$ & $(10,0)$ & $(10,1)$ & $(11,0)$ & $(11,1)$\\
$00$ &729&1&81&9&81&9&81&9\\
$0+$ &2916&0&324&0&324&0&0&36\\
$0-$ &0&4&0&36&0&36&324&0\\
$+0$ &2916&0&324&0&0&36&324&0\\
$++$ &11664&0&1296&0&0&0&0&0\\
$+-$ &0&0&0&0&0&144&1296&0\\
$-0$ &0&4&0&36&324&0&0&36\\
$-+$ &0&0&0&0&1296&0&0&144\\
$--$ &0&16&0&144&0&0&0&0
\end{tabular}
\end{ruledtabular}
\end{table*}
For an event subset $A$, define its score
\begin{equation}
v(A)=\sum_s\max_{\ell\in\{0,1\}}\sum_{r\in A}n_{r,s,\ell}.
\label{eq:subsetscore}
\end{equation}
The Bell-mixture argument in Eq.~(\ref{eq:belloptimum}) shows that $v(A)$ is the best contribution of a merged record class, including optimization over arbitrary terminal CPTP maps.

\subsection{Causal enumeration and a noncausal upper bound}
A causal controller with one persistent state has one event class. For two states, enumerate the three-valued first update $f_1:\{0,+,-\}\to\{0,1\}$ and the second update $f_2:\{0,1\}\times\{0,+,-\}\to\{0,1\}$. There are $2^3 2^6=512$ deterministic rules. Each rule induces a partition of the nine pairs, whose score is the sum of Eq.~(\ref{eq:subsetscore}) over its classes.

For $m\geq3$, the first update can assign the three event values to three distinct persistent states. At the second boundary, each pair $(r_1,r_2)$ can then be mapped to an arbitrary one of at most $m$ labels. Enumerating restricted-growth strings removes label permutations while retaining every partition. Randomized rules do not improve the value: with other kernels and the recovery fixed, the objective is affine in a single stochastic update kernel. Replacing that kernel by an extremal deterministic update does not lower the value. Applying this successively gives a deterministic rule attaining an optimum.

As an independent upper bound, permit an arbitrary, potentially noncausal partition of the nine event pairs into at most $m$ classes. Let $S$ be a set of unassigned events and define
\begin{align}
D_m(\varnothing)&=0,\qquad D_0(S)=-\infty\quad(S\ne\varnothing),\\
D_m(S)&=\max_{\substack{A\subseteq S\\\min S\in A}}
\{v(A)+D_{m-1}(S\setminus A)\}.
\label{eq:capacitydp}
\end{align}
Every partition has exactly one class containing the smallest event of $S$, which proves the recursion by induction. Because this relaxation includes every causal partition, its value is an upper bound. The exhaustive causal constructions attain it. The integer numerators for $m=1,\ldots,5$ are $24300,24680,24900,24936,24972$. Their quotient by $25000$ gives Table~\ref{tab:capacity}. For $m=5$, the map
\begin{equation}
\begin{gathered}
(00,0+,0-,+0,++,+-,-0,-+,--)\\
\mapsto(0,1,2,3,0,2,4,1,2)
\end{gathered}
\end{equation}
is a causal construction attaining the full-record score. No compression can exceed the full-record decoder, proving saturation for every $m\geq5$.

Attaining label maps at every capacity are listed in
Table~\ref{tab:attainingmaps}. For two states the displayed map
retains whether the second event is negative, and is causal without
storing the first event. For three or more states, store the first
ternary event and apply the displayed pair-to-label map at the second
boundary. Together with Table~\ref{tab:integerweights} and the recursion,
these maps specify both constructions and exact upper certificates
within the article.
\begin{table*}[t]
\caption{Causal attaining labels in event order
$(00,0+,0-,+0,++,+-,-0,-+,--)$. Counts refer to deterministic causal update tables for $m=2$ (including repeated induced partitions), and unlabeled partitions into at most $m$ classes for $m\geq3$. The last column is the numerator of the optimal fidelity with denominator 25000.}
\label{tab:attainingmaps}
\begin{ruledtabular}
\begin{tabular}{clrr}
$m$ & Pair-to-label map & Count & Optimal numerator\\
1 & $(0,0,0,0,0,0,0,0,0)$ & 1 & 24300\\
2 & $(0,0,1,0,0,1,0,0,1)$ & 512 & 24680\\
3 & $(0,0,1,0,0,1,2,2,1)$ & 3281 & 24900\\
4 & $(0,0,1,2,0,1,3,3,1)$ & 11051 & 24936\\
5 & $(0,1,2,3,0,2,4,1,2)$ & 18002 & 24972
\end{tabular}
\end{ruledtabular}
\end{table*}

\section{Physical identification and uniform stability}
\label{app:stability}

\subsection{Boundary, basis, and stochastic conventions}

We use an unnormalized Choi matrix,
\begin{equation}
 J(\Lambda)=(\operatorname{id}_{i}\otimes\Lambda)
 (\vket{I_i}\vbra{I_i}),\qquad
 \vket{I_i}=\sum_{a=1}^{d_i}|a\rangle|a\rangle .
 \label{eq:choi-convention}
\end{equation}
Thus a channel obeys $\operatorname{Tr}_{o}J=I_i$ and
$\operatorname{Tr}J=d_i$. The normalized logical Choi state is
$J(\mathcal M)/d_L$.
For a system of dimension $d=2^n$, choose Hermitian unitary Pauli strings
$P_0=I,P_1,\ldots,P_{d^2-1}$ with
$\operatorname{Tr}(P_aP_b)=d\delta_{ab}$. The operation basis and its dual are
\begin{equation}
\begin{gathered}
 \mathcal E_{ab}(X)=P_aXP_b,\\
 \operatorname{Tr}_{\mathrm{L}}(\mathcal E_{ab}^{\dagger}\mathcal E_{cd})
 =d^2\delta_{ac}\delta_{bd},\\
 g_{ab}=d^{-2}\operatorname{Tr}_{\mathrm{L}}
 (\mathcal E_{ab}^{\dagger}\mathcal G).
\end{gathered}
 \label{eq:basis-dual}
\end{equation}
Here $\operatorname{Tr}_{\mathrm L}$ is the matrix trace on Liouville space;
$\mathcal E_{00}=\operatorname{id}$ and
$\Vert\mathcal E_{ab}\Vert_\diamond=1$. These basis elements need not be CP.
In a vectorization in which $\mathcal E_{ab}=P_a\otimes P_b^{\mathsf T}$,
the corresponding Hilbert--Schmidt orthonormal basis is
$\mathbb P_{ab}=\mathcal E_{ab}/d$. Consequently, an $n$-insertion process
coefficient in that convention is $d^{-n}$ times our coefficient, while
each nonidentity instrument coefficient is multiplied by $d$. The same
rescaling applies to the identity coordinate of the generator. Its
contribution is $(da_0)\mathbb P_{00}=a_0\operatorname{id}$, so the
scalar multiplying the identity superoperator, and hence the scalar
exponential below, is unchanged.
Equations below always use Eq.~\eqref{eq:basis-dual}.

We write $\nu=(a,b)\ne(0,0)$ and $q=d^4-1$. Real reference Wiener
increments obey $\mathbb E[dW_\alpha dW_\beta]=\delta_{\alpha\beta}dt$.
For bounded deterministic Hamiltonians and monitored operators, set
\begin{align}
 \mathcal G(X)&=-i[H,X]+\sum_\alpha
 \left(L_\alpha XL_\alpha^\dagger
 -\tfrac12\{L_\alpha^\dagger L_\alpha,X\}\right),\\
 \mathcal B_\alpha(X)&=L_\alpha X+XL_\alpha^\dagger,\\
 a_\nu&=d^{-2}\operatorname{Tr}_{\mathrm L}
 (\mathcal E_\nu^\dagger\mathcal G),\\
 B_{\nu\alpha}&=d^{-2}\operatorname{Tr}_{\mathrm L}
 (\mathcal E_\nu^\dagger\mathcal B_\alpha).
 \label{eq:instrument-dictionary}
\end{align}
The index zero is included when defining $a_0$ and $B_{0\alpha}$.
In particular,
\begin{equation}
\begin{gathered}
 B_{0\alpha}=2\operatorname{Re}\operatorname{Tr}(L_\alpha)/d,
 \\
 a_0=\sum_\alpha\left(
 |\operatorname{Tr}L_\alpha|^2/d^2
 -\operatorname{Tr}(L_\alpha^\dagger L_\alpha)/d\right).
\end{gathered}
 \label{eq:identity-coefficients}
\end{equation}
We impose $\operatorname{Tr}L_\alpha=0$, so $B_{0\alpha}=0$.
Heterodyne monitoring is included by using real Wiener quadratures with
the appropriate factors of $1/\sqrt2$ in the corresponding $\mathcal B$.

For counting, the jump operators need not be traceless. Choose positive
reference intensities $\lambda_\alpha(t)$ and define
\begin{equation}
\begin{gathered}
 \mathcal G_0(X)=-i[H,X]-\tfrac12\sum_\alpha
 \{J_\alpha^\dagger J_\alpha,X\},\\
 \mathcal G=\mathcal G_0+\sum_\alpha\operatorname{Ad}_{J_\alpha},\\
 \mathcal C_\alpha=\lambda_\alpha^{-1}\operatorname{Ad}_{J_\alpha},\qquad
 \mathcal B_\alpha=\mathcal C_\alpha-\operatorname{id},\\
 a_\nu=d^{-2}\operatorname{Tr}_{\mathrm L}
 (\mathcal E_\nu^\dagger\mathcal G),\\
 B_{\nu\alpha}=d^{-2}\operatorname{Tr}_{\mathrm L}
 (\mathcal E_\nu^\dagger\mathcal B_\alpha).
\end{gathered}
 \label{eq:counting-dictionary}
\end{equation}
These sums include the identity coordinate. In particular,
\begin{equation}
\begin{gathered}
 a_0=\sum_\alpha\left(
 \frac{|\operatorname{Tr}J_\alpha|^2}{d^2}
 -\frac{\operatorname{Tr}(J_\alpha^\dagger J_\alpha)}{d}\right),\\
 B_{0\alpha}=\frac{|\operatorname{Tr}J_\alpha|^2}
 {d^2\lambda_\alpha}-1.
\end{gathered}
 \label{eq:counting-identity-coefficients}
\end{equation}
If $N_\alpha$ has reference intensity $\lambda_\alpha$, the density
relative to its common Poisson law has no-click drift
$\mathcal G_0+\sum_\alpha\lambda_\alpha\operatorname{id}$ and click
map $\mathcal C_\alpha$. In compensated form its drift is $\mathcal G$
and its noise map is $\mathcal B_\alpha$ against
$d\widetilde N_\alpha$. Thus the identity noise coefficient and its
corresponding identity insertion are both included in the Poisson active
label count whenever $B_{0\alpha}\ne0$.

For the two-window repetition-code model of Sec.~\ref{sec:capacity}, this
dictionary can be read off without a separate coefficient table. During
window $j$, take $J_{j,\pm}=\sqrt\kappa P_{j,\pm}$ and
$\lambda_{j,\pm}=\kappa$, with only the pair for that window active at a
time. Since $P_{j,\pm}=(I\pm S_j)/2$, the drift is
$\mathcal G=-\kappa\operatorname{id}/2+(\kappa/2)\operatorname{Ad}_{S_j}$,
so $a_0=-\kappa/2$ and $a_{S_jS_j}=\kappa/2$. The normalized jump map
is $\operatorname{Ad}_{P_{j,\pm}}$, and subtracting the identity gives
$B_{00}=-3/4$, $B_{S_j0}=B_{0S_j}=\pm1/4$, and
$B_{S_jS_j}=1/4$; all other coefficients vanish. Across the two
disjoint windows the active-label union is
$\{(0,0),(S_1,0),(0,S_1),(S_1,S_1),(S_2,0),(0,S_2),(S_2,S_2)\}$,
hence $q_{\rm act}^{(\rm P)}=7$. The time-uniform coefficient bounds
are $a^2=\kappa^2/4$ and $b_{\rm P}^2=5\kappa/4$, the latter computed
from the two simultaneously active sign channels within either single
window. The active-label count is global in time, whereas the coefficient
norm is an essential supremum over time.

For the diffusive counterpart of Sec.~\ref{sec:diffusive-memory},
$L_j=\sqrt{\kappa_j}S_j$ gives
$\mathcal G_j=\kappa_j(\operatorname{Ad}_{S_j}-\operatorname{id})$
and $\mathcal B_j(X)=\sqrt{\kappa_j}(S_jX+XS_j)$.
Thus $a_0=-\kappa_j$, $a_{S_jS_j}=\kappa_j$, and
$B_{S_j0}=B_{0S_j}=\sqrt{\kappa_j}$ in window $j$, with all other
coefficients zero. The disjoint-window union has six nonidentity labels:
$(S_j,0),(0,S_j),(S_j,S_j)$ for $j=1,2$. Hence
$q_{\rm act}^{(\rm W)}=6$, $a^2=\max_j\kappa_j^2$, and
$b^2=2\max_j\kappa_j$. Hidden-error intervals can be taken as idle
instrument intervals; their dynamics belongs to the open process.
At equal $\kappa$, the counting drift has half the diffusive
dephasing rate, consistently with the separately specified instruments.
The finite-alphabet feedback theorem applies to any fixed discretized
response interface. The real-readout model of
Sec.~\ref{sec:diffusive-memory} has measurable boundary processing;
the common-realization estimate supplies its deployment certificate
directly, without a uniform finite-alphabet assumption over growing grids.

\subsection{Open physicality and the weighted domain}

An open physical cPT has fixed input and output boundary spaces, and
jointly compatible positive, causal finite marginals for every ordered
nonnegative time smoothing. Compatibility includes insertion of identity
slots: removing such a slot gives precisely the lower-order marginal.
The input is an arbitrary independent input, rather than a single fixed
initial system state. In particular, a common random grid, with one
independent uniform time in each of a set of disjoint ordered cells,
defines one legitimate open finite comb. Its different-order coefficients
are marginals of that same comb. Positivity of independently chosen
averages of different orders would not be a sufficient assumption.

Write $J^{(n)}_{\boldsymbol\nu}(\boldsymbol t)$ for the Choi matrix
obtained by inserting $\mathcal E_{\nu_j}$ at the ordered times.
Finite-comb dilation, insertion of left and right unitaries, and subsequent
partial trace give
\begin{equation}
 \Vert J^{(n)}_{\boldsymbol\nu}(\boldsymbol t)\Vert_2\leq d_i
 \quad\hbox{for almost every }\boldsymbol t.
 \label{eq:pointwise-bound}
\end{equation}
To prove this bound, first dilate one jointly smoothed finite comb.
Each interval is a CPTP map on the system and its retained environment.
An insertion $X\mapsto(P_a\otimes I_E)X(P_b\otimes I_E)$ preserves
the trace norm, including an arbitrary boundary reference. The interval
maps and terminal partial trace contract that norm. The initial operator
is $\vket{I_i}\vbra{I_i}\otimes\sigma_E$,
whose trace norm is $d_i$. Therefore the smoothed response has trace norm
at most $d_i$, and its Hilbert--Schmidt norm is no larger.
The bound is independent of the dilation dimension and of the number
of insertions.

Take normalized nonnegative averages over shrinking ordered rectangles
inside $\Delta_n(T)$. Joint compatibility identifies each average with
the response of the same smoothed finite comb. At Lebesgue points these
averages converge in Hilbert--Schmidt norm; continuity of the trace norm
in finite dimensions preserves the bound $d_i$ in the limit.
This gives Eq.~\eqref{eq:pointwise-bound} almost everywhere for each
label string. There are finitely many strings at fixed order and
countably many orders, so one may choose a common full-measure version.
It is sufficient below to use any common constant $C_\partial\geq d_i$.
For $s\geq1$, define
\begin{equation}
\begin{gathered}
 \Vert\boldsymbol\Upsilon\Vert_{\Gamma_s^\partial}^2
 =\sum_{n\geq0}s^n\sum_{\boldsymbol\nu}
 \int_{\Delta_n(T)}
 \Vert J^{(n)}_{\boldsymbol\nu}(\boldsymbol t)\Vert_2^2d\boldsymbol t,
 \\
 \Vert\boldsymbol\Upsilon^{(n)}\Vert_2^2
 \leq C_\partial^2(qT)^n/n! .
\end{gathered}
 \label{eq:open-tail}
\end{equation}
The vacuum coefficient is an open channel Choi matrix, not the scalar
one. At order $n$ there are $q^n$ label strings, each bounded by
$C_\partial$, and the ordered simplex has volume $T^n/n!$.
Summing these bounds gives the second line of Eq.~\eqref{eq:open-tail}.
In particular, the full weighted squared norm is at most
$C_\partial^2e^{sqT}$ for every finite $s\geq1$.

\subsection{The Wick map is a CP, normalized instrument}

Define the formal insertion field
\begin{equation}
 \xi_\nu(t)=a_\nu(t)+\sum_\alpha B_{\nu\alpha}(t)\dot W_\alpha(t).
\end{equation}
Let $\mathsf T$ denote the matrix-valued Wick series
\begin{equation}
\begin{split}
 \mathsf T\boldsymbol\Upsilon(W)
 ={}&e^{\int_0^Ta_0(t)dt}\sum_{n\geq0}\sum_{\boldsymbol\nu}
 \\&\int_{\Delta_n(T)}J^{(n)}_{\boldsymbol\nu}(\boldsymbol t)
 :\!\prod_{j=1}^n\xi_{\nu_j}(t_j)\!:\,d\boldsymbol t .
\end{split}
 \label{eq:wick-map}
\end{equation}
If $\operatorname{Re}a_0\leq a_0^*$,
$\sum_{\nu\ne0}|a_\nu|^2\leq a^2$ and $\Vert B_\perp\Vert\leq b$, the
Wiener-chaos estimate of the continuum dilation construction, applied
entrywise to the open boundary, gives
\begin{equation}
\begin{gathered}
 \Vert\mathsf T\boldsymbol X\Vert_{L^2(\mu_W;\mathrm{HS})}
 \leq C_0\Vert\boldsymbol X\Vert_{\Gamma_r^\partial},\\
 r=1+b^2,\quad C_0=e^{a_0^*T+a^2T/2}.
\end{gathered}
 \label{eq:weighted-wick-bound}
\end{equation}
For completeness, let $\mathfrak h=L^2([0,T];\mathbb C^q)$ be the
one-particle operation space and let $C$ contract the operation index
with $B_{\nu\alpha}(t)$, so that $\Vert C\Vert\leq b$. The map
$Vf=f\oplus Cf$ has $\Vert V\Vert^2\leq1+b^2$. Its second quantization,
followed by the Fock factorization over this direct sum, obeys
\begin{equation}
 \Vert\Gamma(V)\boldsymbol X\Vert^2
 \leq\sum_{n\geq0}(1+b^2)^n\Vert\boldsymbol X^{(n)}\Vert_2^2.
\end{equation}
Contract the first Fock factor with the exponential vector of
deterministic coefficients (with the conjugation appropriate to
the Hilbert-space inner product), and send the second factor to
Wiener chaos by the Wiener--It\^o isometry. The first contraction
has norm $\exp(\Vert a_\perp\Vert_{L^2}^2/2)\leq e^{a^2T/2}$.
The resulting series is Eq.~\eqref{eq:wick-map}: its terms select
deterministic or stochastic factors at each insertion. This also
controls interference between different insertion orders with the
same stochastic degree. Multiplying by the scalar drift proves
Eq.~\eqref{eq:weighted-wick-bound}, first in finite sectors and then
by completion, with open-boundary matrices as spectator vectors.
The estimate establishes existence of the Wick series; positivity
and normalization follow from the physical identification below.

\textbf{Physical identification lemma.}
For an open physical cPT as defined above,
$J_W=\mathsf T\boldsymbol\Upsilon(W)$ is the Choi density of a CP
instrument and
\begin{equation}
\begin{gathered}
 J_W\geq0\quad\mu_W\hbox{-almost surely},\\
 \int\operatorname{Tr}_{o}J_W\,d\mu_W(W)=I_i .
\end{gathered}
 \label{eq:instrument-normalization}
\end{equation}
Thus every measurable event gives a CP trace-nonincreasing map, and the
whole record space gives a trace-preserving map.

\emph{Proof.}
First let $H,L_\alpha$ be
constant in each cell of a partition with mesh $h$, refining all hardware
response boundaries. In a cell of length $h_j$, put
\begin{align}
 A_j&=-iH_j-\tfrac12\sum_\alpha L_{j\alpha}^\dagger L_{j\alpha},\\
 R_j&=(I+h_j^2A_j^\dagger A_j)^{-1/2},\\
 K_j(\Delta W_j)&=
 \left(I+h_jA_j+\sum_\alpha L_{j\alpha}\Delta W_{j\alpha}\right)R_j .
 \label{eq:exact-cell-kraus}
\end{align}
The increments have their reference Gaussian law. Since
$A_j+A_j^\dagger+\sum L_{j\alpha}^\dagger L_{j\alpha}=0$,
\begin{equation}
 \mathbb E[K_j^\dagger K_j]
 =R_j(I+h_j^2A_j^\dagger A_j)R_j=I.
 \label{eq:exact-cell-tp}
\end{equation}
These instruments are exactly CP and normalized. Apply them to the common randomly smoothed open finite comb.
The resulting density $J_h(W)$ is positive and its integrated output
partial trace is $I_i$. Brownian bridges inside a cell may be included
as independent unused reference records, so every $J_h$ lives on the
same Wiener record space. Bridges of completed cells are revealed by
the next response boundary. They affect no cell Kraus operator and
make the entire past reference path available at that boundary.

We verify convergence to Eq.~\eqref{eq:wick-map}. Expand the cell map
$X\mapsto K_jXK_j^\dagger$ in Hermite polynomials of its Gaussian
increments. The degree is at most two. Its constant and degree-one
parts are, respectively,
\begin{gather}
 \operatorname{id}+h_j\mathcal G_j+O(h_j^2),\\
 \sum_\alpha\bigl(\mathcal B_{j\alpha}+O(h_j)\bigr)
 \Delta W_{j\alpha} .
 \label{eq:hermite-leading}
\end{gather}
Its degree-two part is a bounded linear combination of
$\Delta W_{j\alpha}\Delta W_{j\beta}
-h_j\delta_{\alpha\beta}$ and therefore has $L^2$ norm $O(h_j)$.
All remainder bounds hold in the finite-dimensional coefficient norm
$\sum_\nu|x_\nu|$, including the identity index, uniformly under the
stated bounds on $H,L_\alpha$.

Here is an explicit global estimate that prevents accumulation of the
degree-two terms. For any cell polynomial expanded in an orthonormal
Hermite basis, the sum of the squared coefficient $\ell^1$ norms in
the first chaos is at most $\beta h_j$, and in the second chaos at most
$\gamma h_j^2$. The $\ell^1$ norm of its constant coefficient is at
most $1+\alpha h_j$. The constants $\alpha,\beta,\gamma$ depend only
on the dimension, number of monitored operators and their bounds.
For a specified set of cells carrying nonconstant Hermite factors,
use Eq.~\eqref{eq:pointwise-bound} and sum the other cell coefficients
in absolute value. Orthogonality of different cell Hermite patterns
then shows that the squared $L^2$ norm of all terms containing at
least one second-chaos cell is bounded by
\begin{equation}
\begin{split}
 &C_\partial^2 e^{2\alpha T}
 \Bigl[\prod_j(1+\beta h_j+\gamma h_j^2)\\
 &\hspace{6em}-\prod_j(1+\beta h_j)\Bigr]\\
 &\quad\leq C_\partial^2 e^{(2\alpha+\beta+\gamma h)T}\gamma Th .
\end{split}
 \label{eq:collision-estimate}
\end{equation}
Repeated physical insertions in one cell are thus suppressed by their
total time volume. In the remaining terms, replacing the constant and first-chaos
parts by $\operatorname{id}+h_j\mathcal G_j$ and
$\sum_\alpha\mathcal B_{j\alpha}\Delta W_{j\alpha}$ incurs
$O(\sum_j(h_j^2+h_j^{3/2}))=O(\sqrt h)$ in $L^2$.
To justify this last telescope, fix the replaced cell and use the same
Hermite-pattern estimate in all other cells; their product is bounded
by $e^{(2\alpha+\beta)T}$. Hence the multiplicative constant is
independent of the grid and of the finite-comb environment.

Let $P_h$ average each process kernel on ordered distinct cells and
set it to zero on tuples with two times in the same cell. This is the
orthogonal step-function projection with multiple occupancy removed.
It converges strongly to the identity in every fixed weighted Fock
space: prove this in each finite sector by density of step functions
and vanishing collision volume, then use the summable weighted tail.
The contraction with the remaining Euler cell maps is exactly a
discrete Wiener-chaos contraction of $P_h\boldsymbol\Upsilon$.
More precisely, factor $s_j=1+h_ja_{0,j}$ from each cell; its
nonidentity drift and noise coefficients become $a_{\nu,j}/s_j$
and $B_{\nu\alpha,j}/s_j$. The scalar $\prod_js_j$ tends to
$e^{\int a_0}$, and the divided coefficients tend to $a_\nu,B_{\nu\alpha}$.
There is no identity noise because $B_{0\alpha}=0$. To obtain a
uniform tail estimate for the divided coefficients, put
$M=\operatorname*{ess\,sup}_t|a_0(t)|$. For meshes with $Mh\leq1/2$,
\begin{equation}
 \Vert B_{\perp,j}/s_j\Vert\leq2b,\qquad
 \Vert a_{\perp,j}/s_j\Vert_2\leq2a,\qquad R=1+4b^2.
\end{equation}
Apply Eq.~\eqref{eq:weighted-wick-bound} with this fixed enlarged
weight $R$. The factorial bound in Eq.~\eqref{eq:open-tail} gives a
sector cutoff uniform in all sufficiently fine meshes. For each
fixed number of insertions the discrete It\^o kernels converge in
$L^2$; the common tail estimate then makes the convergence summable.
The enlarged weight is needed only for approximation and does not
alter the sharper bound for the limiting map. Consequently,
\begin{equation}
\begin{gathered}
 J_h\longrightarrow\mathsf T\boldsymbol\Upsilon
 \quad\hbox{in }L^2(\mu_W;\mathrm{HS})\\
 \quad\hbox{and hence in }L^1(\mu_W;\Vert\cdot\Vert_1).
\end{gathered}
 \label{eq:identification-limit}
\end{equation}
Bounded measurable $H,L_\alpha$ follow by bounded step-function
approximation. Their coefficients converge almost everywhere along a
subsequence and are uniformly bounded. Dominated convergence in each
finite It\^o kernel, followed by the same factorial-tail bound, proves
Eq.~\eqref{eq:identification-limit} with the original coefficients.

An $L^1$-convergent subsequence converges almost surely, so the positive
cone is preserved. Integration and partial trace are continuous in
$L^1$, proving Eq.~\eqref{eq:instrument-normalization}. Integrability
gives countable additivity in trace norm. The CP complement of an event
and normalization imply that the event map is trace nonincreasing.
This proves the lemma. For a bounded system--environment generator,
the same expansion is the stochastic Dyson expansion of the usual
linear measurement equation; hence this extension agrees with the
physical continuous-monitoring instrument on process-canonical cMPS.
\hfill$\square$

\subsection{Counting instruments and Poisson stability}
For a counting interface, let $N_\alpha$ be independent Poisson random
measures with deterministic reference intensities $\lambda_\alpha(t)$,
and write $\widetilde N_\alpha=N_\alpha-\lambda_\alpha dt$. Normal
ordering with respect to this reference measure gives the contraction
\begin{equation}
\begin{aligned}
 \mathsf T_{\rm P}\boldsymbol\Upsilon(N)
 &=e^{\int_0^T a_0(t)dt}
 \sum_{n\geq0}\sum_{\boldsymbol\nu}\\
 &\quad\times
 \int_{\Delta_n(T)}J^{(n)}_{\boldsymbol\nu}(\boldsymbol t)
 :\!\prod_{j=1}^n\xi_{\nu_j}(t_j)\!:\,d\boldsymbol t,
\end{aligned}
\label{eq:poisson-wick-map}
\end{equation}
where the last expression is understood as a Poisson multiple integral,
with $a_\nu,B_{\nu\alpha}$ given by
Eqs.~\eqref{eq:counting-dictionary}--\eqref{eq:counting-identity-coefficients}.
The fields $\xi_\nu$ are defined by Eq.~\eqref{eq:poisson-insertion-field};
the labels $\nu_j$ range over the active insertion set, including label
$0$ when its Poisson noise coefficient is nonzero.
The active Poisson insertion set includes label $0$ whenever some
$B_{0\alpha}$ is nonzero, so the factorial-tail count and the Poisson
noise norm use the same set. In the active projection used in the
stability bound, the operation-label sum is restricted to this set. An
identity insertion is a legitimate slot of the same compatible open comb:
removing that slot gives its lower-order marginal, and its pointwise Choi
bound is still $C_\partial$. Thus it contributes one label per slot to
the projected factorial tail; it is not an additional scalar drift.

In the insertion field, the scalar drift has already been factored out.
Accordingly, write
\begin{equation}
\begin{gathered}
 \xi_\nu(t)=\bar a_\nu(t)+
 \sum_\alpha B_{\nu\alpha}(t)\,\dot{\widetilde N}_\alpha(t),\\
 \bar a_0(t)=0,\quad \bar a_\nu(t)=a_\nu(t)\ (\nu\ne0).
\end{gathered}
 \label{eq:poisson-insertion-field}
\end{equation}
In particular, the deterministic coefficient $a_0$ occurs only in the
prefactor $e^{\int a_0dt}$.

The Poisson multiple-integral isometry replaces the Wiener--It\^o
isometry. If $\Vert B_{\rm P}\Vert_{\rm op}\leq b_{\rm P}$, where the
matrix has rows $\nu\in\{0\}\cup\Lambda$ and entries
$\sqrt{\lambda_\alpha}B_{\nu\alpha}$, and
$\sum_{\nu\in\Lambda\setminus\{0\}}|a_\nu|^2\leq a^2$, then, for every open coefficient family,
\begin{equation}
\begin{aligned}
 \Vert\mathsf T_{\rm P}\boldsymbol X\Vert_{L^2(\mu_{\rm P};\mathrm{HS})}
 &\leq C_{{\rm P},0}\Vert\boldsymbol X\Vert_{\Gamma_{r_{\rm P}}^\partial},\\
 r_{\rm P}&=1+b_{\rm P}^2,\qquad
 C_{{\rm P},0}=e^{a_0^*T+a^2T/2}.
\end{aligned}
\label{eq:poisson-weighted-bound}
\end{equation}
Indeed, the Poisson isometry assigns the $n$-fold compensated integral its
$L^2$ norm, while the full operation-index contraction, including its
identity row, has norm at most $b_{\rm P}$. Applying the
second-quantization estimate sector by sector
and contracting the deterministic coefficients with the exponential vector
gives Eq.~\eqref{eq:poisson-weighted-bound}; no Gaussian Wick identity is
used in this step.

\textbf{Common-path physical identification.}
Choose reference rates with
$0<\lambda_-\leq\lambda_\alpha(t)\leq\lambda_+<\infty$.
This is a reference-measure choice; the physical click rates may vanish.
Constant positive reference rates always suffice. We construct every
approximation on this one Poisson probability space.
First take the coefficients and reference rates constant on each cell
$I_j$ of a partition. Write $h_j=|I_j|$ and $h=\max_jh_j$.
Put $A_j=-iH_j-\frac12\sum_\alpha J_{j\alpha}^\dagger J_{j\alpha}$ and use
\begin{equation}
\begin{gathered}
 K_{j,0}=(I+h_jA_j)R_j,\quad
 K_{j,\alpha}=\sqrt{h_j}J_{j\alpha}R_j,\\
 R_j=(I+h_j^2A_j^\dagger A_j)^{-1/2}.
\end{gathered}
 \label{eq:poisson-cell-kraus}
\end{equation}
The identity $A_j+A_j^\dagger+\sum_\alpha J_{j\alpha}^\dagger J_{j\alpha}=0$
gives $K_{j,0}^\dagger K_{j,0}+\sum_\alpha K_{j,\alpha}^\dagger K_{j,\alpha}=I$.
In each cell let $A_{j,0}$ be the event of no reference click and
$A_{j,\alpha}$ the event that its first reference click has mark
$\alpha$. Later clicks and their times remain unused reference
coordinates. With $\ell_j=\sum_\alpha\lambda_{j\alpha}$, their exact
probabilities are
\begin{equation}
 \pi_{j,0}=e^{-\ell_jh_j},\qquad
 \pi_{j,\alpha}=\frac{\lambda_{j\alpha}}{\ell_j}
                      (1-e^{-\ell_jh_j}).
 \label{eq:poisson-cell-reference}
\end{equation}
Define the CP cell density on the full continuous path by
\begin{equation}
\begin{split}
 \mathcal F_j(N)={}&\mathbf1_{A_{j,0}}(N)
                      \frac{\operatorname{Ad}_{K_{j,0}}}{\pi_{j,0}}\\
 &+\sum_\alpha\mathbf1_{A_{j,\alpha}}(N)
                      \frac{\operatorname{Ad}_{K_{j,\alpha}}}{\pi_{j,\alpha}}.
\end{split}
 \label{eq:poisson-common-cell}
\end{equation}
Its reference mean is the complete cell channel. Contracting these
densities with the common randomly smoothed finite comb therefore gives
a positive, normalized Choi density $J_h(N)$ on the same space for every
mesh. At a response boundary the completed-cell path coordinates are
available, including those unused by the cell maps.

We next control the densities in $L^2$, including paths with multiple
clicks. Use the coefficient norm
$\Vert X\Vert_{\rm c}=\sum_\nu|x_\nu|$ for
$X=\sum_\nu x_\nu\mathcal E_\nu$, and put
$\Delta\widetilde N_{j\alpha}=N_\alpha(I_j)-\lambda_{j\alpha}h_j$.
Orthogonal projection in the cell Poisson space gives
\begin{equation}
\begin{split}
 \mathcal F_j={}&\operatorname{id}+h_j\mathcal G_j
 +\sum_\alpha\mathcal B_{j\alpha}\Delta\widetilde N_{j\alpha}\\
 &+D_j+\sum_\alpha E_{j\alpha}\Delta\widetilde N_{j\alpha}+U_j.
\end{split}
 \label{eq:poisson-cell-projection}
\end{equation}
Here $U_j$ is orthogonal to the constant and all first-chaos cell
increments. Uniformly over the partition,
\begin{equation}
\begin{gathered}
 \Vert D_j\Vert_{\rm c}\leq c_1h_j^2,\qquad
 \sum_\alpha\lambda_{j\alpha}\Vert E_{j\alpha}\Vert_{\rm c}^2
       \leq c_2h_j^2,\\
 \mathbb E\Vert U_j\Vert_{\rm c}^2\leq c_3h_j^2.
\end{gathered}
 \label{eq:poisson-cell-remainder}
\end{equation}
The constants depend on the bounded hardware coefficients, reference
rates, and system dimension, and are independent of the comb dilation.
To obtain these estimates, expand $R_j=I+O(h_j^2)$ and
$h_j\lambda_{j\alpha}/\pi_{j,\alpha}=1+O(h_j)$.
On zero-click paths the error from the leading line of
Eq.~\eqref{eq:poisson-cell-projection} is $O(h_j^2)$; on single-click
paths it is $O(h_j)$. On multiple-click paths it is bounded by a
constant times $1+\sum_\alpha N_\alpha(I_j)$.
For a Poisson variable of mean $\ell_jh_j$,
\begin{equation}
 \mathbb E\left[(1+N(I_j))^2\mathbf1_{\{N(I_j)\geq2\}}\right]
       \leq c_4h_j^2.
 \label{eq:poisson-multiple-click-moment}
\end{equation}
Thus the full cell residual has squared $L^2$ norm $O(h_j^2)$.
Its mean is $O(h_j^2)$ because
$\mathbb E\mathcal F_j=\operatorname{id}+h_j\mathcal G_j+O(h_j^2)$.
Its covariance with each compensated increment is $O(h_j^2)$,
which gives the stated first-chaos coefficient estimate after division
by $\lambda_{j\alpha}h_j$. Orthogonal subtraction proves the bound on
$U_j$. These moment estimates control the density on rare paths directly.

Expand the cell factors in orthonormal Poisson bases, starting with the
constant and normalized first-chaos increments. Independence between
cells makes different tensor patterns orthogonal. Their constant
coefficient norms are bounded by $1+\alpha h_j$, their first-chaos
squared coefficient norms by $\beta h_j$, and their residual squared
coefficient norms by $\gamma h_j^2$.
Using Eq.~\eqref{eq:pointwise-bound}, the squared norm of all contracted
terms containing a residual cell is at most
\begin{equation}
\begin{split}
 C_\partial^2e^{2\alpha T}
 \left[\prod_j(1+\beta h_j+\gamma h_j^2)
             -\prod_j(1+\beta h_j)\right]\\
 \leq C_\partial^2e^{(2\alpha+\beta+\gamma h)T}\gamma Th.
\end{split}
 \label{eq:poisson-global-remainder}
\end{equation}
The same pattern bound controls a telescope replacing $D_j$ and
$E_{j\alpha}$ by zero. Its $L^2$ error is at most a constant times
$\sum_j(h_j^2+h_j^{3/2})=O(\sqrt h)$.

It remains to identify the leading cell expansion. Use the process
projection $P_h$ defined in the diffusive proof, with distinct occupied
cells. Factor the scalar drift $1+h_ja_{0,j}$ in every cell.
The other coefficients are divided by this scalar; the identity noise
row is divided as well. For sufficiently small meshes their norms are
bounded by $2a$ and $2b_{\rm P}$. Equation~\eqref{eq:poisson-weighted-bound}
with the fixed enlarged weight $R=1+4b_{\rm P}^2$ therefore bounds the
projected contraction uniformly. The physical factorial tail gives a
sector cutoff uniform in the mesh. At each fixed sector, Poisson
multiple-integral isometry and $P_h\boldsymbol\Upsilon\to\boldsymbol\Upsilon$
give convergence to Eq.~\eqref{eq:poisson-wick-map}.
Taking the sector cutoff after this mesh limit proves
\begin{equation}
 J_h\longrightarrow\mathsf T_{\rm P}\boldsymbol\Upsilon
 \quad\hbox{in }L^2(\mu_{\rm P};\mathrm{HS}).
 \label{eq:poisson-common-path-limit}
\end{equation}
Bounded measurable hardware and reference rates follow by step-function
approximation. Keep the original Poisson measure fixed and average
deterministic kernels with respect to $dt$ and noise kernels with respect
to $\lambda_\alpha(t)dt$. The rate bounds make these norms equivalent
to their step-function versions. Dominated convergence handles each
finite sector, and the enlarged weighted bound handles the common tail.
Equivalently, the first-mark probabilities on a cell are computed from
the original integrated rates in Eq.~\eqref{eq:poisson-common-cell}.
The same local moment estimates hold.

Convergence in Eq.~\eqref{eq:poisson-common-path-limit} implies integrated
Choi trace-norm convergence. A subsequence converges almost surely, so
positivity passes to the limit. The reference average of the output
partial trace remains $I_i$. Integrability gives countable additivity,
and the limiting density is a CP, normalized counting instrument.

The feedback argument is identical to the Wiener case.  A finite-memory
controller partitions the common Poisson path into at most
$N_{\mathbf B}\leq m^K$ frozen-program events.  Applying
Eq.~\eqref{eq:poisson-weighted-bound} on each disjoint event and using
Cauchy--Schwarz gives
\begin{equation}
 \int\left\Vert\sum_c\mathbf1_{E_c}
 \mathsf T_{{\rm P},c}\boldsymbol\Delta\right\Vert_2d\mu_{\rm P}
 \leq\sqrt{N_{\mathbf B}}C_{{\rm P},0}
 \Vert\boldsymbol\Delta_\Lambda\Vert_{\Gamma_{r_{\rm P}}^\partial}.
 \label{eq:poisson-feedback-bound}
\end{equation}
Combining this with the physical factorial tail and the same Choi
trace-norm conversion used for Eq.~\eqref{eq:uniform-operation-bound}
proves the counting part of Theorem~\ref{thm:stability}, including the
optimized fidelity and half-diamond bounds.  If only the intensities are
perturbed on a common finite-dimensional dilation, the hybrid telescope in
Eq.~\eqref{eq:dilation-lipschitz} gives the corresponding realization-level
Lipschitz estimate.

\subsection{Finite-state feedback and optimized performance}

There are $K$ fixed response intervals and at most $m$ persistent
classical states. Each state specifies a bounded deterministic control
program in an interval. A fixed finite alphabet is read at each response
boundary; its Borel grouping uses only the past. Freezing a controller
history gives one of at most $m^K$ deterministic programs. Its occurrence
event $E_{\boldsymbol c}$ is defined on the common reference record
space and is independent of the input and the unknown physical process.
The feedback density is
\begin{equation}
 J^{\mathsf C}(W)=\sum_{\boldsymbol c}
 \mathbf1_{E_{\boldsymbol c}}(W)
 \mathsf T_{\boldsymbol c}\boldsymbol\Upsilon(W)
 \otimes|c_K\rangle\langle c_K|.
 \label{eq:feedback-density}
\end{equation}
To establish causal normalization, implement the updates on the finite-grid instruments of
Eq.~\eqref{eq:exact-cell-kraus}. Each finite-grid feedback instrument
is CP and normalized by conditioning successively on the past,
including the bridges of completed cells revealed at response times.
All frozen histories use the same smoothed finite comb and the same
fixed Borel readout functions on that common reference space.
Its frozen-program decomposition has exactly the same events
$E_{\boldsymbol c}$. Multiplication by their indicators contracts
$L^1$; the finite sum and Eq.~\eqref{eq:identification-limit} prove
that its limit is Eq.~\eqref{eq:feedback-density}, which is therefore
also CP and normalized. This argument permits discontinuous Borel
readout groupings without requiring convergence of posterior states.

Identical frozen contractions can be combined by taking the union of their events. Let $N_{\mathbf B}$ bound the number of resulting programs, so $N_{\mathbf B}\leq m^K$. These events partition the common reference space. The final label is a deterministic function of the record, so attaching it preserves the pointwise matrix norm. Consequently, for $X_{\boldsymbol c}=\mathsf T_{\boldsymbol c}\boldsymbol\Delta$,
\begin{equation}
\begin{split}
 \int\Big\Vert\sum_{\boldsymbol c}\mathbf1_{E_{\boldsymbol c}}
 X_{\boldsymbol c}\Big\Vert_2d\mu_W
 &\leq\left(\sum_{\boldsymbol c}\int_{E_{\boldsymbol c}}
 \Vert X_{\boldsymbol c}\Vert_2^2d\mu_W\right)^{1/2}\\
 &\leq\sqrt{N_{\mathbf B}}C_0
 \Vert\boldsymbol\Delta_\Lambda\Vert_{\Gamma_r^\partial}.
\end{split}
 \label{eq:disjoint-program-bound}
\end{equation}
Here $\Lambda_\chi$ is the active insertion set for the selected modality, including label $0$ on the Poisson side whenever $B_{0\alpha}\ne0$. Projection onto that set commutes with the corresponding Wick contraction, and its cardinality replaces $q$ in the sector bound.

For two open physical processes set
$\delta=\Vert\boldsymbol\Upsilon-
\widetilde{\boldsymbol\Upsilon}\Vert_{\Gamma_1^\partial}$ and
$\boldsymbol\Delta=\boldsymbol\Upsilon-
\widetilde{\boldsymbol\Upsilon}$. Write
$x_n=\Vert\boldsymbol\Delta_{\Lambda_\chi}^{(n)}\Vert_2$ and
$Q=q_{\rm act}^{(\chi)}T$. The triangle inequality at each order gives
$x_n^2\leq4C_\partial^2Q^n/n!$, while
$\sum_nx_n^2\leq\delta^2$. Cauchy--Schwarz across orders yields
\begin{equation}
\begin{aligned}
 \Vert\boldsymbol\Delta_{\Lambda_\chi}\Vert_{\Gamma_r^\partial}^2
 &=\sum_{n\geq0}r^nx_n^2\\
 &\leq\left(\sum_nx_n^2\right)^{1/2}
       \left(\sum_nr^{2n}x_n^2\right)^{1/2}\\
 &\leq2C_\partial e^{r^2Q/2}\delta.
\end{aligned}
 \label{eq:physical-interpolation-squared}
\end{equation}
Taking the square root gives
\begin{equation}
 \Vert\boldsymbol\Delta_{\Lambda_\chi}\Vert_{\Gamma_r^\partial}
 \leq\sqrt{2C_\partial}\,
 e^{r^2q_{\rm act}^{(\chi)}T/4}\sqrt\delta .
 \label{eq:physical-interpolation}
\end{equation}
Let $D=d_id_o$. Since $\Vert X\Vert_1\leq\sqrt D\Vert X\Vert_2$ and the
diamond norm of a map is bounded by the trace norm of its unnormalized
Choi matrix, Eqs.~\eqref{eq:weighted-wick-bound}--\eqref{eq:physical-interpolation}
imply the uniform record-operation bound
\begin{align}
 \sup_{\mathsf C}\frac12
 \Vert\mathcal N_{\boldsymbol\Upsilon,\mathsf C}
 -\mathcal N_{\widetilde{\boldsymbol\Upsilon},\mathsf C}\Vert_\diamond
 &\leq A_{\mathbf B}\sqrt\delta,\\
 A_{\mathbf B}&=\sqrt{DN_{\mathbf B}C_\partial/2}\,C_0
 e^{r^2q_{\rm act}^{(\chi)}T/4}.
 \label{eq:uniform-operation-bound}
\end{align}
The output here includes the classical record and physical system;
passing to the actual accessible coarse record only decreases the norm.
The bound includes a reference entangled with the open input. Encoders,
conditional CPTP decoders, and randomized controller rules cannot
increase it. Therefore, for the identical feasible strategy class,
\begin{equation}
\begin{gathered}
 |F_e^*(\boldsymbol\Upsilon)-F_e^*(\widetilde{\boldsymbol\Upsilon})|
 \leq A_{\mathbf B}\sqrt\delta,\\
 |E_\diamond^*(\boldsymbol\Upsilon)-E_\diamond^*(\widetilde{\boldsymbol\Upsilon})|
 \leq A_{\mathbf B}\sqrt\delta .
\end{gathered}
 \label{eq:optimized-stability}
\end{equation}
Here $C_\partial=d_i$. The constants are uniform in hidden environment dimension and depend on the common boundary, instrument library, and response resources.

\subsection{Computable moduli and realization-level Lipschitz bounds}
The H\"older and optimized cutoff estimates take as input the open-cPT coefficient error in the fixed operation basis. Both use physical factorial tails uniformly over compatible processes and the declared controller class. The realization-level estimate below instead takes an initial bath-state error and a joint-generator error on a common dilation. These inputs give complementary operational guarantees: coefficient data control abstract open-process approximations, while realization data control specified physical perturbations.

The realization-level estimate is independent of the record modality. The
same hybrid telescope applies when the retained field is Wiener, Poisson,
or a finite product of the two, provided the causal instrument and its
record law are held fixed while replacing the underlying joint process
propagator. Changing a monitored jump rate generally changes the visible
record law and is not controlled by a reduced Lindblad-generator difference
alone. Such a perturbation requires a common-record comparison of the
counting instruments, or a joint dilation bound that explicitly includes
the changed instrument maps. This is distinct from the coefficient-norm
stability estimate.

\begin{table}[!htbp]
\caption{Operational stability bounds for the single-qubit instrument: H\"older interpolation, optimized sector cutoff, and the direct Lipschitz estimate of Eq.~\eqref{eq:unweighted-wick-lipschitz}.}
\label{tab:stability-modulus}
\begin{ruledtabular}
\begin{tabular}{cccc}
$\delta$ & H\"older & Cutoff & Lipschitz\\
$10^{-2}$ & $0.53557$ & $0.04077$ & $0.009106$\\
$10^{-3}$ & $0.16936$ & $0.00519$ & $0.000911$\\
$10^{-4}$ & $0.05356$ & $0.00064$ & $0.0000911$
\end{tabular}
\end{ruledtabular}
\end{table}
The same three estimates can be evaluated for a counting instrument. As a
single-qubit scale example, take one unitary jump channel
$J=\sqrt{\kappa}Z$ with $\kappa=0.1$ and $T=1$, with reference rate
$\lambda=\kappa$. Its complete dictionary is
$a_0=-\kappa$, $a_{ZZ}=\kappa$, $B_0=-1$, and $B_{ZZ}=1$.
All other Pauli operation-basis coefficients vanish.
Consequently $q_{\rm act}^{(\rm P)}=2$, $b_{\rm P}^2=2\kappa$, and
$a_0^*=-\kappa$, $a^2=\kappa^2$, giving $r_{\rm P}=1.2$ and
$C_{{\rm P},0}=\exp(-0.095)$. The resulting bounds are
listed in Table~\ref{tab:stability-modulus-poisson}; the table uses the
same boundary dimensions and one frozen program as the Wiener example.

\begin{table}[!htbp]
\caption{Operational stability bounds for the single-qubit Poisson
instrument with $J=\sqrt{0.1}Z$, $T=1$: H\"older interpolation, optimized
sector cutoff, and the unweighted coefficient-norm Lipschitz estimate.}
\label{tab:stability-modulus-poisson}
\begin{ruledtabular}
\begin{tabular}{cccc}
$\delta$ & H\"older & Cutoff & Lipschitz\\
$10^{-2}$ & $0.37365$ & $0.03038$ & $0.009105$\\
$10^{-3}$ & $0.11816$ & $0.003692$ & $0.000911$\\
$10^{-4}$ & $0.03736$ & $0.0004472$ & $0.0000911$
\end{tabular}
\end{ruledtabular}
\end{table}
The factorial tail also yields a cutoff estimate. Separate the sectors at $n$, bound $r^j\leq r^n$ for $j\leq n$, and apply Eq.~\eqref{eq:open-tail} to each of the two physical processes above the cutoff. This gives
\begin{equation}
 \Vert\boldsymbol\Delta_{\Lambda_\chi}\Vert_{\Gamma_r^\partial}^2
 \leq r^n\delta^2+4d_i^2
 \sum_{j>n}\frac{(rq_{\rm act}^{(\chi)}T)^j}{j!}.
 \label{eq:cutoff-proof}
\end{equation}
Taking the infimum and using Eq.~\eqref{eq:disjoint-program-bound} proves Eq.~\eqref{eq:cutoff-modulus}. At each prescribed $\delta$, the cutoff balances the amplified low-sector error against the remaining factorial tail. Its parameters are the union of actually used operation labels $\Lambda$ and the distinct frozen programs of the controller. The response count $K$ is fixed by its decision times and is independent of the numerical propagation mesh. The same proof permits the sharper input $\delta_\Lambda=\Vert\boldsymbol\Delta_\Lambda\Vert_{\Gamma_1^\partial}$ in place of $\delta$. For any prescribed accuracy, first fix a cutoff making the tail small, then take $\delta$ small; thus $\omega(\delta)\to0$. If the weighted process error is known directly, Eq.~\eqref{eq:disjoint-program-bound} instead gives the Lipschitz estimate
\begin{equation}
 d_{\rm rec}\leq\tfrac12\sqrt{DN_{\mathbf B}}C_0
 \Vert\boldsymbol\Delta_\Lambda\Vert_{\Gamma_r^\partial}.
 \label{eq:weighted-lipschitz}
\end{equation}

The Wick estimate admits a one-parameter refinement. For any $\eta>0$, replace the map $V$ in the Fock proof by $V_\eta f=\sqrt\eta f\oplus Cf$, and contract its first factor with the exponential vector of $a_\perp/\sqrt\eta$. The deterministic coefficients are unchanged, while $\Vert V_\eta\Vert^2\leq\eta+b^2$. Consequently,
\begin{equation}
\begin{gathered}
 \Vert\mathsf T\boldsymbol X\Vert_{L^2(\mu_W;\mathrm{HS})}
 \leq C_\eta\Vert\boldsymbol X_\Lambda\Vert_{\Gamma_{r_\eta}^\partial},\\
 r_\eta=\max\{1,b^2+\eta\},\qquad
 C_\eta=\exp\!\left(a_0^*T+\frac{a^2T}{2\eta}\right).
\end{gathered}
 \label{eq:parameterized-wick-bound}
\end{equation}
Thus all preceding weighted, H\"older, and cutoff estimates can be minimized over $\eta$ with $r,C_0$ replaced by $r_\eta,C_\eta$. The choice $\eta=1$ recovers their original constants.

\begin{corollary}[Unweighted Lipschitz bound]
If $b^2<1$ in the fixed operation and time normalization, then
\begin{equation}
\begin{split}
 \sup_{\mathsf C}d_{\rm rec}
 &\leq L_{\mathbf B}\Vert\boldsymbol\Delta_\Lambda\Vert_{\Gamma_1^\partial},\\
 L_{\mathbf B}
 &=\tfrac12\sqrt{DN_{\mathbf B}}\,
 \exp\!\left(a_0^*T+\frac{a^2T}{2(1-b^2)}\right).
\end{split}
 \label{eq:unweighted-wick-lipschitz}
\end{equation}
The change in either optimized objective obeys the same bound.
\end{corollary}
\begin{proof}
Choose $\eta=1-b^2$ in Eq.~\eqref{eq:parameterized-wick-bound}. Then $r_\eta=1$, so Eq.~\eqref{eq:disjoint-program-bound} gives the operational bound directly. CPTP postprocessing and pairing the same strategies give both optimized-value bounds.
\end{proof}

The same coefficient-norm argument applies to counting instruments: use the
Poisson isometry and replace $b$ by $b_{\rm P}$ in the parameterized
contraction estimate, including identity insertions whenever
$B_{0\alpha}\ne0$. In particular, whenever the normalized counting
coefficient satisfies $b_{\rm P}^2<1$, choosing the corresponding split
parameter gives an unweighted Lipschitz bound with the same boundary and
memory prefactor. Thus the Lipschitz statement is a property of the
bounded instrument--process contraction, not of a specifically Gaussian
record.

For an explicit scale evaluation, take a qubit with $H=0$, $L=\sqrt\kappa Z$, $\kappa=0.1$, and $T=1$. In the Pauli left-right basis its coefficients are $a_0=-\kappa$, $a_{ZZ}=\kappa$, and $B_{Z0}=B_{0Z}=\sqrt\kappa$. Therefore $q_{\rm act}=3$, $b^2=0.2$, and $d_i=d_o=2$. Record processing with no adaptive drive uses one frozen program. The choice $\eta=1$ gives $r=1.2$, $C_0=e^{-0.095}$, while $\eta=0.8$ gives $r_\eta=1$ and $L_{\mathbf B}=e^{-0.09375}\simeq0.91051$. Table~\ref{tab:stability-modulus} evaluates all three estimates for the same unweighted process error. All operational bounds may additionally be capped at one.

A common physical realization supplies a second route to quantitative stability. Let two processes have initial bath states $\sigma_E,\widetilde\sigma_E$ and bounded joint system--environment generators $\mathcal L_t,\widetilde{\mathcal L}_t$ on the same finite-dimensional dilation. Use identical causal measurement, control, and terminal interfaces. Then uniformly over those strategies,
\begin{equation}
\begin{split}
 d_{\rm rec}&\leq\varepsilon_{\rm dil},\\
 \varepsilon_{\rm dil}
 &=\tfrac12\Vert\sigma_E-\widetilde\sigma_E\Vert_1
 +\tfrac12\int_0^T\Vert\mathcal L_t-
 \widetilde{\mathcal L}_t\Vert_\diamond dt.
\end{split}
 \label{eq:dilation-lipschitz}
\end{equation}
To prove this, partition time and retain the bath, reference, controller, and measurement record. A hybrid telescope replaces one joint process propagator at a time. Every intervening instrument, including its causal update, is CPTP on this enlarged output and contracts trace distance. Duhamel's formula bounds each replaced propagator by the integral of the generator difference over its cell. The initial-state replacement costs $\frac12\Vert\sigma_E-\widetilde\sigma_E\Vert_1$. Summation and the continuous-instrument limit prove Eq.~\eqref{eq:dilation-lipschitz}. A bounded continuous monitoring evolution follows by splitting the process and instrument in each cell and taking the same normalized limit. Both optimized objectives change by at most $\varepsilon_{\rm dil}$. For Hamiltonian perturbations the generator contribution is at most $\int_0^T\Vert H_t-\widetilde H_t\Vert_{\rm op}dt$. This estimate measures realization error directly, while Eq.~\eqref{eq:cutoff-modulus} measures open cPT coefficient error.

\section{Continuous access gaps and a diffusive witness}
\label{app:accessbridge}

\subsection{A process-level access criterion}
Fix an open physical cPT, an encoding, and a monitoring instrument with
integrable record Choi density $C_y\geq0$. All contractions here retain
the logical reference and the physical output. The record space is
standard Borel and carries a common reference measure $\mu$.
For a terminal CPTP map $\mathcal D$, let
$\ell_{\mathcal D}(C)$ be the entanglement-fidelity contribution of
the branch whose normalized-input Choi matrix is $C$. This functional
is linear and continuous in $C$. The decoder Choi set is compact in
finite dimension, so
\begin{equation}
 h(C)=\max_{\mathcal D\in\mathrm{CPTP}}\ell_{\mathcal D}(C)
 \label{eq:bridge-support}
\end{equation}
is finite, continuous, positively homogeneous, and convex on positive
Choi matrices. It is bounded by $\operatorname{Tr}C$.

For a measurable full-path decoder, the fidelity is
$\int\ell_{\mathcal D_y}(C_y)d\mu(y)$. This is bounded above by
$\int h(C_y)d\mu(y)$. Conversely, choose a countable dense family
of CPTP maps $\{\mathcal D_k\}$. For any $\varepsilon>0$, select
the first $k$ for which
$\ell_{\mathcal D_k}(C_y)\geq h(C_y)-\varepsilon w(y)$,
where $w(y)=1+\operatorname{Tr}C_y$ on a finite reference measure.
Each selection set is measurable: it is the set satisfying the displayed
inequality minus the finitely many sets for earlier indices. Continuity
of $h$ and $\ell_{\mathcal D_k}$ and measurability of $C_y,w(y)$ give
measurability of those sets. Density of the decoder family ensures that
one index exists whenever $w(y)>0$. Integrating the selected score gives
a deficit at most $\varepsilon\int w\,d\mu$. Letting
$\varepsilon\downarrow0$ proves the full-path identity in
Eq.~\eqref{eq:accessjensen}. For Wiener probability measure this
choice of $w$ is integrable; for a general sigma-finite reference
measure one first chooses a strictly positive integrable $w$.

If $E_z$ are finitely many fixed Borel record classes, all paths in
one class share a single decoder. Their Choi contributions add, so
the optimum is $\sum_z h(C_z)$ with
$C_z=\int_{E_z}C_y d\mu(y)$. For a maximizing decoder
$\mathcal D_z$,
\begin{equation}
 h(C_z)=\int_{E_z}\ell_{\mathcal D_z}(C_y)d\mu(y)
 \leq\int_{E_z}h(C_y)d\mu(y).
 \label{eq:bridge-jensen-proof}
\end{equation}
The pointwise difference is nonnegative. Equality of the integrals
therefore holds exactly when that same decoder is optimal for almost
every path in $E_z$. Zero-probability classes have no effect.
For a finite causal interface with fine labels $z$ and a coarser
interface $u=g(z)$, the coarse Choi branch is the sum of the fine
branches in that class. Repeating Eq.~\eqref{eq:bridge-jensen-proof}
with a finite sum gives Eq.~\eqref{eq:finite-access-gap}. The
nonnegative summands vanish precisely when a maximizing coarse
decoder also maximizes every positive-probability fine branch in
its class. Hence one incompatible group makes the gap strict.
For nested resource classes with the same open process, every strategy
in the smaller class is available in the larger class, so the
optimized fidelity gain and the optimized half-diamond-error gain
are nonnegative. Theorem~\ref{thm:stability} bounds the change of each
optimized objective between compatible physical processes by
$A_{\mathbf B}\sqrt\delta$ for a common budget bound. Applying the
triangle inequality to the difference of the two optima gives
Eq.~\eqref{eq:access-gap-stability}. If the reference gain is $G>0$,
the perturbed gain is at least $G-2A_{\mathbf B}\sqrt\delta$.
This proves the stated strict neighborhood for either objective.

\subsection{A continuous diffusive instrument with a strict gap}
Here is a finite-environment, bounded-Hamiltonian open cPT to which the
criterion applies. Encode the unknown logical qubit in the three-qubit
repetition code $V$, and let $X_L=X_1X_2X_3$ denote its logical flip.
Initialize a sensor qubit $A$ in $|0\rangle$. A retained bath bit
$B$ begins in the equal mixture of $|0\rangle$ and $|1\rangle$. For
a preparation time $t_p>0$, apply the bounded Hamiltonian
\begin{equation}
 H_{DAB}=\frac{\pi}{2t_p}|1\rangle\langle1|_B
 \otimes X_L X_A.
 \label{eq:bridge-hamiltonian}
\end{equation}
At the end of this interval, its two bath branches carry respectively
$V\rho V^\dagger\otimes|0\rangle\langle0|_A$ and
$X_LV\rho V^\dagger X_L\otimes|1\rangle\langle1|_A$. Both
branches remain inside the same code space and have identical terminal
syndrome, so the physical data output alone cannot identify the flip.
The relative phase
of the controlled flip is irrelevant because the bath branches are
classical and orthogonal. The bath persists throughout the experiment.

During a separate interval of length $t_m>0$, the sensor undergoes
continuous homodyne monitoring with
$L=\sqrt\kappa Z_A$, $\kappa>0$, and zero Hamiltonian. Its
unnormalized linear propagator relative to a standard Wiener path is
\begin{equation}
 G_{t_m}(y)=\exp(\sqrt\kappa Z_Ay-\kappa t_m I_A).
 \label{eq:bridge-linear-propagator}
\end{equation}
Let $\boldsymbol\Upsilon_{DAB}^{\partial}$ denote the open cPT
defined by the entire bounded Hamiltonian evolution, with its logical
input and data-sensor output left open. If $\mathcal I_y^{Z_A}$ is the
continuous Wiener instrument on the sensor, the accessible record
channel is its open-boundary contraction
\begin{equation}
 \mathcal N_y=\operatorname{Tr}_{AB}\!
 \left[\boldsymbol\Upsilon_{DAB}^{\partial}
 \star\mathcal I_y^{Z_A}\right].
 \label{eq:bridge-cpt-contraction}
\end{equation}
The star means the time-ordered cPT/instrument contraction with
identity operations on unmonitored ports. Its time-resolved
operation-basis expansion is the Wick contraction established in
Appendix~\ref{app:stability}; the bounded realization here evaluates
that contraction exactly through Eq.~\eqref{eq:bridge-linear-propagator}.
The monitoring operator is bounded and traceless on the accessible
data-sensor system, and the measurement acts for strictly positive
time. Contracting this continuous instrument with the open cPT,
then taking the fixed terminal quantum interface to discard $A$ and
$B$, gives Eq.~\eqref{eq:diffusive-witness-instrument}. Indeed,
$G_{t_m}(y)|0\rangle=e^{\sqrt\kappa y-\kappa t_m}|0\rangle$
and
$G_{t_m}(y)|1\rangle=e^{-\sqrt\kappa y-\kappa t_m}|1\rangle$.
Squaring these amplitudes yields $g_+$ and $g_-$, and the Gaussian
moment-generating identity gives
\begin{equation}
 \int g_+(y)d\mu_W(y)=\int g_-(y)d\mu_W(y)=1.
 \label{eq:bridge-normalization}
\end{equation}
Thus the instrument is CP, countably additive, and normalized, with an exact continuous-time Wick/cMPS contraction.

The record-conditioned logical channel is a nonnegative combination
of identity and the logical bit flip. As their Choi states are
orthogonal Bell states, an arbitrary terminal CPTP recovery cannot
exceed the larger branch weight. Conditional $I$ or $X_D$ recovery
attains this bound. With no retained record, the two integrated
weights are equal and the optimal fidelity is $1/2$. Keeping the one
bit $z=\operatorname{sgn}y$ selects $I$ for $y\geq0$ and $X_L$ for
$y<0$. Under the physical conditional law for bath value zero,
$y$ is Gaussian with mean $2\sqrt\kappa t_m$ and variance $t_m$;
the other bath value has the negative mean. Consequently,
\begin{equation}
 F_{\rm bit}=F_{\rm path}=\Phi(2\sqrt{\kappa t_m}),
 \qquad F_{\rm discard}=\tfrac12.
 \label{eq:bridge-fidelities}
\end{equation}
The maximum is already determined by the sign, so keeping the
continuous record cannot improve further in this task. After the
attaining recovery, the residual logical channel is a mixture of
identity and $X_L$ with flip probability $1-F$. The maximally
entangled lower bound and convex-channel upper bound both give
$\tfrac12\Vert\mathcal M-\mathrm{id}\Vert_\diamond=1-F$.
Any alternative decoder has half diamond error at least its
entanglement infidelity. Therefore the values in the main text are
global optima for both objectives, even among non-Pauli decoders.

The witness belongs to a fixed-resource diffusive class with a
finite-dimensional persistent bath, a fixed sensor interface, one
record-readout boundary, one retained classical bit, and bounded
Hamiltonians. In comparing its access classes, keep the encoding,
Hamiltonian, monitoring, and terminal physical interface fixed, with
zero online drive; only record access and terminal recovery vary.
Applying Eq.~\eqref{eq:access-gap-stability} with its
positive gap shows that this access advantage survives compatible
open-cPT perturbations in the stated norm. The bound is uniform over the admissible decoders and controllers.

\subsection{Exact bath-prior perturbations}
Keep the same bounded realization and replace the bath's equal mixture by
\begin{equation}
 \sigma_B(\pi)=\pi|0\rangle\langle0|+(1-\pi)|1\rangle\langle1|,
 \qquad 0<\pi<1.
\end{equation}
The branch map becomes
\begin{equation}
 \mathcal N_y^{(\pi)}(\rho)=\pi g_+(y)V\rho V^\dagger
 +(1-\pi)g_-(y)X_LV\rho V^\dagger X_L.
\end{equation}
\begin{table}[!htbp]
\caption{Exact bath-prior recovery fidelities at $\kappa t_m=1/4$. All physical settings and the sign-bit interface are fixed.}
\label{tab:bath-prior-stability}
\begin{ruledtabular}
\begin{tabular}{cccc}
$\pi$ & Discard & Sign bit & Full path\\
$0.50$ & $0.50000$ & $0.84134$ & $0.84134$\\
$0.51$ & $0.51000$ & $0.84134$ & $0.84139$\\
$0.55$ & $0.55000$ & $0.84134$ & $0.84256$\\
$0.60$ & $0.60000$ & $0.84134$ & $0.84622$
\end{tabular}
\end{ruledtabular}
\end{table}
Equation~\eqref{eq:dilation-lipschitz} gives a common operational and optimized-value error bound $\varepsilon=|\pi-1/2|$, since only the bath preparation changes. Its coefficient is one and is independent of monitoring strength and controller memory. For the displayed uncorrected record channel the bound is attained on a maximally entangled logical input: the two logical Bell branches are orthogonal and $\int(g_++g_-)d\mu_W=2$.

The open coefficient error can also be evaluated exactly on the monitoring interval, with the fixed preparation prefix included in its input boundary. The three active insertions are $\mathcal E_{Z_A0},\mathcal E_{0Z_A},\mathcal E_{Z_AZ_A}$. On each sensor branch these multiply the coefficient by a unit-modulus sign. After tracing the sensor, the two unnormalized logical Choi vectors remain orthogonal with squared norm two. Every order-$n$ coefficient difference therefore has squared Hilbert--Schmidt norm $8\varepsilon^2$, giving
\begin{equation}
 \Vert\boldsymbol\Delta_\Lambda\Vert_{\Gamma_s^\partial}
 =2\sqrt2\,\varepsilon e^{3s t_m/2}.
 \label{eq:prior-exact-coefficient-error}
\end{equation}
The boundary dimensions are $d_i=2$, $d_o=8$. For $t_m=1$, $\kappa=1/4$, and $\varepsilon=10^{-4}$, the unweighted active error is $0.00126761$; Eq.~\eqref{eq:unweighted-wick-lipschitz} gives an operational bound below $0.002102$, while Eq.~\eqref{eq:dilation-lipschitz} gives the exact value $10^{-4}$. This evaluates coefficient-level and realization-level stability on the same physical process pair.

Let $\alpha=2\sqrt{\kappa t_m}$, $Q=\Phi(\alpha)$, and
$a_\pi=(2\alpha)^{-1}\log[(1-\pi)/\pi]$. The Bell-mixture optimization yields
\begin{align}
 F_{\rm discard}(\pi)&=\max\{\pi,1-\pi\},\\
 F_{\rm bit}(\pi)&=\max\{\pi Q,(1-\pi)(1-Q)\}\nonumber\\
 &\quad+\max\{\pi(1-Q),(1-\pi)Q\},\\
 F_{\rm path}(\pi)&=\pi\Phi(\alpha-a_\pi)
 +(1-\pi)\Phi(\alpha+a_\pi).
 \label{eq:prior-exact-recovery}
\end{align}
Here the bit remains the fixed sign of $y$, while the full-path decoder compares the two posterior weights and uses the threshold $y/\sqrt{t_m}=a_\pi$. In the interval $1-Q\leq\pi\leq Q$, the sign-bit value stays exactly $Q$. Consequently its advantage is
\begin{equation}
 F_{\rm bit}(\pi)-F_{\rm discard}(\pi)
 =Q-\tfrac12-|\pi-\tfrac12|.
 \label{eq:prior-exact-gap}
\end{equation}
Table~\ref{tab:bath-prior-stability} evaluates the three access interfaces at $\kappa t_m=1/4$. A bath preparation error of $0.1$ leaves a strict sign-bit gain of $0.24134$. Optimal half diamond errors again equal $1-F$ at every prior.

\section{Fixed-process limits and implementable discretization}
\label{app:limits}

\subsection{Complete-record approximation at a fixed process}
On a finite-dimensional system and fixed finite environment, the unnormalized linear measurement propagator satisfies
\begin{equation}
dG_t=(-iH_t-\tfrac12\sum_\alpha L_\alpha^\dagger L_\alpha)G_tdt
+\sum_\alpha L_\alpha G_t dY_\alpha(t)
\end{equation}
under a common reference Wiener measure. For uniformly bounded predictable Hamiltonians and bounded monitoring operators, It\^o estimates and Gronwall's inequality give mean-square convergence of $G_T^{(n)}$ whenever $\mathbb E\int_0^T\norm{H_t^{(n)}-H_t}^2dt\to0$. To handle the correlation of a feedback approximation with $G_t$, first truncate to $\norm{G_t}\leq R$; uniform fourth moments bound the complement uniformly, and then $R\to\infty$.

For a pure open logical input with reference, the unnormalized trajectory Choi density is a rank-one outer product before inaccessible boundaries are traced. The trace norm of the difference of two such outer products is bounded by the product of the vector difference and the sum of vector norms. Uniform second and fourth moments therefore give convergence of the open Choi densities in integrated trace norm. Partial trace over bath and inaccessible fields is contractive. If $\mathcal F_n$ are increasing finite sigma algebras generating the visible path, the finite-record density is the conditional expectation $\mathbb E[J\mid\mathcal F_n]$. Martingale convergence in $L^1$ then gives convergence in the record-operation norm.

For paired record instruments, every conditional recovery is a CPTP postprocessing. Its entanglement-fidelity difference is bounded by the record-operation distance. Taking the supremum over the same recovery class proves convergence of the optimized recovery fidelity. If the same record partition is used on both processes, the result applies directly to compressed records. This fixed-process approximation permits record precision to grow; Theorem~\ref{thm:stability} separately controls process-model approximation over a fixed-resource feedback class.

The discretization mechanism is modality-independent at the level of the
global estimate. For diffusive records, local weak Euler instruments use
Gaussian increments; for counting records, the local update consists of a
no-click propagator together with click Kraus operators. Under bounded
coefficients and a common finite-dimensional realization, each local CP,
trace-preserving approximation converges to its continuous instrument, and
a telescoping contraction argument converts the local error into a uniform
global record-operation bound. The two local expansions have different
collision terms, but the resulting optimized objectives are controlled by
the same triangle inequality and CPTP contractivity. Thus the fixed-grid
and process-approximation decomposition in Eq.~\eqref{eq:combined-grid-process}
applies to both diffusive and counting interfaces.

\subsection{Uniform convergence for a compact finite-state class}
Fix finite system and environment dimensions, a controller with at most $m$ states, a finite number of channels, and time-independent bounded parameters in a compact set $\Theta$. Each time step first applies the state-conditioned unitary evolution, then a weak-counting instrument
\begin{equation}
K_0=\left(I-h\sum_jJ_j^\dagger J_j\right)^{1/2},
\qquad K_j=\sqrt h J_j,
\end{equation}
and finally a classical transition with stochastic matrix $I+hQ$. For sufficiently small $h$, all three stages are CPTP uniformly on $\Theta$. Their product agrees with the exact finite-dimensional semigroup to local diamond error at most $C h^2$, with $C$ uniform on the compact parameter class. Telescoping $T/h$ CPTP factors, each of diamond norm one, gives
\begin{equation}
\sup_{\vartheta\in\Theta}\tfrac12
\norm{\mathcal M_{\vartheta,h}-\mathcal M_\vartheta}_\diamond
\leq CTh.
\end{equation}
For any paired strategy, entanglement fidelity and half diamond error change by at most this distance. Taking the supremum or infimum over the same class proves convergence of the constrained optimal values.

\subsection{Combined process and implementation errors}
Suppose a common admissible class has an additional uniform grid error bound $e_{\mathbf B}(h)$ for the instrument used in Theorem~\ref{thm:stability}. The triangle inequality gives
\begin{equation}
\begin{split}
\tfrac12\Vert\mathcal M_{\U_n,\mathsf C,h}
-\mathcal M_{\U,\mathsf C}\Vert_\diamond
&\leq e_{\mathbf B}(h)\\
&\quad+A_{\mathbf B}\Vert\U_n-\U\Vert_{\Gamma_1^\partial}^{1/2}.
\end{split}
\label{eq:combined-grid-process}
\end{equation}
The first term controls implementation discretization and the second controls open-process approximation, uniformly over the paired strategies. The same bound applies to either optimized objective over that common class.

\section{Numerical implementation}
\label{app:numerics}

\subsection{Storage controller and open-channel propagation}
Both storage benchmarks use the three-qubit repetition code and thirteen-state controller of Sec.~\ref{sec:numerics}. Time is measured in units with $T=1$. The two waiting stages have total mean duration $0.08$, and the two pulse stages have total mean duration $0.16$. During the selected pulse stages, the applied Hamiltonian is $11X_i$; the no-drive comparisons set this amplitude to zero. These parameters remain fixed as the monitoring strength varies.

The bath calculation propagates the logical reference, system, four-qubit environment, and controller blocks in Eq.~\eqref{eq:contracted-controller-blocks}. The environment is traced at the terminal boundary. The independent-flip calculation propagates analogous open Choi blocks with the three Lindblad generators. Classical fourth-order Runge--Kutta integration uses 128 steps for the bath and 256 for independent flips. The same independent-flip branches supply the Petz and Pauli-frame comparisons in Table~\ref{tab:petzcase}. All propagations sum inaccessible trajectories deterministically.

\subsection{Decoder solver and feasibility repair}
A complex Hermitian SDP optimizes the complete physical-output decoder for each visible branch, using the conjugated Choi objective in Eq.~\eqref{eq:SDP}. The $16\times16$ decoder Choi matrices have logical-output--physical-input ordering. The programs are formulated in CVXPY and solved with CLARABEL using warm starts. Absolute and relative duality-gap tolerances and feasibility tolerance are all $10^{-10}$. A branch of mass $\mu_r=\Tr C_r$ uses the unit-mass objective $\widehat W_r=W_r/\mu_r$, with its value subsequently rescaled by $\mu_r$. Branches below mass $10^{-12}$ are omitted from the solves; their total positive mass $\varepsilon_{\rm skip}$ is added to the upper value.

The returned decoder Choi matrix is made positive by replacing eigenvalues below $10^{-14}$ with $10^{-14}$, giving $D_r^+$. Its output partial trace $M_r=\Tr_LD_r^+$ is positive definite. Congruence normalization then gives the CPTP decoder
\begin{equation}
 D_r^{\rm feas}=(I_L\otimes M_r^{-1/2})D_r^+
 (I_L\otimes M_r^{-1/2}).
 \label{eq:decoder-primal-repair}
\end{equation}
Its original, unscaled objective $\Tr(W_rD_r^{\rm feas})$ is evaluated directly and compared with the feasible syndrome-conditioned Pauli-frame baseline; the larger value supplies the branch lower value.

For the dual, Hermitize the trace-preserving constraint multiplier to obtain $Y_r$ and form $S_r=I_L\otimes Y_r-\widehat W_r$. Define
\begin{equation}
\begin{split}
 \xi_r&=\max\{0,-\lambda_{\min}(S_r)\}+10^{-11},\\
 Y_r^{\rm feas}&=Y_r+\xi_r I_B.
\end{split}
 \label{eq:decoder-dual-repair}
\end{equation}
The branch upper value is $\mu_r\Tr Y_r^{\rm feas}$. Summation and addition of $\varepsilon_{\rm skip}$ give the total upper value. Its difference from the summed feasible lower values defines the primal--dual gap.

\subsection{Strength scans and exact event contractions}

The storage scan uses eleven matched strengths $\kappa=0,1,2,3,4,6,8,10,12,14,16$, with feasible CPTP decoder values plotted at each point. Feedback percentages refer to reductions in entanglement infidelity. The capacity scan uses the exact event model at 41 equally spaced values of $K$ from 0 to 4, together with $K=\log5$. At each strength it evaluates five causal memory budgets and the full-record benchmark.

For spacetime interrogation, exact cMPS contraction gives three event Kraus operators per window and nine open Choi branches. The three access interfaces are applied before terminal recovery optimization. The scan uses $p=0.05,0.1,0.25$ and $K=0,0.25,0.5,0.75,1,1.5,2,3,4,6,8$, together with the exact projective endpoint. These give 108 recovery problems. Table~\ref{tab:spacetime} reports the analytic values of Theorem~\ref{thm:records}, attained by the contracted decoder.

\subsection{Gate-channel propagation}
\label{app:gate-numerics}
The gate calculations use dimensionless time $u=t/T$ and strength $k=\kappa T$. On each of $n$ equal cells, the measured involutions are frozen at the midpoint. Their exact dephasing channels are
\begin{equation}
\mathcal Q_{g,h}(X)=\frac{1+e^{-2kh}}2X
+\frac{1-e^{-2kh}}2gXg,\qquad h=1/n.
\label{eq:gate-frozen-step}
\end{equation}
The two factors commute because the rotated stabilizers commute. Their composition is an exact frozen monitoring channel and converges under smooth midpoint freezing at each fixed $k$.

With noise, each step composes an environmental half step, midpoint monitoring, and an environmental half step. Independent flips use exact Pauli mixture channels. The bath Hamiltonian is split into bath-$Z$ and mutually commuting $X$-string parts, using the symmetric $Z/2$, $X$, $Z/2$ unitary product. During the hold window, the frozen counting instrument is integrated exactly before environmental splitting, retaining all three branches.

Both noisy gate panels use the ten positive strengths $k=1,4,16,32,64,128,256,512,768,1024$ per environment. Fixed paths use at least 8192 cells, increasing as $32k$; adaptive paths use 8192 or 16384 cells. The original fixed-speed loop and the paused readout protocol have their respective time schedules defined in Sec.~\ref{sec:holonomic} and Appendix~\ref{app:noisyholonomy}. At zero monitoring, the bath and independent-flip target fidelities are $0.815489$ and $0.833789$. For path selection, adjoint suffix propagation evaluates the scores $f(y,b)$ of Eq.~\eqref{eq:gate-policy-optimum}.

\subsection{Numerical precision}
\label{app:numerical-precision}
The numerical results have three sources of precision. Counting-task
values follow from the exact branch weights and finite causal enumeration.
Diffusive full-record and quantizer scores follow from Gaussian
probabilities. Their evaluation uses floating-point arithmetic without
time discretization. Storage and gate channels are propagated on time
grids. For each propagated storage channel, the repaired primal and dual
SDP solutions bracket its decoder optimum; their gap is below $10^{-9}$
for the selected storage-table comparisons.

Table~\ref{tab:numerical-precision} reports changes between the stated
time grids. These quantify observed refinement sensitivity, rather than
an upper bound on the error relative to the continuous-time limit.
The SDP brackets concern the channel supplied to the solver and do not
include propagation error. For gate control,
Eq.~\eqref{eq:gate-policy-optimum} is an exact finite enumeration for
specified branch scores; its reported values use the propagated scores.
Displayed digits identify those computed values. The storage entries
below concern $\kappa=8$; the noisy gate entry covers all twenty displayed
controller comparisons.
\begin{table}[t]
\caption{Entanglement-fidelity changes between the indicated time grids at fixed model parameters.}
\label{tab:numerical-precision}
\begin{ruledtabular}
\begin{tabular}{lcc}
Calculation & Cell counts & $|\Delta F|$\\
Storage, bath & $128,\ 256$ & $<10^{-9}$\\
Storage, flips & $256,\ 512$ & $<10^{-12}$\\
Noiseless gate, $k=256$ & $16384,\ 32768$ & $3.48\times10^{-6}$\\
Noisy path control & $n/2,\ n$ & $<3.2\times10^{-4}$
\end{tabular}
\end{ruledtabular}
\end{table}
Every displayed one-bit gate gain is positive on both compared grids. For the four benchmarks of Table~\ref{tab:holonomic-memory}, define the score-matrix difference
\begin{equation}
\epsilon_{\rm grid}=\sum_y\max_b|f_{16384}(y,b)-f_{8192}(y,b)|.
\end{equation}
For either grid, changing every branch score by at most its corresponding
summand changes any policy value by at most $\epsilon_{\rm grid}$.
Taking maxima preserves this bound. Thus
Eq.~\eqref{eq:gate-policy-optimum} bounds the optimized-value difference
between these two grids by $\epsilon_{\rm grid}$ and their access-gap
difference by $2\epsilon_{\rm grid}$. The largest benchmark gap difference
bound is below $8.7\times10^{-5}$, compared with a smallest reported
one-bit gain of $0.035779$. The fixed-path grid changes at the examined
noisy benchmarks are below $1.8\times10^{-5}$.

\section{Fixed-target access and holonomic path control}
\label{app:targetholonomy}
\subsection{Proof of the target-aware corollary}
For a normalized-input branch Choi density $C$, set
\begin{equation}
\ell_{U,\mathcal D}(C)=\Tr[\ket{\Phi_U}\bra{\Phi_U}(\id\otimes\mathcal D)(C)].
\end{equation}
This is linear in $C$ and continuous in the finite-dimensional decoder Choi matrix. A nonempty compact $\mathfrak R$ therefore attains the maximum. Its supremum of linear functionals is convex and positively homogeneous even when $\mathfrak R$ is not convex. A measurable maximizing decoder can be selected for a measurable finite-dimensional Choi density. For a coarse event $E_z$, let $\mathcal D_z$ maximize the integrated branch. Then
\begin{equation}
\begin{split}
&\int_{E_z}h_{U,\mathfrak R}(C_y)\,d\mu(y)
-h_{U,\mathfrak R}\!\left(\int_{E_z}C_y\,d\mu(y)\right)\\
&\quad=\int_{E_z}[h_{U,\mathfrak R}(C_y)-\ell_{U,\mathcal D_z}(C_y)]\,d\mu(y).
\end{split}
\end{equation}
The integrand is nonnegative; its integral vanishes exactly when $\mathcal D_z$ is optimal almost everywhere on that event. Summing over events proves the access inequality and equality criterion. The finite-label form follows identically.

For any paired admissible strategy on two processes, the recovered channels differ in half diamond norm by at most $d_{\rm rec}$. Expectation against the projector $\ket{\Phi_U}\bra{\Phi_U}$ changes by at most this amount, since the integrated Choi difference has zero trace. The reverse triangle inequality gives the same bound for $E_U$. Taking suprema or infima over the common feasible class proves the two optimized bounds. Applying each bound to both classes gives the factor two for access gaps.

The monitoring extension in Sec.~\ref{sec:stability} fixes the same reference quadratures and detector readout for both compared processes. On a frozen state history, each selected $L_{a,j,c}(t)$ and $H_{j,c}(t)$ is deterministic. The local Kraus normalization and weighted Wick estimates hold with the common coefficient bounds. Multiplication by the same causal event indicators is contractive in the integrated norm; there are still at most $m^K$ histories. These estimates prove the extension for the stated bounded program class and fixed response clock.

\subsection{Moving-code geometry and endpoint identification}
Write the horizontal encoding as $F(t)=W(t)Vu(t)$, where $F^\dagger F=I_L$ and $F^\dagger\dot F=0$. It obeys
\begin{equation}
\dot u(t)=-A(t)u(t),\qquad u(0)=I_L,
\end{equation}
with $A(t)=V^\dagger W^\dagger\dot W V$. For a closed projector loop, $F(T)$ lies in the initial code and its logical action is $V^\dagger F(T)$, proving Eq.~\eqref{eq:holonomy}. A frame change $W(t)V\mapsto W(t)Vq(t)$ changes the connection and endpoint factor together, leaving the physical gate unchanged when the initial logical coordinates are fixed. Keeping only the connection in a nonperiodic frame would omit a physical endpoint identification.

For Eq.~\eqref{eq:minimal-path}, $V^\dagger RV=V^\dagger HRV=0$ and $V^\dagger HV=Z_L$. Direct differentiation gives
\begin{align}
V^\dagger W_\theta^\dagger\partial_\phi W_\theta V
 &=i\frac{\theta}{2\pi}\cos(2\phi)Z_L,\\
u(\phi)&=\exp[-i\theta\sin(2\phi)Z_L/(4\pi)].
\end{align}
Thus $u(2\pi)=I_L$, while $V^\dagger W_\theta(2\pi)V=e^{i\theta Z_L}$. This is the horizontal lift of the repetition-code construction in Ref.~\cite{lanka_holonomic}. It gives $e^{-i\pi/8}T_L$ for $\theta=-\pi/8$, without applying $W_\theta$ as a physical Hamiltonian evolution.

The ideal unitary is a Zeno limit, not the finite-strength branch map. With independent real reference Wiener paths the actual linear unnormalized propagator satisfies
\begin{equation}
dG_t=-\tfrac12\sum_aL_a(t)^\dagger L_a(t)G_t\,dt
+\sum_a L_a(t)G_t\,dY_a(t).
\end{equation}
The physical Kraus density is $G_TV$. Averaging its record-conditioned CP maps over the reference measure gives Eq.~\eqref{eq:holonomic-lindblad}. Adding environmental dynamics gives the contracted cPT--instrument evolution.

\subsection{Fixed decoder and finite-strength channel}
Let $\mathcal E_X=\{I,X_1,X_2,X_3\}$ and $P_s=E_sP_0E_s^\dagger$, with $E_s\in\mathcal E_X$. The decoder has Kraus operators
\begin{equation}
D_s=V^\dagger E_s^\dagger P_s,\qquad
\mathcal D_{\rm std}(X)=\sum_s D_sXD_s^\dagger.
\label{eq:fixed-gate-decoder}
\end{equation}
The syndrome projectors resolve $I_S$, so $\sum_sD_s^\dagger D_s=I_S$. On the code this decoder acts as $V^\dagger$ in the fixed logical coordinates, independently of $\theta$.

The four reference blocks $C_{ab}(0)=V\ket a\bra bV^\dagger/2$ evolve under Eq.~\eqref{eq:holonomic-lindblad}. Their open Choi state is $C=\sum_{ab}\ket a\bra b\otimes C_{ab}$, with logical output $(\id\otimes\mathcal D_{\rm std})(C)$. Appendix~\ref{app:gate-numerics} gives the CPTP midpoint propagation.

The moving-code calculation also identifies the correctability condition for the path. The cross term for the physical $X_3$ error obeys
\begin{equation}
V^\dagger W_\theta(\phi)^\dagger X_3W_\theta(\phi)V
=\sin(\theta\phi/\pi)\sin(2\phi)Z_L,
\end{equation}
which is generally non-scalar. The ancilla and error-syndrome structure of
Ref.~\cite{lanka_steering} supplies the corresponding correctable moving-code
extension. In this way the minimal loop isolates the geometric gate mechanism,
while the same cPT framework accommodates the full correctable path.

\subsection{Noisy gates and causal path programs}
\label{app:noisyholonomy}
\paragraph{Environmental dynamics and physical interfaces.}
The bath calculation uses exactly $H_4$, its parameters, and the initial state $\ket{0000}$ from Sec.~\ref{sec:numerics}, with $T=1$. The initial logical-reference state is independent of that bath. The joint generator is
\begin{equation}
\dot\rho_{SE}=-i[H_4,\rho_{SE}]
+\kappa\sum_{a=1}^2\mathcal D[g_a(t)\otimes I_E](\rho_{SE}).
\end{equation}
Here $\mathcal D[L](X)=LXL^\dagger-\{L^\dagger L,X\}/2$. The independent-flip calculation replaces the Hamiltonian commutator by $\sum_i0.1(X_i\rho_S X_i-\rho_S)$. Both gate protocols have zero applied Hamiltonian. The bath is traced at the final boundary, before the fixed $\mathcal D_{\rm std}$ of Eq.~\eqref{eq:fixed-gate-decoder}. The rotating measured operators enter the full generator throughout the path.

\paragraph{The finite-strength readout instrument.}
Use dimensionless time $u=t/T$, $\alpha=7/16$, $\beta=9/16$, and $\theta=-\pi/8$. For the controlled protocol set
\begin{equation}
(\phi,a)=
\begin{cases}
(\pi u/\alpha,\theta u/(2\alpha)),&0\le u\le\alpha,\\
(\pi,\theta/2),&\alpha\le u\le\beta.
\end{cases}
\end{equation}
At every time $W=e^{iaH}e^{i\phi R}$, $g_1=Z_1Z_2$, and $g_2=WZ_2Z_3W^\dagger$. Throughout the readout window $g_2=Z_2Z_3$. Unravel detector 2 by Eq.~\eqref{eq:gate-click}; the total counting rate is $2\kappa$, since $\sum_sJ_{2,s}^\dagger J_{2,s}=2\kappa I$. The retained symbol $y$ is no click, first positive click, or first negative click from this finite-rate instrument. Later clicks and the entire parity-1 record are hidden.

Let $\rho_0,\rho_+,\rho_-$ be the three unnormalized joint blocks during the window, with $\rho_0(\alpha)=\rho(\alpha)$ and $\rho_\pm(\alpha)=0$. If $\mathcal L_{\rm env}$ denotes the environmental generator, their exact continuous equations are
\begin{align}
\dot\rho_0&=\mathcal L_{\rm env}(\rho_0)
+\kappa\mathcal D[g_1](\rho_0)-2\kappa\rho_0,\\
\dot\rho_s&=\mathcal L_{\rm env}(\rho_s)
+\kappa\sum_a\mathcal D[g_a](\rho_s)
+2\kappa P_{2,s}\rho_0P_{2,s}.
\label{eq:gate-record-blocks}
\end{align}
Adding the three branch equations gives the original averaged generator, with no-click mass $\exp[-2k(\beta-\alpha)]$, $k=\kappa T$. The shared acquisition latch has three states; the budget $m$ counts labels persisting after $\beta$, separately from the common clock and acquisition latch.

\paragraph{Bounded geometric continuations.}
The selected suffix, for $v=(u-\beta)/(1-\beta)$, is
\begin{align}
\phi_b(v)&=\pi(1+v),\\
a_b(v)&=\theta(1+v)/2+b\sin(2\pi v),\\
\mathcal B&=\{-4,-3,\ldots,4\}/16.
\label{eq:gate-suffix-library}
\end{align}
The amplitudes are dimensionless radians. Every suffix is continuous with the common prefix and ends at the original projector. The known response boundary permits a change in path velocity; all operators and velocities on each segment are bounded. The detector family, duration, library, and decoder are the same for every memory budget.

The suffix holonomy depends on $b$. For the no-excursion code sector,
\begin{equation}
\begin{aligned}
A&=i\dot a\cos(2\phi)Z_L,\quad \int_0^T A\,dt=i\pi bZ_L,\\
U_{\rm geo}(b)&=e^{i(\theta-\pi b)Z_L}.
\end{aligned}
\label{eq:suffix-holonomy}
\end{equation}
Thus $b=0$ is the intended ideal $T$ loop; other members generate physical geometric compensations when chosen from the record. At finite strength, the complete branch channel rather than its ideal holonomy is scored against the same fixed $T_L$.

For the three Choi branches $C_y$ at $\beta$, let $\mathcal S_b$ be the joint suffix channel. The score used in Eq.~\eqref{eq:gate-policy-optimum} is
\begin{equation}
f(y,b)=\Tr\!\left[\ket{\Phi_{T_L}}\bra{\Phi_{T_L}}
(\id\otimes\mathcal D_{\rm std}\circ\Tr_E\circ\mathcal S_b)(C_y)\right],
\end{equation}
with $\Tr_E$ omitted for independent flips. Fix the at most $m$ library actions assigned to retained labels. Each record symbol can be sent to the label giving its largest score. Maximizing over the action set proves Eq.~\eqref{eq:gate-policy-optimum}; linearity makes deterministic assignments sufficient. The resulting optimum is global over the declared finite library and single response boundary.

At $k=64,256$ in both environments, the optimal two-state map chooses $b=0$ after no click or a positive click, and $b=-1/4$ after a negative click. The best blind choice is $b=-1/16$, except for the bath at $k=256$, where it is $b=0$. At $256$, two and three persistent states select the same policy. Without environmental noise, $(F_1^{\mathrm{gate},*},F_2^{\mathrm{gate},*})=(0.822278,0.892426)$ at $64$ and $(0.925635,0.961346)$ at $256$. The record therefore diagnoses finite-strength geometric excursions even in the absence of bath memory.

\section{A compact dictionary for processes, instruments, and contractions}
\label{app:continuum-primer}

\subsection{The cPT and its cMPS and Choi representations}
The operational definition is the smoothed-marginal definition of Ref.~\cite{dilation}, with the input and output left open. Use the operation basis $\mathcal E_0=\id$, $\mathcal E_\nu$ of Eq.~\eqref{eq:basis-dual} and $q=d_S^4-1$. An open cPT is a boundary-valued vector
\begin{equation}
 \U^\partial\in
 \Gamma\bigl(L^2([0,T])\otimes\mathbb C^q\bigr)
 \otimes\mathcal B(\cH_i\otimes\cH_o),
 \label{eq:primer-fock}
\end{equation}
with the square-summability norm in Eq.~\eqref{eq:opennorm}. Its coefficients $J^{(n)}_{\bm\nu}(\bm t)$ are the Choi matrices of the response to ordered insertions. Identity labels are restored by removing their slots. For every finite ordered set of slots, smoothing their joint coefficients against a nonnegative normalized temporal density must give a positive, normalized, causal comb. These marginals must be mutually compatible, including after identity insertion. Causality means that future deterministic operations cannot change earlier outcome probabilities. Positivity is complete positivity with respect to an arbitrary boundary reference. Fixing an input state and tracing the output gives the scalar process and closed norm in Eq.~\eqref{eq:closednorm}, whose vacuum coefficient is one. Retaining the boundaries gives the no-intervention channel Choi operator in the zero sector, with output partial trace $I_i$.

For a bounded system--environment generator $\mathcal L_{SE}(t)$, let $\mathcal U(t,s)$ be its propagator and $\iota_\sigma(X)=X\otimes\sigma_E$. A regular realization has coefficients
\begin{equation}
 \begin{split}
 \mathcal K_{\bm\nu}(\bm t)={}&\Tr_E\circ\mathcal U(T,t_n)
 \circ\mathcal E_{\nu_n}^{SE}\circ\cdots\\[-2pt]
 &\circ\mathcal E_{\nu_1}^{SE}\circ\mathcal U(t_1,0)
 \circ\iota_\sigma,\qquad
 J^{(n)}_{\bm\nu}=J(\mathcal K_{\bm\nu}),
 \end{split}
 \label{eq:primer-coefficients}
\end{equation}
where the omitted factors are the propagators between successive insertions and $\mathcal E_\nu^{SE}=\mathcal E_\nu\otimes\id_E$. Individual insertion maps need not be CP; physicality concerns their complete comb. These ordered products are precisely the coefficients of the cMPS
\begin{equation}
 \begin{gathered}
 |\Phi[Q,\{R_\nu\},B]\rangle=
 \Tr_{\rm aux}\!\left[B\,\mathcal T e^{\int_0^T K(t)dt}\right]|\Omega\rangle,\\
 K(t)=Q(t)\otimes I+\sum_\nu R_\nu(t)\otimes\psi_\nu^\dagger(t).
 \end{gathered}
 \label{eq:primer-cmps}
\end{equation}
For this process realization, the auxiliary space is the SE Liouville space, with $\vket X=\sum_{ab}X_{ab}|a\rangle\otimes|\overline b\rangle$ and $\operatorname{vec}(AXB^\dagger)=(A\otimes\overline B)\vket X$. For Hamiltonian dynamics its dimension is $(d_Sd_E)^2$ and
\begin{equation}
 Q=\mathbb H_{SE}(t)=-iH_{SE}(t)\otimes I_{SE}
       +iI_{SE}\otimes H_{SE}(t)^*.
 \label{eq:primer-liouville-drift}
\end{equation}
Here $H_{SE}$ is the physical Hermitian Hamiltonian and $Q$ is its Liouville generator. More generally, $Q$ is the matrix of $\mathcal L_{SE}$, while $R_\nu$ is the matrix of $\mathcal E_\nu^{SE}$. The boundary $B$ encodes initial preparation and final contraction; leaving those boundaries open gives Eq.~\eqref{eq:primer-coefficients}. Bounded Hamiltonian SE dynamics gives the process-canonical subclass of Refs.~\cite{continuum,dilation}. Basis rescalings relative to those references are specified in Appendix~\ref{app:stability}.

To construct the continuum Choi operator following Ref.~\cite{dilation}, split $\nu=(a,b)$ into left and right Pauli actions and put $\gamma=\Gamma(L^2([0,T])\otimes\mathbb C^{d_S^2-1})$. Its boundary-valued kernel on $\gamma$ is
\begin{equation}
 [\widehat\U^\partial]^{(k;k')}_{\bm a;\bm b}(\bm t;\bm s)
 =J^{(k+k')}_{(\bm a,0),(0,\bm b)}(\bm t,\bm s),
 \label{eq:primer-choi-kernel}
\end{equation}
where the two time strings are chronologically merged. Single-contour labels determine the kernel; Choi duality reconstructs the cross-species data on physical processes~\cite{dilation}. For regular finite-bond realizations this gives the continuum Choi/cMPO kernel by reorganizing the insertion coefficients. Density-operator vectorization used in numerical propagation is a separate representation step. The dual instrument is generally a kernel form, with an operator-trace pairing when bounded.

\subsection{Continuous instruments and their contraction}
A continuous instrument assigns CP interventions to measurable record events, is countably additive, and sums to a deterministic intervention. With quantum boundaries open it produces a CP-map-valued measure; its trace gives the outcome probabilities. For dominated records write $\mathcal N_y^S=\U^\partial\star\mathcal I_y$. Then
\begin{equation}
 C_y=\frac{(\id_R\otimes\mathcal N_y^S\circ\operatorname{Ad}_V)
 (\vket{I_L}\vbra{I_L})}{d_L},
 \qquad p(y)=\Tr C_y.
 \label{eq:primer-open-branch}
\end{equation}
Thus contraction leaves a reference-assisted channel. When the process and the dual instrument effect admit regular cMPS parameters $(Q_p,R_{p\nu},B_p)$ and $(Q_i,R_{i\nu},B_i)$, their overlap is computed using
\begin{align}
 \mathbb W_{pi}&=Q_p\otimes I+I\otimes\overline{Q_i}
 +\sum_\nu R_{p\nu}\otimes\overline{R_{i\nu}},\\
 \langle\Phi_i|\Phi_p\rangle
 &=\Tr\!\left[(B_p\otimes\overline{B_i})
 \mathcal T e^{\int_0^T\mathbb W_{pi}(t)dt}\right].
 \label{eq:primer-overlap}
\end{align}
An insertion at time $t$ places $R_{p\nu}(t)\otimes I$ between the transfer propagators on either side, with $\xi_0(t)=I$ and $\xi_\nu(t)=\psi_\nu(t)$ for $\nu\ne0$. Leaving the boundary indices uncontracted gives Eq.~\eqref{eq:primer-open-branch}. This is the open-boundary form of the cMPS overlap calculus in Eqs.~(14) and (F15) of Ref.~\cite{continuum}. It pairs two different Fock-space objects, a process and an instrument effect. The auxiliary tensor product in $\mathbb W_{pi}$ is the product of their bond spaces.

For counting monitoring, separate the instrument's no-click generator and click maps,
\begin{align}
 \mathcal G_0(X)&=-i[H_c,X]-\tfrac12\sum_x\{L_x^\dagger L_x,X\},\\
 \mathcal J_x(X)&=L_xXL_x^\dagger,\qquad
 \mathcal J_x=\sum_{\nu=0}^q j_{x\nu}\mathcal E_\nu.
 \label{eq:primer-jump}
\end{align}
Expand $\mathcal G_0=\sum_{\nu=0}^q g_\nu\mathcal E_\nu$ using the dual basis in Eq.~\eqref{eq:basis-dual}. The no-click effect is $\langle\mathcal I_0|=\langle\Phi[\overline{g_0},\{\overline{g_\nu}\}_{\nu\ne0},1]|$; the conjugates convert ket parameters into the bra's coefficients. A record $y=((x_1,t_1),\ldots,(x_n,t_n))$ has effect
\begin{equation}
 \langle\mathcal I_y|=\langle\mathcal I_0|
 \prod_{j=1}^n\left[\sum_{\nu=0}^q j_{x_j\nu}\xi_\nu(t_j)\right].
 \label{eq:primer-click-effect}
\end{equation}
Its contraction propagates with $\mathcal L_{SE}+\mathcal G_0$ between prescribed clicks and $\mathcal J_{x_j}$ at each click. The process generator $\mathcal L_{SE}$ supplies hidden noise; $g_\nu$ and $j_{x\nu}$ specify the measurement generator and click maps. Equations~(85)--(87) and (F16)--(F17) of Ref.~\cite{continuum} give the closed-boundary formulas. Their section ``Continuous measurements: jump and diffusive cases'' constructs the corresponding dual effects in the operation-label Fock space.

For homodyne monitoring use the linear unnormalized instrument relative to a common Wiener measure,
\begin{equation}
 d\widetilde\rho=(\mathcal L_{SE}+\mathcal G)\widetilde\rho\,dt
 +\sum_\alpha\mathcal B_\alpha\widetilde\rho\,dW_\alpha.
 \label{eq:primer-linear-instrument}
\end{equation}
Here $\mathcal G$, $\mathcal B_\alpha$ and their dual-basis coefficients are given in Eq.~\eqref{eq:instrument-dictionary}. In the operation-field expansion the instrument pairs an occupied slot with
\begin{equation}
 d\ell_\nu(t)=a_\nu(t)dt+\sum_\alpha B_{\nu\alpha}(t)dW_\alpha(t),
 \qquad \nu\ne0.
 \label{eq:primer-wick-slots}
\end{equation}
Ordered products are Wick/It\^o integrals; for the traceless monitors used here the identity coordinate contributes $e^{\int a_0dt}$. This gives the reference-measure interpretation of formulas (91)--(92), (F37), and (F39)--(F40) of Ref.~\cite{continuum}. The white-noise current is interpreted distributionally. Appendix~\ref{app:stability} establishes the weighted stochastic pairing and its CP normalization on the physical class.

Normalizing the branch map gives $\rho_y=\mathcal N_y(\rho)/\Tr\mathcal N_y(\rho)$. On a Markovian embedding retaining the persistent bath, its evolution under the physical record law is
\begin{align}
d\rho_{\rm cond}&=\mathcal L(\rho_{\rm cond})dt
+\sum_\alpha\mathcal H[L_\alpha](\rho_{\rm cond})dW_\alpha,\\
dY_\alpha&=\Tr[(L_\alpha+L_\alpha^\dagger)\rho_{\rm cond}]dt+dW_\alpha,
\end{align}
where $\mathcal L$ includes unconditional monitoring and
\begin{equation}
\mathcal H[L](\rho)=L\rho+\rho L^\dagger-\Tr[(L+L^\dagger)\rho]\rho.
\end{equation}
Here $W_\alpha$ are innovations under the physical law, distinct from the reference Wiener coordinates used for the linear instrument. For record-independent $\mathcal L$, integration over records gives $d\bar\rho=\mathcal L(\bar\rho)dt$. Finite-state feedback instead propagates the joint controller blocks, retaining the correlations needed for future controls.

\subsection{Feedback and finite retained records}
Feedback is encoded by choosing future instrument coefficients from the past record. For example, a click $x$ at $t$ followed by $\operatorname{Ad}_{U_x}=\sum_\nu u_{x\nu}\mathcal E_\nu$ at $t+\epsilon$ inserts the additional factor $\sum_\nu u_{x\nu}\xi_\nu(t+\epsilon)$ in Eq.~\eqref{eq:primer-click-effect}, as in Eq.~(95) of Ref.~\cite{continuum}. Bounded delayed pulses give the finite-duration version. In our finite-state architecture the retained label updates causally at the response boundaries and selects the next frozen instrument segment. Equation~\eqref{eq:transfer} gives the instrument-only Liouville generator; Eq.~\eqref{eq:contracted-controller-blocks} adds the cPT process generator on the shared system before propagating the record-conditioned open boundary. The same separation applies to time-dependent path control.

Finally, retain a label $r=f(y)$ by summing the unnormalized branches,
\begin{equation}
 C_r=\int_{f^{-1}(r)}C_y\,d\mu(y),\qquad
 \mathcal M=\int\mathcal D_{f(y)}\circ\mathcal N_y\,d\mu(y).
 \label{eq:primer-coarse-record}
\end{equation}
Discarding the record sets $f$ constant; retaining a persistent state preserves the corresponding classical blocks. Deterministic propagation of those blocks sums hidden trajectories without erasing accessible information. Equations~\eqref{eq:primer-open-branch}--\eqref{eq:primer-coarse-record} are the computational bridge from the cPT to the recovery objectives used throughout the paper.

\section{Stabilizer decisions, ties, and causal realizability}
\label{app:general-memory}
\subsection{Optimal recovery with a terminal syndrome}
Choose a Pauli representative $T_s$ of each terminal syndrome. Every Pauli
error restricts to $T_s\overline P_\ell V$ on the code, up to a stabilizer
and an irrelevant phase. Fixed QND monitoring multiplies this operator by
a record-dependent scalar along each classical error history. Summing the
inaccessible histories therefore gives
\begin{equation}
 \mathcal N_y(\rho)=\sum_{s,\ell}w_{s,\ell}(y)
 T_sV P_\ell\rho P_\ell^\dagger V^\dagger T_s^\dagger.
 \label{eq:pauli-branch-general}
\end{equation}
The orthogonal syndrome blocks may be measured at recovery without changing
this channel. Let $D=2^k$ and let $K_b$ be Kraus operators of a logical
CPTP recovery on one block. For the $D^2$ logical Paulis,
\begin{equation}
 f_\ell=D^{-2}\sum_b|\Tr(K_bP_\ell)|^2\geq0,
 \qquad \sum_\ell f_\ell=1.
\end{equation}
The second identity follows from orthogonality of the Pauli operator basis
and $\sum_bK_b^\dagger K_b=I$. Hence
$\sum_\ell w_\ell f_\ell\leq\max_\ell w_\ell$, attained by a Pauli inverse.
Summing the blocks and records proves Eq.~\eqref{eq:general-pauli-score}.
The corrected channel is a logical Pauli channel with identity weight $F$.
Its half diamond distance from the identity is $1-F$: the Pauli convex
decomposition gives the upper bound and a maximally entangled input gives
the matching lower bound. Every other recovery has half diamond error at
least its entanglement infidelity. This proves both optimal objectives.

\subsection{Common maximizers and terminal-label counting}
For a cell $A$ and syndrome $s$, choose a maximizer $\ell_0$ of the
integrated weights. Its score deficit is
\begin{equation}
 \int_A\left[\max_\ell w_{s,\ell}(y)-w_{s,\ell_0}(y)\right]d\mu(y).
\end{equation}
The integrand is nonnegative. It vanishes exactly when $\ell_0$ belongs
to the maximizing set almost everywhere in $A$. This proves the
common-optimal-correction criterion including all ties and zero weights.
Pairwise overlap of maximizing sets is insufficient; one maximizing class
must be common to the entire cell for each syndrome.

There are $L=4^k$ possible decisions at each of $S=2^{n-k}$ syndromes.
Selecting a maximizer measurably, with a fixed rule at ties, partitions
the record space into at most $L^S$ profiles, so
\begin{equation}
 m_{\rm label}^*\leq (4^k)^{2^{n-k}}.
 \label{eq:coarse-label-bound}
\end{equation}
With unique maximizing profiles on positive-probability sets, no lossless
cell can contain two different profiles. Their number is therefore the
exact terminal-label count. With ties, the exact answer is the smallest
number of cells admitting the above common selections. No equivalence
relation is inferred from pairwise sharing of an optimum.

\subsection{Finite-horizon causal state count}
For finite window alphabets, let $y=(y_1,\ldots,y_J)$ and let
$\mathcal A(y)$ be the set of all syndrome-conditioned maximizing profiles
at that record. At zero-probability records all profiles are allowed.
Choose any selection $D(y)\in\mathcal A(y)$. At layer $j$, two prefixes
$h,h'$ are equivalent when their residual profile functions agree:
\begin{equation}
 h\equiv_{j,D}h'
 \quad\Longleftrightarrow\quad
 D(hz)=D(h'z)\quad\forall z.
 \label{eq:residual-decision-equivalence}
\end{equation}
Let $N_j(D)$ be the number of residual functions, including the single
initial function at $j=0$. The exact minimal persistent-state count, with
time-dependent updates and the terminal syndrome available only to the
decoder, is
\begin{equation}
 m_{\rm causal}^*=\min_{D(y)\in\mathcal A(y)}\max_{0\leq j\leq J}N_j(D).
 \label{eq:causal-residual-count}
\end{equation}
To prove necessity, prefixes reaching the same state of a deterministic
attaining controller have the same final decoder after every continuation;
their selected profiles therefore agree as in
Eq.~\eqref{eq:residual-decision-equivalence}. For sufficiency, store the
residual-function class. Appending a symbol defines a well-defined next
class, since equality of residual functions is preserved under restriction
to that symbol. At the final layer the state specifies $D(y)$, which is
optimal in each syndrome. The same state labels can be reused across
layers, proving the maximum in Eq.~\eqref{eq:causal-residual-count}.

Randomization does not lower this finite-alphabet count. The score is
affine in each stochastic update kernel with the other kernels and decoder
fixed; replacing kernels successively by deterministic extreme points
cannot lower it. Compactness of the finite product of simplices and CPTP
decoder sets gives attainment. A deterministic optimal decoder can then
be selected syndrome by syndrome from its common maximizing classes.
This also connects the exact count to the supremum definition in
Eq.~\eqref{eq:three-memory-resources}.

\subsection{CSS factorization}
Under Eq.~\eqref{eq:css-posterior-factorization}, maximization over the two
logical components separates, and integration gives $F_XF_Z$. Uniqueness
makes every full decision profile an ordered pair of component profiles.
Product record support realizes every such pair, proving
Eq.~\eqref{eq:css-label-product}. Updating the ordered pair of component
persistent states gives the causal product construction. For achieved
component fidelities $\widehat F_X,\widehat F_Z$,
\begin{equation}
 F_XF_Z-\widehat F_X\widehat F_Z
 =(F_X-\widehat F_X)F_Z+
 \widehat F_X(F_Z-\widehat F_Z)
 \leq\varepsilon_X+\varepsilon_Z.
\end{equation}
This establishes the approximate product bound without requiring equality
of minimal causal state counts.

\section{Proof of the diffusive memory theorem}
\label{app:diffusive-memory}
\subsection{Sufficient statistics and nine terminal labels}
For a fixed stabilizer eigenvalue $\sigma$, the normalized current obeys
$dY_t=2\sqrt\kappa\sigma\,dt+dW_t$. Its path likelihood with respect to
Wiener measure is $\exp(2\sqrt\kappa\sigma Y_\tau-2\kappa\tau)$.
Thus $Z=Y_\tau/\sqrt\tau$ is sufficient, giving
Eq.~\eqref{eq:diffusive-two-window}. The three hidden flips leave the
stabilizer eigenvalues fixed within the two respective windows.

For syndrome $s=(s_1,s_2)$, its $e_2=0$ representative is
$(s_1,0,s_2)$ and the competing history is its bitwise complement. Their
prior weights are $a_s=p^{s_1+s_2}q^{3-s_1-s_2}$ and
$b_s=p^{3-s_1-s_2}q^{s_1+s_2}$. Their likelihood ratio gives
Eq.~\eqref{eq:nine-decision-odds}. In the invertible coordinates
$A=x+y$ and $B=y-x$, the two decisions with $s_1=0$ change at
$A=-3h_p,-h_p$, while those with $s_1=1$ change at $B=-h_p,h_p$.
Each coordinate has three decision intervals, giving exactly $3\times3=9$
profiles. The joint Gaussian mixture has strictly positive density on the
whole plane, so every region has positive probability and ties have
measure zero. Theorem~\ref{thm:general-pauli-memory} proves that these
nine labels are necessary and sufficient for terminal compression.

The two Gaussian means for a fixed syndrome are opposite vectors of norm
$M=\sqrt{\mu_1^2+\mu_2^2}$. Projection on their common axis gives the
binary success $S(a_s,b_s,M)$. The $s=00$ block contributes
$S(q^3,p^3,M)$. Each other block contributes $S(pq^2,p^2q,M)$;
$S$ is symmetric in its first two arguments. This proves
Eq.~\eqref{eq:diffusive-full-score}.

\subsection{A uniform lower bound for all finite-state controllers}
Let $\varphi$ be the standard normal density. Consider only syndrome
$00$, with prior weights $a=q^3$, $b=p^3$. Its first-readout density
and posterior are
\begin{align}
 g(z)&=a\varphi(z-\mu_1)+b\varphi(z+\mu_1),\nonumber\\
 \theta(z)&=\frac{a\varphi(z-\mu_1)}{g(z)},\nonumber\\
 z(\theta)&=\frac{\log(\theta/(1-\theta))-\log(a/b)}{2\mu_1}.
\end{align}
The unnormalized posterior density is
\begin{equation}
 \rho(\theta)=\frac{g(z(\theta))}{2\mu_1\theta(1-\theta)}.
 \label{eq:posterior-density-lower}
\end{equation}
After the first readout, the optimal success with the full second readout
is the binary value
\begin{align}
 V(\theta)&=\int\max\{\theta\varphi(v-\mu_2),
                    (1-\theta)\varphi(v+\mu_2)\}\,dv,\nonumber\\
 t(\theta)&=\frac{\log((1-\theta)/\theta)}{2\mu_2},\nonumber\\
 V''(\theta)&=\frac{\varphi(\mu_2-t(\theta))+
                        \varphi(\mu_2+t(\theta))}
                    {2\mu_2\theta(1-\theta)}>0.
 \label{eq:binary-value-curvature}
\end{align}

Fix $I=[1/4,3/4]$ and let $\rho_0=\min_I\rho>0$ and
$\nu_0=\min_I V''>0$. Grant the decoder the exact second readout and
the terminal syndrome, as well as a flag for $\theta\in I$ and the exact
$\theta$ outside $I$. This enlarged interface can only improve any
$m$-state controller. On $I$, write its randomized first update as
$K_c(\theta)$, $c=1,\ldots,m$, with $\sum_cK_c=1$. The deficit for this
enlarged interface is the Jensen gap of $V$ in each persistent state.
If $a_c$ denotes that state's posterior mean restricted to $I$, strong
convexity gives
\begin{align}
 \mathrm{loss}&\geq\frac{\nu_0}{2}\sum_c
      \int_I\rho(\theta)K_c(\theta)(\theta-a_c)^2d\theta\nonumber\\
 &\geq\frac{\nu_0\rho_0}{2}
      \int_I\min_c(\theta-a_c)^2d\theta
 \geq\frac{\nu_0\rho_0|I|^3}{24m^2}.
 \label{eq:uniform-diffusive-loss-proof}
\end{align}
For the last inequality, the nearest-center cells on a line are intervals.
An interval of length $l$ has squared-error integral at least $l^3/12$;
summing at most $m$ cells and convexity of the cube yield $|I|^3/(12m^2)$.
This argument allows arbitrary measurable randomized update cells.
All other syndrome deficits are nonnegative. The Pauli reduction bounds
every CPTP recovery by the same binary score, proving the lower bound
with $c=\nu_0\rho_0|I|^3/24$ uniformly over the entire controller class.
It therefore remains strict after taking its supremum.
For $p=0.1$ and $\mu_1=\mu_2=2$, the interval $I$ maps to
$z\in[-1.92258,-1.37326]$. Bounding each Gaussian density by its
endpoint minimum and using $\theta(1-\theta)\leq1/4$ gives
$\rho_0\geq0.00046038$ and $\nu_0\geq0.0600357$.
One may therefore take $c=1.43\times10^{-7}$ in
Eq.~\eqref{eq:diffusive-memory-bounds}.

\subsection{A matching constructive upper bound}
Set $u=(1+e^{-2\mu_1z})^{-1}$. For syndrome $s$, its first-readout
posterior is a fractional-linear function $\theta_s(u)$; for $s_1=0$,
\begin{equation}
 \theta_s(u)=\frac{a_su}{a_su+b_s(1-u)},
\end{equation}
and for $s_1=1$ replace $u$ by $1-u$. In either case its Lipschitz
constant is bounded by $L_s=\max\{a_s/b_s,b_s/a_s\}<\infty$.
The curvature in Eq.~\eqref{eq:binary-value-curvature} tends to zero as
$\theta\to0,1$: its Gaussian dependence on the logarithmic odds dominates
the reciprocal factor $1/[\theta(1-\theta)]$. Hence
$M_V=\sup_{0<\theta<1}V''(\theta)<\infty$.
An explicit finite bound follows by writing $v=\log(\theta/(1-\theta))$:
\begin{align}
 V''(\theta)&=\frac{4\varphi(\mu_2)}{\mu_2}
     e^{-v^2/(8\mu_2^2)}\cosh^3(v/2),\nonumber\\
 M_V&\leq\frac{4}{\mu_2\sqrt{2\pi}}e^{4\mu_2^2}.
 \label{eq:explicit-curvature-upper}
\end{align}
Here $\cosh(v/2)\leq e^{|v|/2}$ and completing the square gives the bound.

Quantize $u$ into $B$ equal bins. In each bin the oscillation of
$\theta_s$ is at most $L_s/B$, so its conditional variance is at most
$L_s^2/(4B^2)$. The upper Taylor bound on the Jensen gap gives
\begin{equation}
 F_{\rm path}^{\rm diff}-F_{B,\rm quant}
 \leq\frac{M_V}{8B^2}\sum_s(a_s+b_s)L_s^2.
 \label{eq:diffusive-constructive-upper}
\end{equation}
The first boundary stores the bin. At the second boundary, let
$\alpha_{s,i}$ and $\beta_{s,i}$ be the bin probabilities under first
means $(-1)^{s_1}\mu_1$ and its opposite. The decision for syndrome $s$
compares $Z_2$ with
\begin{equation}
 t_{s,i}=\frac{\log[b_s\beta_{s,i}/(a_s\alpha_{s,i})]}{2\mu_2}.
 \label{eq:quantized-second-threshold}
\end{equation}
For $s=00,01$ the likelihood ratio $\alpha_{s,i}/\beta_{s,i}$ is the
same and $a_{00}/b_{00}>a_{01}/b_{01}$, giving a fixed order of their
two thresholds. The same holds for the pair $10,11$.
Thus the second update always outputs one of the same nine ordered
decision profiles; at most $\max\{B,9\}$ states suffice across both layers.
For $m\geq9$, take $B=m$ in
Eq.~\eqref{eq:diffusive-constructive-upper}. This proves the matching
upper bound with the finite constant
$C=M_V\sum_s(a_s+b_s)L_s^2/8$ and completes the theorem.

\subsection{Achievable scores and fixed-controller deployment}
For any first-stage interval grid, including the two plotted schemes,
Gaussian differences compute $\alpha_{s,i},\beta_{s,i}$ exactly in
terms of $\Phi$. Under the nominal model the achieved fidelity is
\begin{equation}
 F_{B,\rm quant}=\sum_{s,i}\left[
 a_s\alpha_{s,i}\Phi(\mu_2-t_{s,i})+
 b_s\beta_{s,i}\Phi(\mu_2+t_{s,i})\right].
 \label{eq:quantized-gaussian-score}
\end{equation}
For deployment at a perturbed hidden-flip probability $p'$, keep every
grid edge and threshold $t_{s,i}$ fixed at its nominal design. Replace
only $a_s,b_s$ outside the Gaussian probabilities by their true prior
weights. This evaluates the same controller on the true process, rather
than redesigning it at each plotted point. The finite-history mixture
also gives the exact uniform total-variation perturbation modulus in
Sec.~\ref{sec:certified-deployment}.

\bibliographystyle{apsrev4-2}
\makeatletter
\immediate\write\@auxout{\string\citation{strategicCodeBibControl}}
\makeatother
\bibliography{references}

@CONTROL{strategicCodeBibControl,
  title = {0}
}

@misc{lanka_holonomic,
  author = {Lanka, Anirudh and Garcia-Nila, Juan and Brun, Todd A.},
  title = {Continuous measurement-based holonomic quantum computation},
  year = {2025},
  eprint = {2510.06725},
  archivePrefix = {arXiv},
  primaryClass = {quant-ph},
  doi = {10.48550/arXiv.2510.06725}
}

@misc{lanka_steering,
  author = {Lanka, Anirudh and Garcia-Nila, Juan and Brun, Todd A.},
  title = {Steering paths mid-flight for fault-tolerance in measurement-based holonomic gates},
  year = {2026},
  eprint = {2603.02552},
  archivePrefix = {arXiv},
  primaryClass = {quant-ph},
  doi = {10.48550/arXiv.2603.02552}
}

@misc{lanka_bosonic,
  author = {Garcia-Nila, Juan and Lanka, Anirudh and Brun, Todd A.},
  title = {Holonomic quantum gates via continuous measurement in bosonic codes: GKP and cat states},
  year = {2026},
  eprint = {2608.11369},
  archivePrefix = {arXiv},
  primaryClass = {quant-ph},
  doi = {10.48550/arXiv.2608.11369}
}

@misc{strategic,
  author = {Tanggara, Andrew and Gu, Mile and Bharti, Kishor},
  title = {Strategic Code: A Unified Spatio-Temporal Framework for Quantum Error-Correction},
  year = {2024},
  eprint = {2405.17567},
  archivePrefix = {arXiv},
  primaryClass = {quant-ph},
  doi = {10.48550/arXiv.2405.17567}
}

@misc{continuum,
  author = {Wassner, Clara and Fuksa, Jon{\'a}{\v s} and Eisert, Jens and White, Gregory A. L.},
  title = {An Operational Continuum Limit of Quantum Combs},
  year = {2026},
  eprint = {2601.16974},
  archivePrefix = {arXiv},
  primaryClass = {quant-ph},
  doi = {10.48550/arXiv.2601.16974}
}

@misc{dilation,
  author = {Fuksa, Jon{\'a}{\v s} and Wassner, Clara and Eisert, Jens and White, Gregory A. L.},
  title = {Dilation Theorem for Continuum Quantum Stochastic Processes},
  year = {2026},
  eprint = {2609.35403},
  archivePrefix = {arXiv},
  primaryClass = {quant-ph},
  doi = {10.48550/arXiv.2609.35403}
}

@misc{temporalaccess,
  author = {Wang, He},
  title = {Temporal-Access Resource Theory for Continuous Process Tensors},
  year = {2026},
  eprint = {2610.13110},
  archivePrefix = {arXiv},
  primaryClass = {quant-ph},
  doi = {10.48550/arXiv.2610.13110}
}

@misc{cprc,
  author = {Wang, He},
  title = {Continuous-Process Randomized Compilation for Quantum Process Tensors},
  year = {2026},
  eprint = {2610.01521},
  archivePrefix = {arXiv},
  primaryClass = {quant-ph},
  doi = {10.48550/arXiv.2610.01521}
}

@article{ddft,
  author = {Ng, Hui Khoon and Lidar, Daniel A. and Preskill, John},
  title = {Combining Dynamical Decoupling with Fault-Tolerant Quantum Computation},
  journal = {Physical Review A},
  volume = {84},
  pages = {012305},
  year = {2011},
  doi = {10.1103/PhysRevA.84.012305}
}

@misc{approximate_dynamical,
  author = {Basak, Nirupam and Tanggara, Andrew and Mohan, Ankith and Paul, Goutam and Bharti, Kishor},
  title = {Approximate Dynamical Quantum Error-Correcting Codes},
  year = {2025},
  eprint = {2502.09177},
  archivePrefix = {arXiv},
  primaryClass = {quant-ph},
  doi = {10.48550/arXiv.2502.09177}
}

@article{knill1997,
  author = {Knill, Emanuel and Laflamme, Raymond},
  title = {Theory of Quantum Error-Correcting Codes},
  journal = {Physical Review A},
  volume = {55},
  pages = {900--911},
  year = {1997},
  doi = {10.1103/PhysRevA.55.900}
}

@article{barnum2002,
  author = {Barnum, Howard and Knill, Emanuel},
  title = {Reversing Quantum Dynamics with Near-Optimal Quantum and Classical Fidelity},
  journal = {Journal of Mathematical Physics},
  volume = {43},
  pages = {2097--2106},
  year = {2002},
  doi = {10.1063/1.1459754},
}

@article{fletcher2007,
  author = {Fletcher, Andrew S. and Shor, Peter W. and Win, Moe Z.},
  title = {Optimum Quantum Error Recovery Using Semidefinite Programming},
  journal = {Physical Review A},
  volume = {75},
  pages = {012338},
  year = {2007},
  doi = {10.1103/PhysRevA.75.012338}
}

@article{beny2010,
  author = {B{\'e}ny, C{\'e}dric and Oreshkov, Ognyan},
  title = {General Conditions for Approximate Quantum Error Correction and Near-Optimal Recovery Channels},
  journal = {Physical Review Letters},
  volume = {104},
  pages = {120501},
  year = {2010},
  doi = {10.1103/PhysRevLett.104.120501}
}

@article{ng2010,
  author = {Ng, Hui Khoon and Mandayam, Prabha},
  title = {Simple Approach to Approximate Quantum Error Correction Based on the Transpose Channel},
  journal = {Physical Review A},
  volume = {81},
  pages = {062342},
  year = {2010},
  doi = {10.1103/PhysRevA.81.062342}
}

@article{oreshkov2007,
  author = {Oreshkov, Ognyan and Brun, Todd A.},
  title = {Continuous Quantum Error Correction for Non-Markovian Decoherence},
  journal = {Physical Review A},
  volume = {76},
  pages = {022318},
  year = {2007},
  doi = {10.1103/PhysRevA.76.022318}
}

@article{pollock2018,
  author = {Pollock, Felix A. and Rodr{\'i}guez-Rosario, Carlos and Frauenheim, Thomas and Paternostro, Mauro and Modi, Kavan},
  title = {Non-Markovian Quantum Processes: Complete Framework and Efficient Characterization},
  journal = {Physical Review A},
  volume = {97},
  pages = {012127},
  year = {2018},
  doi = {10.1103/PhysRevA.97.012127}
}

@article{combs2009,
  author = {Chiribella, Giulio and D'Ariano, Giacomo Mauro and Perinotti, Paolo},
  title = {Theoretical Framework for Quantum Networks},
  journal = {Physical Review A},
  volume = {80},
  pages = {022339},
  year = {2009},
  doi = {10.1103/PhysRevA.80.022339}
}

@article{milz2021,
  author = {Milz, Simon and Modi, Kavan},
  title = {Quantum Stochastic Processes and Quantum Non-Markovian Phenomena},
  journal = {PRX Quantum},
  volume = {2},
  pages = {030201},
  year = {2021},
  doi = {10.1103/PRXQuantum.2.030201}
}

@article{stinespring2008,
  author = {Kretschmann, Dennis and Schlingemann, Dirk and Werner, Reinhard F.},
  title = {A Continuity Theorem for Stinespring's Dilation},
  journal = {Journal of Functional Analysis},
  volume = {255},
  pages = {1889--1904},
  year = {2008},
  doi = {10.1016/j.jfa.2008.07.023},
}

@book{wiseman2014,
  author = {Wiseman, Howard M. and Milburn, Gerard J.},
  title = {Quantum Measurement and Control},
  publisher = {Cambridge University Press},
  year = {2009},
  edition = {1},
  doi = {10.1017/cbo9780511813948},
}

@article{rosal2025,
  author = {Rosal, Alberto J. B. and Potts, Patrick P. and Landi, Gabriel T.},
  title = {Deterministic Equations for Feedback Control of Open Quantum Systems},
  journal = {Physical Review Letters},
  volume = {137},
  pages = {050401},
  year = {2026},
  doi = {10.1103/pg2p-h529},
  eprint = {2507.01934},
  archivePrefix = {arXiv}
}

@article{berk2021,
  author = {Berk, G. D. and Garner, A. J. P. and Yadin, B. and Modi, K. and Pollock, F. A.},
  title = {Resource theories of multi-time processes: A window into quantum non-Markovianity},
  journal = {Quantum},
  volume = {5},
  pages = {435},
  year = {2021},
  doi = {10.22331/q-2021-04-20-435},
}

@article{berk2023,
  author = {Berk, G. D. and Milz, S. and Pollock, F. A. and Modi, K.},
  title = {Extracting quantum dynamical resources: Consumption of non-Markovianity for noise reduction},
  journal = {npj Quantum Information},
  volume = {9},
  pages = {104},
  year = {2023},
  doi = {10.1038/s41534-023-00774-w},
}

@article{zambon2024,
  author = {Zambon, G.},
  title = {Process tensor distinguishability measures},
  journal = {Physical Review A},
  volume = {110},
  pages = {042210},
  year = {2024},
  doi = {10.1103/physreva.110.042210},
}

@article{dowling2024,
  author = {Dowling, N. and Modi, K. and Mu{\~n}oz, R. N. and Singh, S. and White, G. A. L.},
  title = {Capturing long-range memory structures with tree-geometry process tensors},
  journal = {Physical Review X},
  volume = {14},
  pages = {041018},
  year = {2024},
  doi = {10.1103/physrevx.14.041018},
}

@article{white2020,
  author = {White, G. A. L. and Hill, C. D. and Pollock, F. A. and Hollenberg, L. C. L. and Modi, K.},
  title = {Demonstration of non-Markovian process characterisation and control on a quantum processor},
  journal = {Nature Communications},
  volume = {11},
  pages = {6301},
  year = {2020},
  doi = {10.1038/s41467-020-20113-3},
}

@article{white2022,
  author = {White, G. A. L. and Pollock, F. A. and Hollenberg, L. C. L. and Modi, K. and Hill, C. D.},
  title = {Non-Markovian quantum process tomography},
  journal = {PRX Quantum},
  volume = {3},
  pages = {020344},
  year = {2022},
  doi = {10.1103/prxquantum.3.020344},
}

@book{barchielli2009,
  author = {Barchielli, A. and Gregoratti, M.},
  title = {Quantum Trajectories and Measurements in Continuous Time: The Diffusive Case},
  publisher = {Springer},
  series = {Lecture Notes in Physics},
  volume = {782},
  year = {2009},
  doi = {10.1007/978-3-642-01298-3},
}

@misc{costa2025,
  author = {Costa, F. and Yang, J.},
  title = {Continuous operations on non-Markovian processes},
  year = {2025},
  eprint = {2512.05884},
  archivePrefix = {arXiv},
  primaryClass = {quant-ph}
}

@article{ahn2002,
  author = {Ahn, Charlene and Doherty, Andrew C. and Landahl, Andrew J.},
  title = {Continuous quantum error correction via quantum feedback control},
  journal = {Physical Review A},
  volume = {65},
  pages = {042301},
  year = {2002},
  doi = {10.1103/physreva.65.042301}
}

@article{sarovar2004,
  author = {Sarovar, Mohan and Ahn, Charlene and Jacobs, Kurt and Milburn, Gerard J.},
  title = {Practical scheme for error control using feedback},
  journal = {Physical Review A},
  volume = {69},
  pages = {052324},
  year = {2004},
  doi = {10.1103/physreva.69.052324}
}

@article{chase2008,
  author = {Chase, Bradley A. and Landahl, Andrew J. and Geremia, JM},
  title = {Efficient feedback controllers for continuous-time quantum error correction},
  journal = {Physical Review A},
  volume = {77},
  pages = {032304},
  year = {2008},
  doi = {10.1103/physreva.77.032304}
}

@article{mabuchi2009,
  author = {Mabuchi, Hideo},
  title = {Continuous quantum error correction as classical hybrid control},
  journal = {New Journal of Physics},
  volume = {11},
  pages = {105044},
  year = {2009},
  doi = {10.1088/1367-2630/11/10/105044}
}

@article{mohseninia2020,
  author = {Mohseninia, Razieh and Yang, Jing and Siddiqi, Irfan and Jordan, Andrew N. and Dressel, Justin},
  title = {Always-On Quantum Error Tracking with Continuous Parity Measurements},
  journal = {Quantum},
  volume = {4},
  pages = {358},
  year = {2020},
  doi = {10.22331/q-2020-11-04-358}
}

@article{atalaya2021,
  author = {Atalaya, J. and Zhang, S. and Niu, M. Y. and Babakhani, A. and Chan, H. C. H. and Epstein, J. M. and Whaley, K. B.},
  title = {Continuous quantum error correction for evolution under time-dependent Hamiltonians},
  journal = {Physical Review A},
  volume = {103},
  pages = {042406},
  year = {2021},
  doi = {10.1103/physreva.103.042406}
}

@article{livingston2022,
  author = {Livingston, William P. and Blok, Machiel S. and Flurin, Emmanuel and Dressel, Justin and Jordan, Andrew N. and Siddiqi, Irfan},
  title = {Experimental demonstration of continuous quantum error correction},
  journal = {Nature Communications},
  volume = {13},
  pages = {2307},
  year = {2022},
  doi = {10.1038/s41467-022-29906-0}
}

@article{denhez2012,
  author = {Denhez, Gabrielle and Blais, Alexandre and Poulin, David},
  title = {Quantum-error-correction benchmarks for continuous weak-parity measurements},
  journal = {Physical Review A},
  volume = {86},
  pages = {032318},
  year = {2012},
  doi = {10.1103/physreva.86.032318}
}

@article{halaski2026,
  author = {Halaski, Anton and Koch, Christiane P.},
  title = {Obstacles to continuous quantum error correction via parity measurements},
  journal = {Physical Review Applied},
  volume = {26},
  pages = {034074},
  year = {2026},
  doi = {10.1103/8bkm-b48b}
}

@misc{vanhandel2005,
  author = {van Handel, Ramon and Mabuchi, Hideo},
  title = {Optimal Error Tracking via Quantum Coding and Continuous Syndrome Measurement},
  year = {2005},
  eprint = {quant-ph/0511221},
  archivePrefix = {arXiv}
}
\end{document}